\documentclass[aip,jcp,reprint,nobibnotes,superscriptaddress,floatfix, 
showpacs,citeautoscript,showkeys]{revtex4-1}

\usepackage{amsmath}

\DeclareMathAlphabet{\mathsf}{OT1}{phv}{b}{n}

\newcommand{\Vector}[1]{\ensuremath{\mathbf{#1}}}
\newcommand{\Tensor}[1]{\ensuremath{\mathsf{#1}}}

\newcommand{\rmd}{{\mathrm{d}}}

\newcommand{\crossVorg}{\ensuremath{%
         \setbox0=\hbox{$V$}
        V \kern-\wd0{\raise.3ex\hbox{$\relbar$}}}}

\newcommand{\crossVxx}[2]{%
	{\setbox0=\hbox{$#1#2V$}
         \setbox1=\hbox{$#1#2$}
         \setbox2=\hbox{$#1V$}
         \dimen1=\wd0
	 \advance\dimen1-\wd1
         \raise.2\ht0\hbox{$#1#2$}\kern-.4\wd0}}

\newcommand{\ie}{i.e.}

\usepackage{pifont}

\usepackage{graphicx}

\usepackage[version=4]{mhchem}

\usepackage{latexsym}
\usepackage{textcomp}
\usepackage{amssymb}
\usepackage{amsbsy}
\usepackage{amsmath}
\usepackage{epstopdf}
\usepackage{amsmath}
\usepackage{bm}
\usepackage{longtable}
\usepackage{subfigure}
\usepackage{epsfig}
\usepackage{dcolumn}

\usepackage{latexsym}
\usepackage{textcomp}
\usepackage{bm}
\usepackage[Euler]{upgreek}

\usepackage{longtable}
\usepackage{multirow}

\newcommand{\fnu}{\ensuremath{\Vector{F}_{\mathrm{\nu}}}}

\newcommand{\etas}{\ensuremath{\eta_{\mathrm{s}}}}

\DeclareMathAlphabet\mathbfcal{OMS}{cmsy}{b}{n}
\DeclareMathAlphabet{\mathbfsf}{\encodingdefault}{\sfdefault}{bx}{sl}

\newcommand{\bcal}[1]{\bf{\mathbfcal{#1}}}

\newcommand{\bOmega}{\ensuremath{\boldsymbol{\varOmega}}}
\newcommand{\bdelta}{\ensuremath{\boldsymbol{\delta}}}

\newcommand{\br}{\ensuremath{\bm{\Vector{r}}}}

\begin{document}
\title{Size, shape and diffusivity of a single Debye-H\"{u}ckel polyelectrolyte chain  in solution}
\date{\today}
\author{W Chamath Soysa}
\affiliation{Department of Chemical Engineering, Monash University,
Melbourne, VIC 3800, Australia}
\author{B. D\"{u}nweg}
\affiliation{Department of Chemical Engineering, Monash University, Melbourne, VIC 3800, Australia}
\affiliation{Max Planck Institute for Polymer Research, Ackermannweg 10, 55128 Mainz, Germany}
\affiliation{Condensed matter physics, TU Darmstadt, Karolinenplatz 5, 64289 Darmstadt, Germany}
\author{J. Ravi Prakash}
\email[Electronic mail: ]{ravi.jagadeeshan@monash.edu}
\affiliation{Department of Chemical Engineering, Monash University,
Melbourne, VIC 3800, Australia}


\begin{abstract}
Brownian dynamics simulations of a coarse-grained bead-spring chain model, with Debye-H\"{u}ckel electrostatic interactions between the beads, are used to determine the root-mean-square end-to-end vector, the radius of gyration, and various shape functions (defined in terms of eigenvalues of the radius of gyration tensor) of a weakly-charged polyelectrolyte chain in solution, in the limit of low polymer concentration. The long-time diffusivity is calculated from the mean square displacement of the centre of mass of the chain, with hydrodynamic interactions taken into account through the incorporation of the Rotne-Prager-Yamakawa tensor.  Simulation results are interpreted in the light of the OSFKK blob scaling theory~(R. Everaers, A. Milchev, and V. Yamakov, Eur. Phys. J. E 8, 3 (2002)) which predicts that all solution properties are determined by just two scaling variables---the number of electrostatic blobs $X$, and the reduced Debye screening length, $Y$. We identify three broad regimes, the ideal chain regime at small values of $Y$, the blob-pole regime at large values of $Y$, and the crossover regime at intermediate values of $Y$, within which the mean size, shape, and diffusivity exhibit characteristic behaviours. In particular, when simulation results are recast in terms of blob scaling variables, universal behaviour independent of the choice of bead-spring chain parameters, and the number of blobs $X$, is observed in the ideal chain regime and in much of the crossover regime, while the existence of logarithmic corrections to scaling in the blob-pole regime leads to non-universal behaviour.
\end{abstract}



 
\maketitle

\section{\label{sec:intro}Introduction}

Many synthetic polymers and most biopolymers are polyelectrolytes. Their use in a range of  industrial and biological applications makes a thorough understanding of their behaviour highly desirable.  While the scaling of many important observable properties of neutral polymers in solution has been successfully predicted by blob theories~\cite{degennes,Rubinstein2003}, much is yet to be understood with this approach even for the simple case of dilute polyelectrolyte solutions at equilibrium. The electrostatic blob, which sets the length scale at which the energy of electrostatic interactions is of order $k_BT$ (where $k_B$ is the Boltzmann constant and $T$ is the temperature), provides the basis for scaling theories of such a system~\cite{Degennes1976}. A commonly used scaling theory for polyelectrolyte solutions is the OSFKK scaling picture~\cite{Odijk1977,Skolnick1977,Khokhlov1982,Everaers2002} (after Odjik, Skolnick, Fixman, Khokhlov, and Khachaturian), which predicts the scaling of the mean chain size in different scaling regimes based on the dominating physics. Even though simulations  have examined the predictions of OSFKK scaling theory in detail~\cite{Everaers2002}, many questions remain unanswered and not many studies have applied the blob picture to properties other than the root-mean-square end-to-end vector.  The aim of this work is to carry out Brownian dynamics (BD) simulations in order to investigate the dependence of several  static and dynamic properties of dilute polyelectrolyte solutions on the intrinsic parameters that govern their behaviour. In particular, in addition to the mean size, we use the OSFKK scaling picture to interpret simulation predictions of the shape and diffusivity of polyelectrolyte chains. A variety of shape functions, defined in terms of the eigenvalues of the radius of gyration tensor, are used to examine changes in polymer shape in the different regimes of conformational phase space. Further, hydrodynamic interactions have been incorporated via the Rotne-Prager-Yamakawa tensor~\cite{Rotne1969,Yamakawa1970}, in order to obtain an accurate prediction of the chain diffusivity in various scaling regimes. The OSFKK scaling picture ignores the presence of logarithmic corrections in the blob-pole regime of the conformational phase diagram, which are predicted to be important by more refined scaling theories~\cite{Dobrynin2005}. For all the properties examined here, we explore whether logarithmic corrections can account for departures from OSFKK scaling in the blob-pole regime.

In most scaling theories, polyelectrolyte molecules are represented as freely-jointed chains with $N_\text{K}$ Kuhn steps, each of length $b_\text{K}$. The number of charges per Kuhn step, $f = \alpha  N_\text{m,K}$, where, $N_\text{m,K}$ is the number of monomers per Kuhn step, and $\alpha$ is the degree of ionization per chain, is used to characterize the extent of ionic group dissociation. Counterions and salt ions are not modelled explicitly, and the screening of monomer charges due to the presence of free ions is quantified via the Debye screening length, $l_\text{D}$. Another important length scale is the Bjerrum length, $l_\text{B}$, which is the distance at which the Coulomb energy between two unit elementary charges in the solvent is equal to the thermal energy $k_\text{B}T$, and is defined by, 
\begin{equation}
\label{eq:lb}
l_\text{B}=\frac{e^2}{4\pi \varepsilon_{0} \varepsilon_\text{r} k_\text{B}T}
\end{equation}
where $e$ is the charge of an electron, $\varepsilon_{0}$ is the vacuum permittivity, and $\varepsilon_\text{r}$ is the relative dielectric constant of the solvent. Such a representation of a dilute polyelectrolyte solution is referred to here as the bare-model, and in what follows, it is always assumed that the macromolecules are dissolved in a theta solvent (which is simulated by switching off excluded volume interactions). 

\begin{figure}[t]
\centering
\includegraphics[width=8cm,height=!]{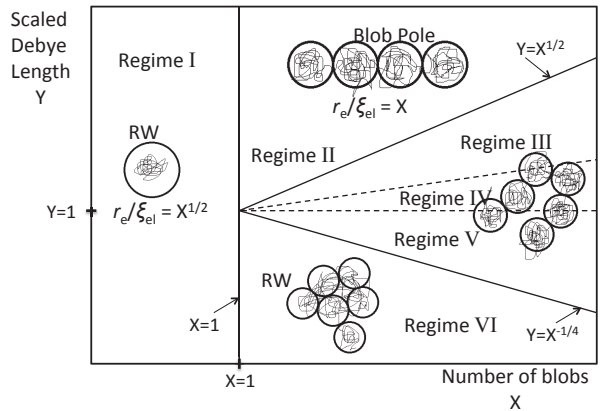}  
\caption{The phase diagram in the OSFKK scaling picture~\cite{Everaers2002}. Various regimes are marked I-VI, with scaling predictions for the end-to-end vector $r_\text{e}$, in each regime, listed in Table~\ref{table1}. Equations governing the boundaries between regimes are indicated in the figure, and the abbreviation ``RW'' implies random walk statistics are obeyed.}
\label{fig:f1}
\end{figure}

\begin{table*}[t]
\caption{\label{table1} Scaling expressions for the end-to-end vector in the various regimes shown schematically in Fig.~\ref{fig:f1}, in terms of bare and blob parameters (the value of the Flory excluded volume exponent has been approximated to be $3/5$). The variables $X$ and $Y$ are defined in Eqs.~(\ref{eq:x}) and (\ref{eq:y}), respectively. }
\begin{center}
\begin{tabular}{ccc}
 \hline \hline {Regime } &  \hspace{0.5in} $r_\text{e}/\xi_\text{el} $ {(blob parameters) \hspace{0.5in}} & $r_\text{e}/b_\text{K}$ (bare parameters) \\ 
 \hline I&  $X^{1/2}$ & $ N_\text{K}^{1/2}$ \\ 
 II & $X$ & $f^{2/3} \, {\hat l}_\text{B}^{1/3} \, N_\text{K}$ \\ 
 III & $Y X^{1/2}$ & $f^{2/3} \, {\hat l}_\text{B}^{1/3} \, {\hat l}_\text{D} \, N_\text{K}^{1/2}$ \\
 IV & $ Y^{3/5}X^{3/5}$ & $f^{8/15} \, {\hat l}_\text{B}^{4/15}  \,  {\hat l}_\text{D}^{3/5}  \, N_\text{K}^{3/5}$ \\ 
 V & $ Y^{2/5}X^{3/5}$ & $f^{2/5} \,  {\hat l}_\text{B}^{1/5} \, {\hat l}_\text{D}^{2/5} \,  
 N_\text{K}^{3/5} $ \\  
 VI & $ X^{1/2}$ & $ N_\text{K}^{1/2}$ \\ 
 \hline \hline
\end{tabular}
\end{center}
\end{table*}

The electrostatic blob denotes the length scale below which the conformation of a polyelectrolyte chain is practically unaffected by electrostatic interactions, since the electrostatic energy of the sub-chain within a blob is less than the thermal energy. This length scale can be used to divide the chain into a number blobs, $X$, of diameter, $\xi_\text{el}$, with random walk statistics followed within the blob, while the conformation of the chain of blobs as a whole depends on the electrostatic interactions between the blobs. In addition, within the OSFKK scaling picture,  the screening of electrostatic interactions is accounted for by the scaled variable $Y$, which is the ratio of the Debye length to the blob size. In terms of the bare-model parameters, the OSFKK expressions for the blob scaling variables are,~\cite{Everaers2002}
\begin{equation}
\label{eq:x}
X= \frac{N_\text{K}}{{\hat \xi}_\text{el}^{2}}
\end{equation}
and,
\begin{equation}
\label{eq:y}
Y=\frac{{\hat l}_\text{D}}{{\hat \xi}_\text{el}}
\end{equation}
where, ${\hat l}_\text{D} \equiv l_\text{D}/b_\text{K}$, and the scaled blob size ${\hat \xi}_\text{el} \equiv \xi_\text{el}/b_\text{K}$, is given by,
\begin{equation}
\label{eq:xi}
{\hat \xi}_\text{el} = \left( f^2 \, {\hat l}_\text{B} \right)^{-\frac{1}{3}}
\end{equation}
with, ${\hat l}_\text{B} \equiv l_\text{B}/b_\text{K}$. Note, for future reference, that it is sufficient to prescribe the reduced set of bare-model parameters $\{ N_\text{K}, f, {\hat l}_\text{B}, {\hat l}_\text{D} \}$ in order to determine the scaling variables $X$ and $Y$. 
The OSFKK scaling picture is briefly summarized in Fig.~\ref{fig:f1}, which is a phase diagram in $(X, Y)$ space, highlighting the different scaling regimes. Table~\ref{table1} lists the scaling expressions for the chain size, in terms of both blob and bare parameters, in each of the regimes. In regime I, when $X\ll 1$ (the blob is much larger than the chain), $r_\text{e}/\xi_\text{el} \propto X^{1/2}$, where $r_\text{e}$ is the root-mean-square end-to-end vector. The Debye length is irrelevant in this regime as thermal energy completely dominates the electrostatic interactions between the segments. When $X\gg 1$ (when the chain is composed of many electrostatic blobs) the $XY$ plane is divided into different regimes according to how large $Y$ is compared to $X$. The blob-pole conformation occurs in regime II, when $Y\rightarrow \infty$, and electrostatic interactions between blobs are unscreened. In this regime, $r_\text{e}/\xi_\text{el} \propto X$. When $Y$ decreases below $X^{1/2}$, electrostatic interactions act as short ranged repulsive forces and a crossover regime follows, which, according to Khokhlov and Khachaturian~\cite{Khokhlov1982} is decomposed into regimes III-V. Here the confirmations are governed by an intricate interplay of chain stiffness and excluded volume interactions (both of electrostatic origin). Since our numerical investigations lack the resolution to study the details of these regimes we collectively refer to them as the crossover regime. Finally when $Y\ll X^{-1/4}$ the Debye length is sufficiently small for the scaling to return to that of an ideal chain.  

\citet{Everaers2002} have carried out extensive single chain Monte Carlo simulations with Debye-H\"{u}ckel electrostatic interactions and their results support the {OSFKK} scaling picture for the end-to-end vector.\ ~\citet{ravip2008} used single chain Brownian dynamics simulations with Debye-H\"{u}ckel electrostatic interactions in order to investigate the scaling of the end-to-end vector and viscometric functions in simple shear flow. They showed that when Brownian dynamics simulation data were represented in terms of blob scaling variables, predictions independent of the choice of parameters in the bead-spring chain model were obtained. As in the present study, these Brownian dynamics simulations were unable to distinguish between the different regimes in the crossover region due to the computational cost of simulating the extremely long chains required to verify OSFKK scaling. By including a characteristic non-dimensional shear rate as an additional scaling variable in shear flow,\  ~\citet{ravip2008} were able to show that the equilibrium electrostatic blob model provides a framework to obtain universal scaling even for non-equilibrium properties. Both these simulation studies, however, did not attempt to describe the scaling of other static properties, or the diffusivity in the different regimes, nor did they examine the occurrence of logarithmic corrections in the blob-pole regime. 

The appropriateness of using a Debye-H\"{u}ckel potential to represent electrostatic interactions in Brownian dynamics simulations has been examined by ~\citet{stoltz2007}  \!\!\!\!\!\!\!\!{.} \, They compared the predictions of chain size as a function of Bjerrum length (at various concentrations in the dilute regime) of a bead-spring chain model with explicit counterions and pair-wise Coulomb interactions between charges, with those of a model with pair-wise Debye-H\"{u}ckel electrostatic interactions between beads on chains. They found that the results of both models are nearly identical for values of Bjerrum length roughly equal to or less than the distance between the beads on a chain (all of which were assumed charged). For larger values of Bjerrum length, in accord with Manning's theory~\cite{Manning1969}, they observed the onset of counterion condensation. We have adopted a Debye-H\"{u}ckel potential in this work, and as discussed in greater detail subsequently, chosen parameter values in the bead-spring chain model to ensure that the simulations always remain in the regime where this approximation is valid, and there is no counterion condensation. 

Property predictions from Brownian dynamics simulations of bead-spring chains depend on several model parameters, such as the number of beads, the charge on the beads, the finite extensibility parameter for the springs, and so on. A key issue is a rational choice of values for these parameters. The earlier BD simulations of mean size and viscometric functions by ~\citet{ravip2008} has shown the advantage of using blob scaling variables to interpret results of simulations since this leads to a description independent of the level of coarse-graining, which is extremely useful for comparing simulation results with experiments. In this study, we investigate the shape and diffusivity of polyelectrolyte chains in the various regimes of the phase diagram, by recasting BD simulation results in terms of blob scaling variables, and  examine their independence from the specific choice of bead-spring chain model parameters.

Many studies have shown that the shape of a neutral polymer chain, even at equilibrium, is not spherical about the centre of mass of the chain~\cite{Kuhn1934,Koyama1968,Solc1971,Solc1971a,Yoon1974,Kranbuehl1977,Theodorou1985,Aronovitz1986,Bishop1986,Rudnick1987,Bishop1989,Wei1990,Bishop1991,Zifferer1999,Wei1997,Haber2000,Steinhauser2005}. The nature of the asymmetry in chain shape has been examined in terms of a number of different quantities, such as the \textit{degree of prolateness}, the \textit{asphericity},  the \textit{acylindricity},  the \textit{shape anisotropy}, and so on, which are functions of the eigenvalues of the radius of gyration tensor, since the breaking of symmetry is reflected in the three eigenvalues differing from each other. The symmetry or otherwise of polyelectrolyte chain shapes, particularly in the different scaling regimes, has not yet been systematically investigated. Here, we examine if a universal description of polyelectrolyte chain shapes can be obtained, when BD simulation results are represented in terms of electrostatic blob scaling variables. 

While the OSFKK scaling picture ignores the presence of logarithmic corrections to the scaling of chain size with degree of polymerisation in the blob-pole regime, their existence has been derived in a number of different ways, ranging from Flory type energy minimisation arguments~\cite{Dobrynin2005}, to refined scaling theories that account for the nonuniform stretching of polyelectrolyte chains along the elongation axis~\cite{degennes,Liao2003}. In Appendix~\ref{sec:App} (for the sake of completeness), we have used a Flory type argument to show how logarithmic corrections arise in regime II. In particular, it can be shown that $r_\text{e}$ obeys the following scaling expression in terms of the blob scaling variable $X$,\begin{equation}
\label{eq:relogX}
\frac{r_\text{e}}{\xi_\text{el}} \sim X \left[  \ln X \right]^{\frac{1}{3}}
\end{equation} 
~\citet{Liao2003}  have carried out molecular dynamic simulations of bead-spring chains with explicit counterions, and have shown in terms of bare-model parameters, that for sufficiently long chains, the scaling of the end-to-end vector does indeed exhibit logarithmic corrections  in the blob-pole regime. Here, we examine whether results of BD simulations in the blob-pole regime exhibit logarithmic corrections as described by Eq.~(\ref{eq:relogX}).

A common assumption in the various theoretical descriptions of the blob-pole regime is that electrostatic interactions lead to chain stretching along one direction, while leaving the chain conformation unperturbed in directions perpendicular to the stretching direction~\cite{Dobrynin2005}. In terms of blob scaling variables, this implies that chain dimensions  lateral to the stretching direction are expected to scale as $\xi_\text{el} \, X^{\frac{1}{2}}$.  By examining the eigenvalues of the gyration tensor, we verify if chain dimensions perpendicular to the stretching direction indeed obey ideal chain scaling laws. 

The concept of the Zimm diffusivity of a blob has been successfully used to develop scaling relations for the diffusivity of neutral polymer chains, both in the dilute concentration regime (in terms of thermal blobs~\cite{Rubinstein2003}), and in the semidilute regime (in terms of correlation blobs~\cite{Rubinstein2003,JainPRL2012}). Here, we examine if the scaling of the diffusivity of polyelectrolyte chains in the various regimes of the phase diagram, becomes independent of bead-spring chain parameters, when the Zimm diffusivity of an electrostatic blob is used to interpret simulation results. In particular, since the blob-pole is reminiscent of the shish-kebab model for rodlike polymers~\cite{Doi1986} (with blobs taking the place of beads), we examine if the diffusivity of a polyelectrolyte chain in regime II can be understood in terms of  the translational diffusivity of rodlike polymers.

The paper is structured as follows. In section II we describe the bead-spring chain model used to represent polyelectrolyte chains, and the governing equations for the time evolution of the position vectors for the beads. Section III, which summarises our results and the relevant discussions, is subdivided into four sections; the first looks at the scaling of the chain size and the extent of logarithmic corrections in the blob-pole regime, the second and third consider the scaling of various functions that describe the shape of the chain, while the fourth examines the scaling of chain diffusivity and relaxation time.  Finally, the key findings of this work are summarised  in section IV.

\section{\label{sec:SM} The bead-spring chain model}

\subsection{\label{sec:beadspring}  Governing equations and Brownian dynamics simulations}

A polyelectrolyte chain is modelled in the BD simulations by a coarse-grained version of the bare-model, i.e., by a bead-spring chain consisting of $N_\text{b}$ beads of radius $a$, connected linearly by ($N_\text{b}-1$) finitely extensible non-linear elastic (FENE) springs, with spring constant $H$ and maximum stretch $Q_0$. The beads act as centres of frictional resistance, with a Stokes friction coefficient, $\zeta = 6 \pi \etas \, a$ (where $\etas$ is the solvent viscosity). The total charge on the bead-spring chain is set equal to that of a chain in the bare-model, distributed uniformly along the length of the chain with each bead having an identical charge $q$, given by 
\begin{equation}
\label{eq:cbal}
q=\dfrac{f N_\text{K}}{N_\text{b}}
\end{equation}

The time evolution of the position vector
${\Vector{r}}_{\mu}(t)$ of bead $\mu$, is described by the non-dimensional stochastic differential equation~\cite{Ottinger1996}
\begin{equation}
\rmd \Vector{r}_{\mu} = \frac{1}{4} \, \sum_{\nu}
\Tensor{D}_{\mu\nu} \cdot \fnu \,
\rmd t + \frac{1}{\sqrt{2}} \, \sum_{\nu} \Tensor{B}_{\mu\nu} \cdot
\rmd \Vector{W}_{\nu}
\label{eq:sde}
\end{equation}
where, the length scale $l_H=\sqrt{k_\text{B}T/H}$ and time scale ${\lambda}_H=\zeta/4H$ have been used for non-dimensionalization. 
The dimensionless diffusion tensor
$\bm{\Tensor{D}}_{\mu\nu}$ is a $3 \times 3$ matrix for a fixed pair of beads $\mu$ and
$\nu$. It is related to the hydrodynamic interaction tensor, as
discussed further subsequently. The sum of all the non-hydrodynamic forces on bead $\nu$
due to all the other beads is represented by ${\bm{\Vector{F}}}_{\nu}$, the quantity $\bm{\Vector{W}}_{\nu}$ is a Wiener process, and  $\bm{\Tensor{B}}_{\mu\nu }$ is a non-dimensional tensor whose presence leads to multiplicative noise~\cite{Ottinger1996}. 
Its evaluation requires the decomposition of the diffusion tensor. Defining the matrices
$\bcal{D}$ and $\bcal{B}$ as block matrices consisting of $N \times N$
blocks each having dimensions of $3 \times 3$, with the $(
\mu,\nu)$-th block of $\bcal{D}$ containing the components of the
diffusion tensor $\bm{\Tensor{D}}_{\mu\nu }$, and the corresponding
block of $\bcal{B}$ being equal to $\bm{\Tensor{B}}_{ \mu\nu}$, the
decomposition rule for obtaining $\bcal{B}$ can be expressed as
\begin{gather}
\bcal{B} \cdot {\bcal{B}}^\textsc{t} = \bcal{D} \label{decomp}
\end{gather}
The non-hydrodynamic forces on a bead $\mu$ are comprised of the non-dimensional spring forces ${\bm{\Vector{F}}}_{\mu}^{\text{s}}$ and non-dimensional electrostatic
interaction forces ${\bm{\Vector{F}}}_{\mu}^{\text{es}}$, \ie,
${\bm{\Vector{F}}}_{\mu} = {\bm{\Vector{F}}}_{\mu}^{\text{s}} +
{\bm{\Vector{F}}}_{\mu}^{\text{es}}$. The entropic
spring force on bead $\mu$ due to adjacent beads can be expressed as
${\bm{\Vector{F}}}_{\mu}^{\text{s}} =
{\bm{\Vector{F}}}^c({\bm{\Vector{Q}}}_{\mu}) -
{\bm{\Vector{F}}}^c({\bm{\Vector{Q}}}_{\mu - 1})$ where
${\bm{\Vector{F}}}^c({\bm{\Vector{Q}}}_{\mu - 1})$ is the force
between the beads $\mu -1$ and $\mu$, acting in the direction of the
connector vector between the two beads ${\bm{\Vector{Q}}}_{\mu - 1} =
{\bm{\Vector{r}}}_{\mu} - {\bm{\Vector{r}}}_{\mu - 1}$. Specifically, the spring force in the FENE springs used here is given by,
\begin{equation}
{\bm{\Vector{F}}}^\text{c}({\bm{\Vector{Q}}}_{\mu}) =
\dfrac{{\bm{\Vector{Q}}}_{\mu}}{1-({\Vector{Q}}^2_{\mu}/b)}
\end{equation}
where $b=HQ_0^2/k_\text{B}T$ is the dimensionless finite extensibility parameter.  
The vector $\bm{\Vector{F}}_{\mu}^{\text{es}}$ is given in terms of the dimensionless electrostatic potential $U^\text{es}_{\mu\nu}$ between the beads $\mu$ and $\nu$ of the chain,
\begin{equation}
{\bm{\Vector{F}}}_{\mu}^{\text{es}} = - \sum_{\substack{ \nu = 1\\ \nu \ne \mu}}^{N_\text{b}} \, \frac{\partial }{ \partial\br_{\mu}} \, U^\text{es}_{\mu\nu}
\label{esforce}
\end{equation}
The OSFKK scaling theory assumes Debye-H\"{u}ckel electrostatic interactions~\cite{Debye1923} between charges, which is justified for weakly charged polyelectrolyte chains in the absence of Manning counterion condensation~\cite{Manning1969,Stevens1995,Stevens1996,stoltz2007}. We adopt a Debye-H\"{u}ckel potential in this work, 
\begin{equation}
\label{eq:debpt}
U^\text{es}_{\mu\nu}= \frac{l^{*}_\text{B}q^2}{r_{\mu\nu}}\exp{\left( - \dfrac{r_{\mu\nu}}{l^{*}_\text{D}}\right ) }
\end{equation}
with $r_{\mu\nu}=\left\vert \textbf{r}_{\mu\nu} \right\vert$, where $\br_{\mu \nu} = \br_{\mu}-\br_{\nu}$, is the vector between beads $\nu$ and $\mu$, and $l^{*}_\text{B}$ and $l^{*}_\text{D}$ are the nondimensional Bjerrum and Debye lengths, respectively.

The non-dimensional diffusion tensor $\bm{\Tensor{D}}_{\nu \mu}$ is related to the non-dimensional hydrodynamic interaction tensor $\bOmega$ through
\begin{equation}
\label{eq:Domega}
{\bm{\Tensor{D}}}_{\mu \nu} = \delta_{\mu \nu} \, \bdelta
+ (1 - \delta_{\mu \nu}) \, \bOmega (\bm{r}_\nu - \bm{r}_\mu)
\end{equation}
where $\bdelta$ and $\delta_{\mu \nu}$ represent a unit
tensor and a Kronecker delta, respectively, while
$\bOmega$ represents the effect of the motion of a
bead $\mu$ on another bead $\nu$ through the disturbances carried by
the surrounding fluid. The hydrodynamic interaction tensor
${\bOmega}$ is assumed to be given by the
Rotne-Prager-Yamakawa (RPY) regularisation of the Oseen function
\begin{equation}
\label{eq:RPY}
{\bOmega}({\bm{\Vector{r}}})
= {\varOmega}_1 \, \bdelta
+ {\varOmega}_2 \frac{{\bm{\Vector{r}}} {\bm{\Vector{r}}}} {{{{r}}}^2}
\end{equation}
where for $r \equiv \vert {\bm{\Vector{r}}} \vert \ge 2 a^{*}$,
\begin{equation}
\label{eq:b1}
{\varOmega}_1 = \frac{3}{4} \, \frac{a^{*}}{r} \left( 1 + \frac{2}{3} \, \frac{{a^*}^2}{r^{2}} \right)
\, \, \, \, \text{and} \, \, \, \,
{\varOmega}_2 = \frac{3}{4} \, \frac{a^{*}}{r} \left( 1 - 2 \, \frac{{a^*}^2}{r^{2}} \right)
\end{equation}
while for $0 < r \le 2a^{*}$,
\begin{equation}
\label{eq:b2}
{\varOmega}_1 = 1 - \frac{9}{32} \,
\frac{r}{a^*}
\, \, \, \, \text{and} \, \, \, \,
{\varOmega}_2 = \frac{3}{32} \,
\frac{r}{a^*}
\end{equation}
Here, $a^{*}$ is the non-dimensional bead radius, which is related to the conventionally defined~\cite{BirdVol2}  hydrodynamic interaction parameter, $h^{*}$, by $a^{*} = \sqrt{\pi} h^{*}$. We have set $a^{*} = 0.5$ in all simulations in which hydrodynamic interactions are incorporated. A choice of $h^{*}$ close to 0.25 ensures that the non-draining limit is reached (and consequently universal predictions obtained), at relatively small values of $N_\text{b}$~\cite{Osa72,Ott87d,RaviBookchapter1999,Kroger2000,SuntharMacro2005,Sunthar2006}.

The spatial configuration of the chain at any time $t$, i.e., ${\Vector{r}}_{\mu} (t)$ for all beads $\mu = 1, \ldots , N_{\text{b}}$, is obtained by integrating Eq.~(\ref{eq:sde}) using a semi-implicit predictor-corrector scheme proposed by Prabhakar and Prakash~\cite{PraRavi04}. In the presence of fluctuating HI, the problem of the computational intensity of calculating the Brownian term is reduced by the use of a Chebyshev polynomial representation for the Brownian term~\cite{Fixman1986,jendrejack:jcp-00}. We have adopted this strategy, and the details of the exact algorithm followed here are given in Ref.~\citenum{PraRavi04}.

While the majority of results reported here have been obtained with the ``single-chain'' BD algorithm described in Ref.~\citenum{PraRavi04}, the predictions of eigenvalues and shape functions have been obtained with the ``multi-chain'' BD algorithm described in Ref.~\citenum{Jain2012}. The most significant difference between the two algorithms is that the latter (in which multiple chains are simulated in a box with periodic boundary conditions), accounts for \textit{inter-particle} hydrodynamic and electrostatic interactions in addition to 
\textit{intra-particle} interactions. Essentially, $N_\text{c}$ bead-spring chains are simulated in a cubic simulation box of length $L$, such that the concentration $c=N_\text{b}N_\text{c}/L^{3}$. With the overlap concentration $c^*$ defined by, $c^*=N_\text{b}/[(4/3)\pi r_\text{g,eq}^3]$, where $r_\text{g,eq}$ is the radius of gyration of a chain at equilibrium, we maintain a scaled concentration $c/c^* = 10^{-5}$, to ensure that the system is in the dilute limit. Since the typical distance between chains at these concentrations is much greater than the Debye screening lengths considered here, the strength of short-ranged inter-molecular Debye-H\"{u}ckel electrostatic interactions is effectively zero. The equivalence of the use of either of the algorithms has been verified by comparison of predictions of the end-to-end vector and the radius of gyration, and ensuring that no difference was observed. Plots which contain data from the multi-chain BD algorithm are identified as such in the figure captions, and unless explicitly stated, most plots report data obtained using the single chain BD algorithm. 


\subsection{\label{sec:par}Size, shape and diffusivity}

The two static properties examined here are: (i) the end-to-end distance, $r_\text{e} \equiv \sqrt{\langle r_\text{e}^2 \rangle}$, with,
\begin{equation}
\label{re}
\langle r_\text{e}^2\rangle= \left< (\textbf{r}_{\text{N}_\text{b}}-\textbf{r}_\text{1})\cdot(\textbf{r}_{\text{N}_\text{b}}-\textbf{r}_\text{1})\right> 
\end{equation}
where, $\langle  (.)  \rangle$ represents an ensemble average, and $\textbf{r}_{\text{N}_\text{b}}$ and $\textbf{r}_\text{1}$ are the positions of the two beads at either end of the chain, and, (ii) the radius of gyration of the chain, $r_\text{g} \equiv \sqrt{\langle r_\text{g}^2 \rangle}$, with,
\begin{equation}
\label{rg}
\langle r_\text{g}^2\rangle=\langle\lambda^2_1\rangle+\langle\lambda^2_2\rangle+\langle\lambda^2_3\rangle
\end{equation}
where, $\lambda^2_1$, $\lambda^2_2$, and $\lambda^2_3$ are eigenvalues of the gyration tensor $\textbf{G}$ (arranged in ascending order), with,
\begin{equation}
\label{eq:gy}
\textbf{G} =  \frac{1}{2N_\text{b}^{2}}\sum_{\mu=1}^{N_\text{b}}\sum_{\nu=1}^{N_\text{b}} \textbf{r}_{\mu\nu} \textbf{r}_{\mu\nu}
\end{equation}
Note that, $\textbf{G}$, $\lambda^2_1$, $\lambda^2_2$, and $\lambda^2_3$ are calculated for each trajectory in the simulation before the ensemble averages are evaluated. 

\begin{table*}[t]
\centering
\caption{\label{table2} Definitions of shape functions in terms of eigenvalues of the gyration tensor, $\textbf{G}$. Note that, $I_{1} = \lambda^2_1 + \lambda^2_2 + \lambda^2_3$, and, $I_{2}= \lambda^2_1 \lambda^2_2 +  \lambda^2_2 \lambda^2_3 +  \lambda^2_3 \lambda^2_1 $, are invariants of $\textbf{G}$.}
\begin{center}
\begin{tabular}{lc}
 \hline \hline  Shape function & Definition \\ 
 \hline 
Asphericity~\cite{Theodorou1985} &  
\parbox{6cm}{
\begin{equation}
 \label{eq:sph}
 B = \langle \lambda^2_{3}\rangle - \frac{1}{2} \left[ \langle \lambda^2_{1}\rangle + \langle \lambda^2_{2}\rangle \right] 
 \end{equation}}  \\
 Acylindricity~\cite{Theodorou1985} &   
 \parbox{6cm}{
 \begin{equation}
 \label{eq:cyl}
 C =  \langle \lambda^2_{2}\rangle - \langle \lambda^2_{1}\rangle 
 \end{equation}}  \\ 
\multirow{2}{*}[-25pt]{Degree of prolateness~\cite{Bishop1986,Zifferer1999} } &   
\parbox{9cm}{
 \begin{equation}
 \label{eq:pro2}
 S^{*} =  \left\langle \frac{ (3\lambda_{1}^2 -  I_{1}^{2}) (3\lambda_{2}^2 -  I_{1}^{2})(3\lambda_{3}^2 -  I_{1}^{2}) }{ \left( I_{1}^{2} \right)^{3} } \right\rangle
 \end{equation} } \\ 
&   
\parbox{9cm}{
 \begin{equation}
 \label{eq:pro1}
 S = \frac{ \left\langle (3\lambda_{1}^2 -  I_{1}^{2}) (3\lambda_{2}^2 -  I_{1}^{2})(3\lambda_{3}^2 -  I_{1}^{2})\right\rangle}{\left\langle \left( I_{1}^{2} \right)^{3} \right\rangle}
 \end{equation} }   \\
\multirow{2}{*}[-25pt]{Relative shape anisotropy~\cite{Theodorou1985,Bishop1986,Zifferer1999} }    &  
 \parbox{6cm}{
 \begin{equation}
 \label{eq:kappa2}
{ \kappa^{2}}^{*}  = 1 -  3 \left\langle \frac{I_{2} }{ I_{1}^{2} } \right\rangle
 \end{equation}}  \\ 
&  
 \parbox{6cm}{
 \begin{equation}
 \label{eq:kappa1}
 \kappa^{2}  = 1 - 3 \frac{\left\langle I_{2} \right\rangle}{\left\langle I_{1}^{2}  \right\rangle}
 \end{equation}}  \\  
 \hline \hline
\end{tabular}
\end{center}
\end{table*} 

The asymmetry in equilibrium chain shape has been studied previously in terms of various functions defined in terms of the eigenvalues of the gyration tensor~\cite{Kuhn1934,Koyama1968,Solc1971,Solc1971a,Yoon1974,Kranbuehl1977,Theodorou1985,Aronovitz1986,Bishop1986,Rudnick1987,Bishop1989,Wei1990,Bishop1991,Zifferer1999,Wei1997,Haber2000,Steinhauser2005}. Apart from $\lambda^2_1$, $\lambda^2_2$, and $\lambda^2_3$, themselves, we have examined the following \textit{shape functions}: the asphericity ($B$), the acylindricity  ($C$), the degree of prolateness ($S$ and $S^{*}$), and the shape anisotropy ($\kappa^{2}$ and ${\kappa^{2}}^{*}$), as defined in Table~\ref{table2}. As is evident from the Table, the latter two quantities are evaluated in terms of two different definitions. Typically, it is easier to evaluate $S$ and $\kappa^{2}$ for analytical calculations, rather than $S^{*}$ and ${\kappa^{2}}^{*}$, which require the evaluation of averages of ratios of fluctuating quantities~\cite{Steinhauser2005}.  

The only dynamic property that is directly calculated here is the long-time self-diffusion coefficient $D$, defined by, 
\begin{equation}
\label{eq:diffcon}
D=\lim\limits_{t\rightarrow \infty}  \left< \frac{|\textbf{r}_\text{c}(t)-\textbf{r}_\text{c}(0)|^2}{6t}\right> 
\end{equation}
where, $\textbf{r}_\text{c}=\dfrac{1}{N_\text{b}}\sum_{\mu=1}^{N_\text{b}}\textbf{r}_\mu$, is the position vector of the center of mass of the chain. 

The calculation of the radius of gyration and the diffusivity of a chain, enables an estimation of the longest relaxation time from the expression,
\begin{equation}
\label{eq:relax}
\tau = \frac{\langle r_\text{g}^2\rangle}{D} 
\end{equation}

\begin{table*}[t]
\caption{\label{table3} Values for the bead-spring chain parameters 
$\{N_\text{b}, b, q, l_\text{B}^*\}$, the resultant scaling variables $X$ and $\xi_\text{el}^*$, and the corresponding values for the bare model parameters $\{N_\text{K}, f, {\hat l}_\text{B} \}$, used in Brownian dynamics simulations. Note that the listed values of the scaling variable $Y$ are specified independently of other parameters by suitably choosing $l_\text{D}^*$ (not tabulated here). $L^{*}$ is the non-dimensional contour length of the bead-spring chain, and $\gamma_{0}$ is the Manning parameter. }  
\vskip10pt
\centering       
{
\begin{tabular}{c | c | c | c | c | c | c }    
\hline\hline                          
  \multicolumn{3}{c}{Blob model}  \vline&  \multicolumn{2}{c}{Bead-spring chain model}  \vline &  \multicolumn{2}{c}{Bare-model} \\
\hline  
 $X$ &$Y$ & $\xi_\text{el}^*$ & $\{N_\text{b}, b, q, l_\text{B}^*\}$ &$L^*$ &   $\{N_\text{K}, f, {\hat l}_\text{B} \}$ & $\gamma_{0}$ \\ 
\hline          
20.3 & 30 & 2.07 & $\{  32, 72.42, 1.1, 0.69  \}$   & 272  & $\{  800, 0.044, 2.09   \}$  &  0.092 \\
\hline  
 \multirow{3}{*}{24} &   \multirow{3}{33.5mm}{0.01,0.05,0.1,0.5,1,1.6, 3,4.8,10,20,30,100}  
 & 1.77 &  $\{ 28, 61.67, 1.79, 0.41 \} $ & 220  & $\{ 600, 0.0833, 1.15 \}$  & 0.096 \\
 \cline{3-7}
  & & 1.90 &  $\{ 32, 72.42, 1.25, 0.69 \} $  & 272  & $\{ 800, 0.05, 2.08  \}$ & 0.104   \\
 \cline{3-7}  
 & & 2.21 & $\{ 40, 61.67, 1.38, 0.4   \}$  & 314  &  $\{ 867, 0.0638, 1.13   \}$  &  0.072  \\
\hline
27.6 &  \multirow{4}{*}{30}  & 1.77 &  $\{ 32, 61.67, 1.79, 0.41  \}$   & 251 &  $\{ 689, 0.0831, 1.16  \}$   &  0.096 \\ 
\cline{1-1} \cline{3-7}  
30.74 &   & 1.68 &  $\{ 32, 72.42, 1.5, 0.69  \}$  & 272   & $\{  800, 0.06, 2.09  \}$ &  0.182   \\ 
\cline{1-1} \cline{3-7}    
34.83 &   & 1.72 &  $\{ 38, 72.42, 1.45, 0.69  \}$  & 323   &  $\{ 955, 0.0577, 2.09 \}$ &   0.170   \\ 
\cline{1-1} \cline{3-7}  
37.67  &   & 1.70  &  $\{ 40, 72.42, 1.48, 0.69   \}$  & 340    &  $\{ 1006, 0.0588, 2.09  \}$  &  0.125\\ 
\hline    
 \multirow{2}{*}{48} &  \multirow{2}{33.5mm}{0.01,0.05,0.1,0.5,1,1.6, 3,4.8,10,20,30,100} 
 & 1.73 & $\{  52, 72, 1.81, 0.45 \} $ & 441   & $\{  1310, 0.0718, 1.36 \} $  &  0.098
  \\
  \cline{3-7}  
& & 1.85   & $\{  60, 61.67, 1.70, 0.41 \}$   & 471   & $\{  1310, 0.0777, 1.16  \}$   &  0.090 \\
\hline  
 \multirow{2}{*}{72}   & \multirow{2}{33.5mm}{0.01,0.05,0.1,0.5,1,1.6, 3,4.8,10,20,30,100} 
 & 1.70  & $\{ 75, 72, 1.95, 0.41  \}$ & 636 & $ \{ 1900, 0.0771, 1.24  \}$ & 0.096  \\ 
  \cline{3-7}  
 & & 1.80   & $\{ 85, 61.67, 1.61, 0.5  \}$  & 668  & $\{ 1900, 0.0732, 1.41  \}$ &  0.103 \\
\hline                        
\hline    
\end{tabular}
}
\end{table*}

An ensemble averaging method has been used to estimate all the properties, details of which are given in Appendix A of Ref.~\citenum{JainThesis}. We find that trajectories level off after roughly one relaxation time, but a true stationary state from which equilibrium properties can be estimated, generally occurs after tens of relaxation times. Typically, more than $\mathcal{O} (10^{6})$ data points are used in the estimation of averages and standard errors of mean, where a data point is a saved property value at some time after the beginning of the stationary state. For properties evaluated with the single chain code, a time step size of 0.005 was used, and data was saved roughly after every 200 time steps. Averages were carried out over 1000 trajectories, with 1000 data points in each trajectory. For properties evaluated with the multi-chain code, a time step size of 0.005 was used, and data was saved roughly after every 10 to 20 time steps. Averages were carried out over 64 trajectories, each with 15 chains in the simulation box, and 2000 data points in each trajectory. 
 

\subsection{\label{sec:bdv} Mapping between bead-spring chain and blob variables}

The concept of an electrostatic blob enables the representation of experimental and simulation data for dilute polyelectrolyte solutions in terms of a significantly reduced number of scaling variables ($\{ X, Y \}$), when compared to the number of bare-model parameters ($\{ N_\text{K}, f, {\hat l}_\text{B}, {\hat l}_\text{D} \}$). Further, the universal nature of property predictions when expressed in terms of blob scaling variables has been established previously by simulations, which have shown that various combinations of bare-model parameters lead to identical results, provided values of the scaling variables are kept fixed~\cite{Everaers2002}. In order to achieve a similar demonstration for BD simulations of bead-spring chains, it is necessary to map the bead-spring chain parameters $\{ N_\text{b}, b, q,  l_\text{B}^{*},  l_\text{D}^{*} \}$ onto the blob scaling variables $\{ X, Y \}$. This issue has been discussed previously by ~\citet{ravip2008} \!\!\!\!\!\!\!\!{.} \, For the sake of completeness, we summarise their arguments here.

The key first step is to map bead-spring chain parameters onto bare-model parameters, after which the mapping onto blob scaling variables is straight-forward. This is achieved by equating the end-to-end vectors and the contour lengths of the bare-model and bead-spring chains. If $L$ is the contour length, then, 
\begin{equation}
\label{eq:L}
L = N_\text{K} \, b_\text{K} = (N_\text{b}-1) \, Q_0 = (N_\text{b} - 1) \, \sqrt{b} \, l_\text{H}
\end{equation}
Further, using the analytical result for the end-to-end vector of FENE chains~\cite{BirdVol2}, we get,
\begin{equation}
\label{eq:etoe}
\langle r_\text{e}^2\rangle = N_\text{K} \, b_\text{K}^{2} =  (N_\text{b} - 1) \, \frac{3 \, b}{b+5} \,  l_\text{H}^{2}
\end{equation}
Equations~(\ref{eq:L}) and~(\ref{eq:etoe}) can be solved for $N_\text{K}$, and $b_\text{K}^{*} = b_\text{K} /  l_\text{H}$, to give,  
\begin{equation}
\label{eq:nkbk}
N_\text{K}  = (N_\text{b}-1) \, \frac{b+5}{3} \, ; \quad b_\text{K}^{*}  = \frac{3 \, \sqrt{b}}{b+5}
\end{equation}
Since ${\hat l}_\text{D} =  l_\text{D}^{*}/b_\text{K}^{*}$, ${\hat l}_\text{B} =  l_\text{B}^{*} / b_\text{K}^{*}$, and ${\hat \xi}_\text{el} = \xi_\text{el}^{*}/b_\text{K}^{*}$, substituting from Eq.~(\ref{eq:nkbk}) for $N_\text{K}$ and $b_\text{K}^{*}$ into Eqs~(\ref{eq:x}) and~(\ref{eq:y}),  and using Eq.~(\ref{eq:cbal}), leads to the following equations for the blob scaling variables in terms of bead-spring chain parameters,
\begin{align}
\label{eq:BSv}
X & =  (N_\text{b}-1)  \left[ \frac{3 \, b}{b+5} \right] \frac{1}{{\xi_\text{el}^{*}}^{2}} \\ 
Y & = \frac{l_\text{D}^*}{\xi_\text{el}^*}
\end{align}
where,
\begin{equation}
\label{eq:BSxi}
\xi_\text{el}^*=\left[\frac{9 \, (N_\text{b}-1)^2 \, b^2}{N_\text{b}^2 \, q^2 \, l_\text{B}^* \, 
(b+5)^2}\right]^{1/3}
\end{equation}

It is clear from the expressions above that various combinations of bead-spring chain parameters can result in identical values for $X$ and $Y$. Table~\ref{table3} displays the various sets of values of bead-spring chain parameters, and the resultant sets of scaling variables $X$ and $Y$, that have been used in the simulations reported here, and the corresponding bare-model parameter values. Two to three different values of $N_\text{b}$ are typically used for each value of $X$. Note that $Y$ can be varied independently of $X$ by varying $ l_{D}^{*}$ (and keeping all other bead-spring chain parameters fixed). As a result, the dependence of properties of dilute  polyelectrolyte solutions on $X$ and $Y$ can be explored by carrying out BD simulations, and their extent of universality assessed. 

In terms of bare-model parameters, the Manning parameter, $\gamma_{0}$, which is defined as the number of elementary charges per Bjerrum length~\cite{Dobrynin2005}, is given by, $\gamma_{0} = f  {\hat l}_{B}$. The OSFKK scaling theory assumes that $\gamma_{0} \ll 1$, since counterion condensation, which occurs for $\gamma_{0} \gtrsim 1$, is not taken into account~\cite{Dobrynin2005,stoltz2007}. The last column in Table~\ref{table3} shows that the Manning parameter is significantly smaller than one in all the simulations reported here.

\begin{figure}[t]
\begin{center}
\begin{tabular}{c}
\includegraphics[width=8cm,height=!]{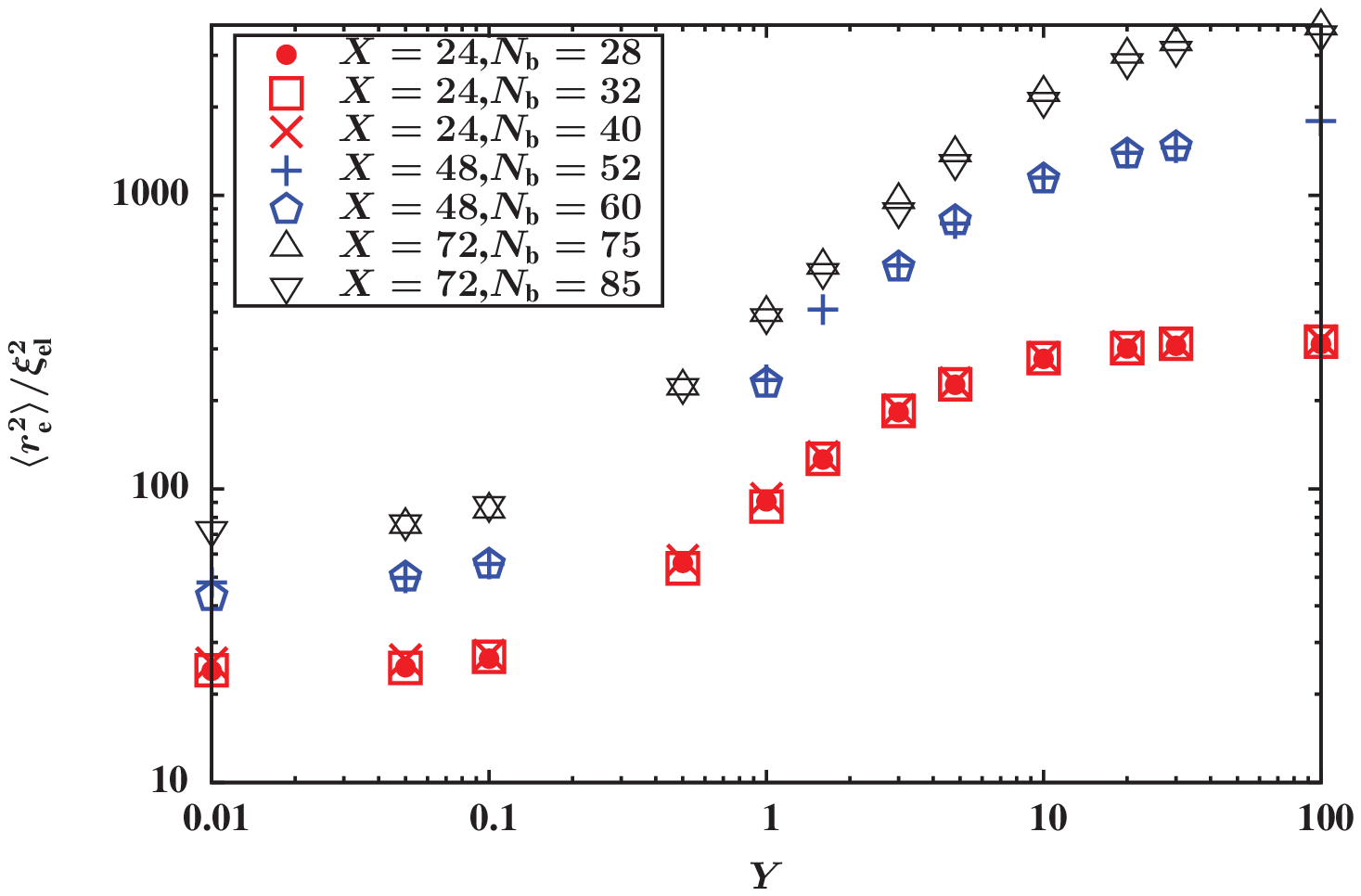} \\
(a) \\ 
\includegraphics[width=8cm,height=!]{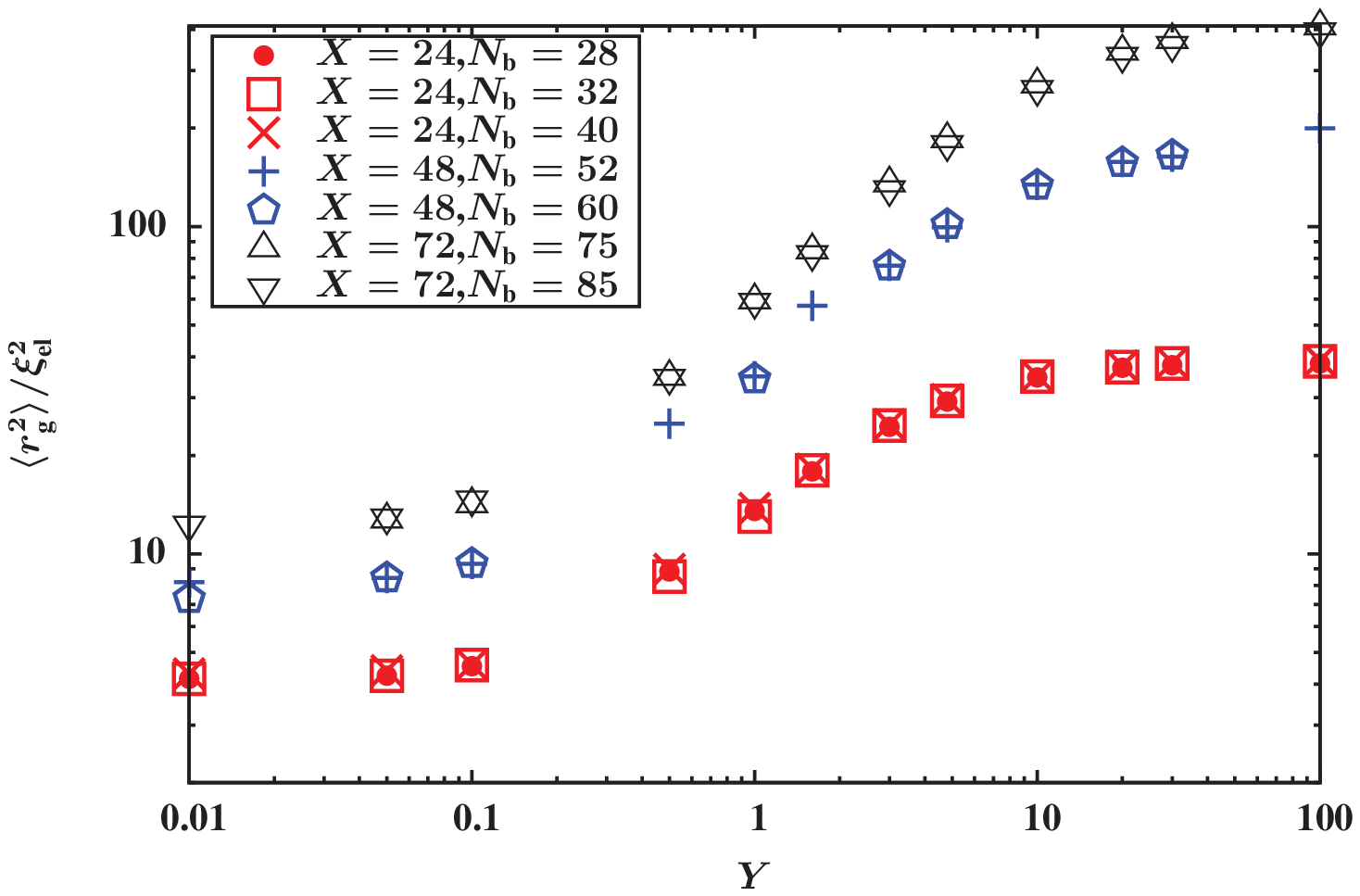} \\
(b)  \\
\end{tabular}
\end{center}
\caption{(Color online) Measures of mean chain size versus the reduced screening length, $Y$, at various values of the number of blobs, $X$: (a) scaled end-to-end vector, $\langle r_\text{e}^2\rangle/\xi_\text{el}^2$, (b) scaled radius of gyration, $\langle r_\text{g}^2\rangle/\xi_\text{el}^2$. Simulation data acquired at different values of chain length, $N_\text{b}$, are displayed  at each value of $X$, to demonstrate parameter-free data collapse. The various combinations of bead-spring chain parameters that correspond to the displayed values of $N_\text{b}$, are shown in Table \ref{table3}. Error bars are of the order of symbol size.}
\label{fig:rvsy}
\end{figure}

\begin{figure}[ht]
\begin{center}
\begin{tabular}{c}
\includegraphics[width=8cm,height=!]{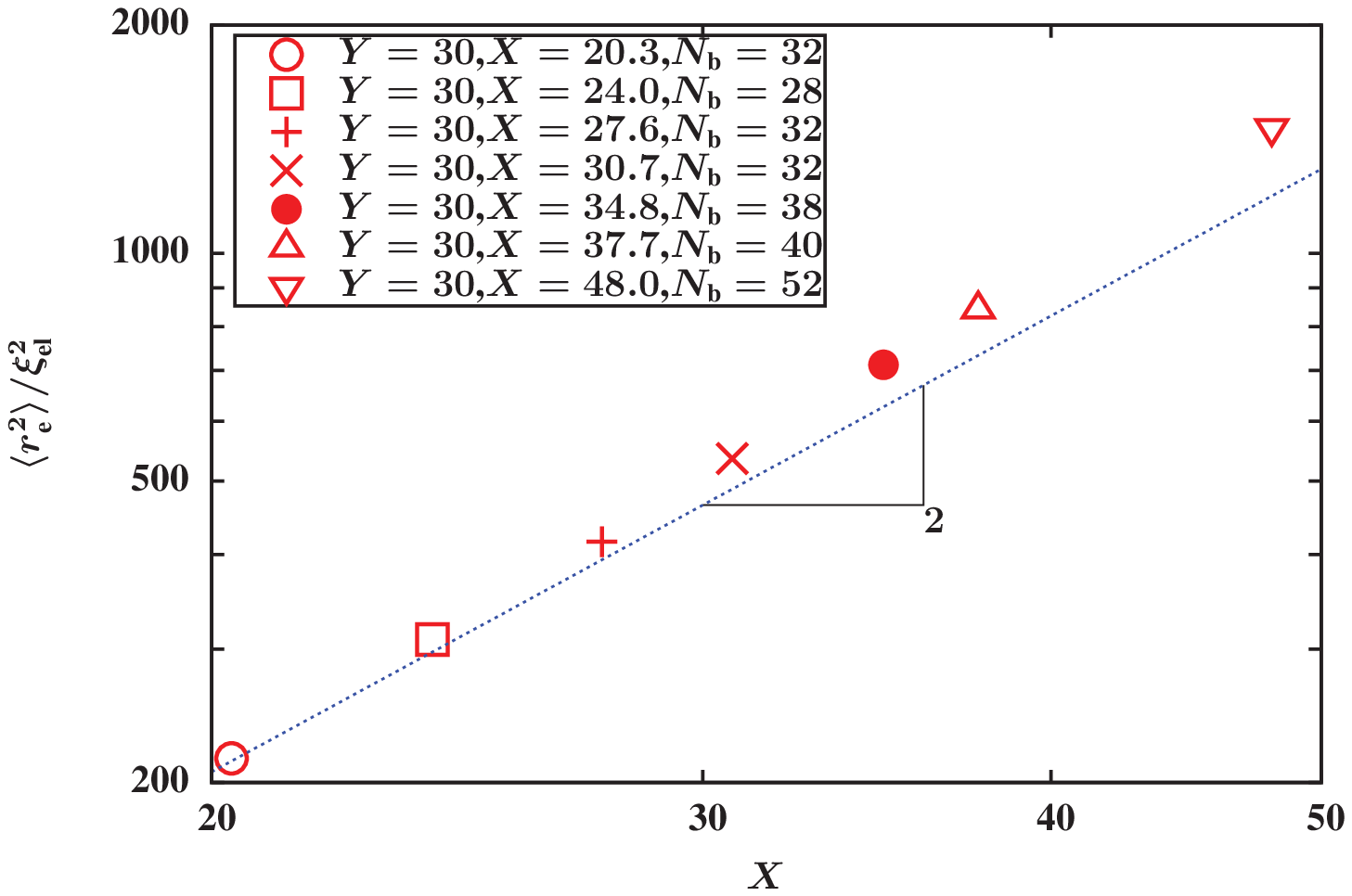} \\ 
(a) \\
\includegraphics[width=8cm,height=!]{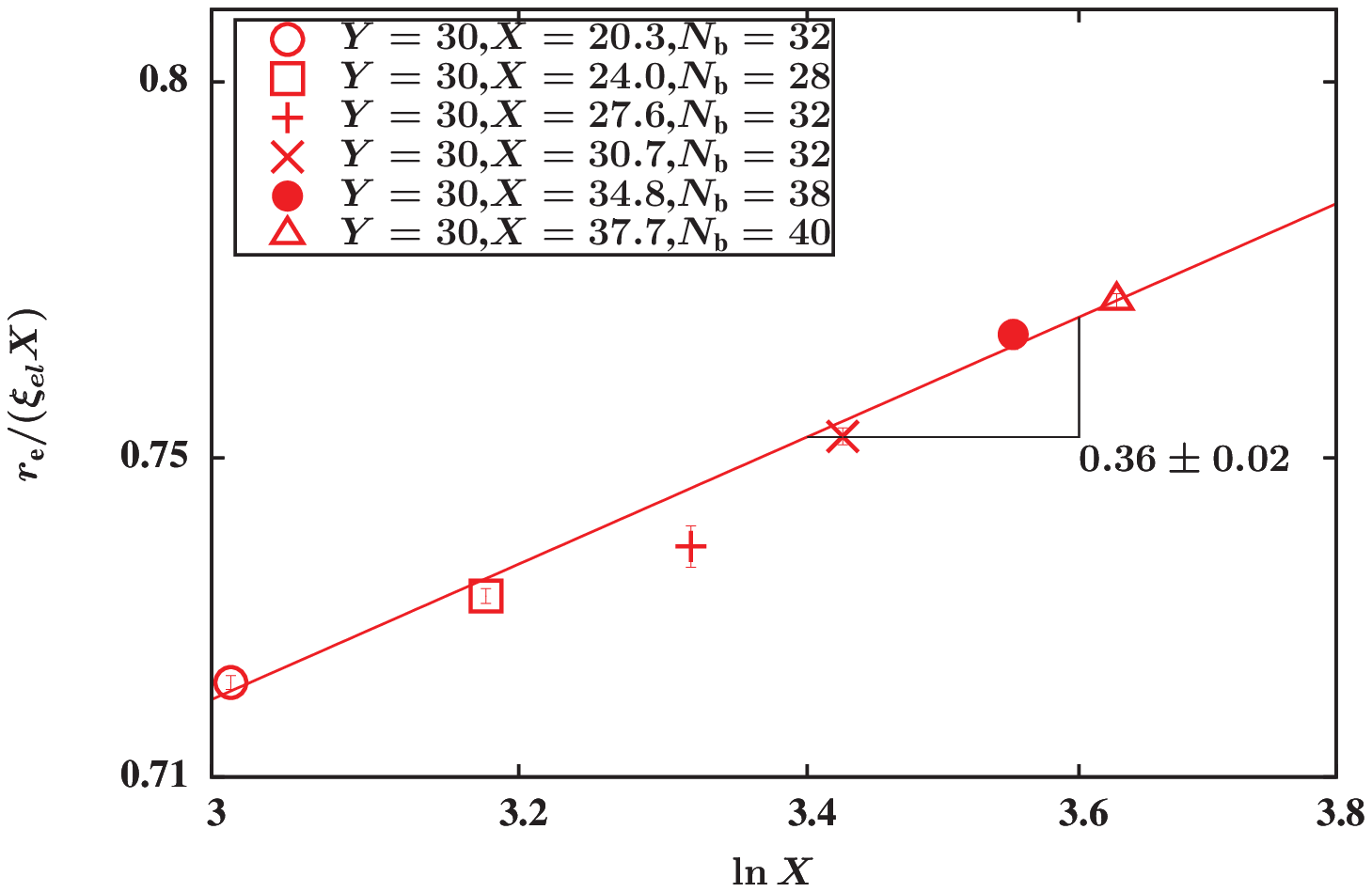} \\
(b)  \\
\end{tabular}
\end{center}
\caption{(Color online) Demonstration of the existence of logarithmic corrections to scaling at $Y=30$ (in the blob-pole regime) for the scaled end-to-end vector. (a) The dotted line shows that $\langle r_\text{e}^2\rangle/\xi_\text{el}^2$ departs from the $X^2$ scaling predicted by OSFKK theory for $X \gtrsim 30$.  (b) A power law fit through the data for  ${r_\text{e}}/{(\xi_\text{el}X)}$ versus $\ln X$ (solid line) shows that the logarithmic corrections have the expected exponent of $1/3$ (within error bars). The various combinations of bare parameters corresponding to the value of $N_\text{b}$ are shown in Table \ref{table3}. Error bars are of the order of symbol size.} 
\label{fig:rvsx}
\end{figure}


\subsection{\label{sec:exptoblob} Mapping from experiments to blob variables}

While the results of the simulations reported here have not been compared directly with experimental observations, it is worth outlining how experimental data on dilute polyelectrolyte solutions can be mapped onto blob variables, to facilitate future comparisons. In this context, the various equations for evaluating the blob variables $X$ and $Y$ for an arbitrary polyelectrolyte solution are derived in Appendix~\ref{sec:AppB}. 

In the current simulation, the behaviour of various static properties and the diffusivity of dilute polyelectrolyte solutions has been explored for a range of blob variable values, $20 < X < 80$ and $0.01 \leq Y \leq 100$. In Appendix~\ref{sec:AppB}, the specific example of an aqueous solution of sodium poly(styrene sulfonate) is considered, and it is shown how an appropriate choice of temperature, polymer molecular weight, degree of ionization per chain, and salt and monomer concentration can lead to blob variables that span this range. Further, parameter values are chosen to ensure that the Manning parameter $\gamma_{0}  \lessapprox 1$ in order to avoid counterion condensation.


\section{\label{sec:DHs}Results}

\subsection{\label{sec:size} Mean chain size as a function of $X$ and $Y$}

From the OSFKK scaling picture depicted schematically in Fig.~\ref{fig:f1}, we expect the chain size to be independent of $Y$ in the limits of $Y\rightarrow 0$ (regime~VI) and $Y\rightarrow \infty$ (regime~II) for $X\gg 1$, while at  intermediate values of $Y$ (in the crossover regimes~III, IV, and~V), we expect the chain size to depend on $Y$. This expectation has been verified by several prior simulations, which have examined the scaling of the end-to-end vector~\cite{ravip2008,Liao2003,Everaers2002}. Figs.~\ref{fig:rvsy}~(a) and~\ref{fig:rvsy}~(b) show that in the current simulations as well, both the scaled end-to-end vector, $\langle r_\text{e}^2\rangle/\xi_\text{el}^2$, and the scaled radius of gyration, $\langle r_\text{g}^2\rangle/\xi_\text{el}^2$, asymptotically approach constant values at the two extremes of $Y$. Note that the asterisk in ${\xi_\text{el}^{*}}^{2}$ has been dropped for simplicity of notation, since the non-dimensional character of the blob size is clear from the context. In the  crossover regime, the chain size increases monotonically with $Y$ as the chain starts to swell due to increasing electrostatic repulsion with increasing Debye screening length. As mentioned previously, our BD simulations are unable to distinguish between the different regimes in the crossover region due to the computational cost of simulating the extremely long chains that would be required for such a purpose. 

\begin{figure}[t]
\begin{center}
\begin{tabular}{c}
\includegraphics[width=8cm,height=!]{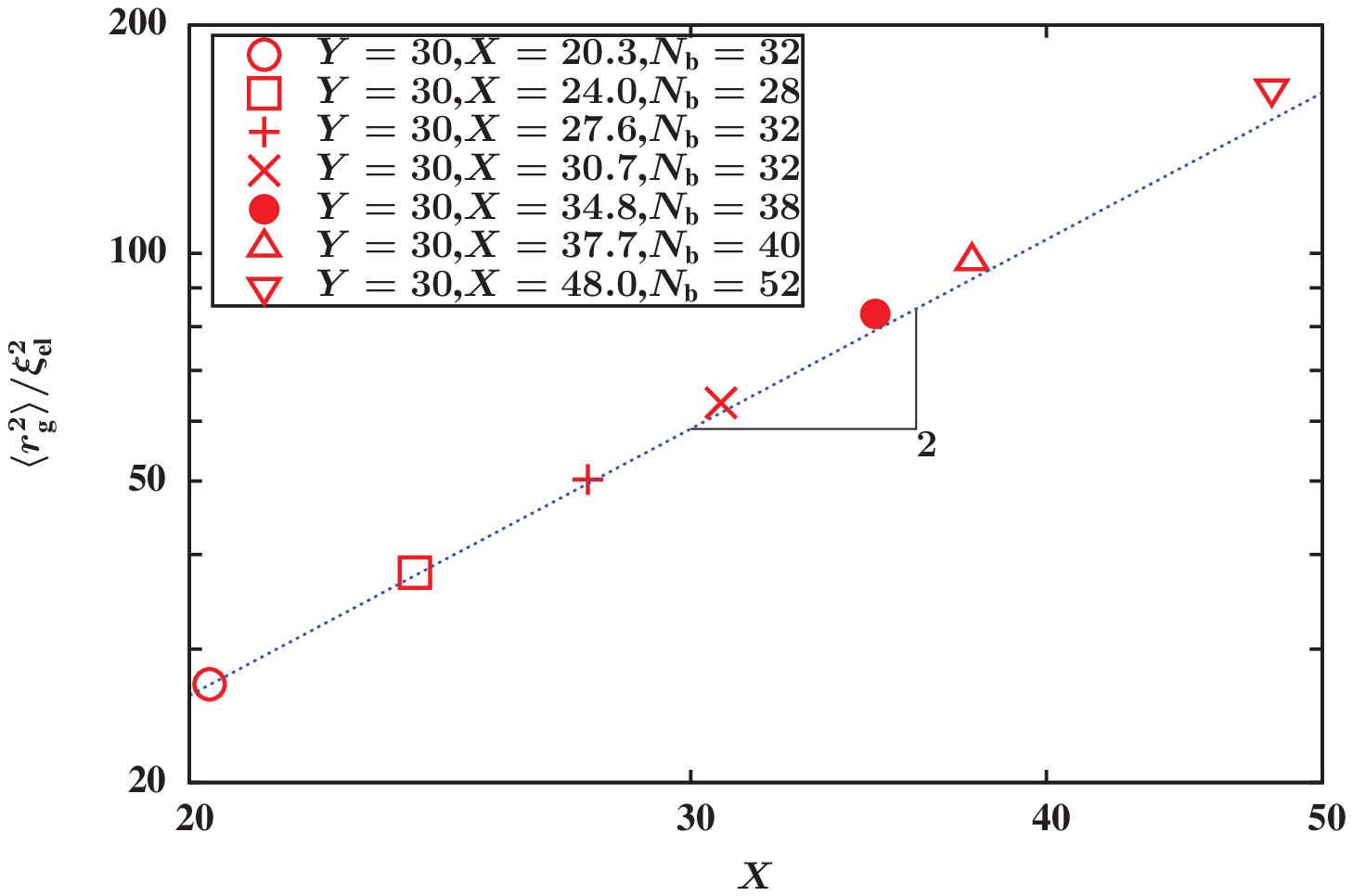} \\
(a) \\
\includegraphics[width=8cm,height=!]{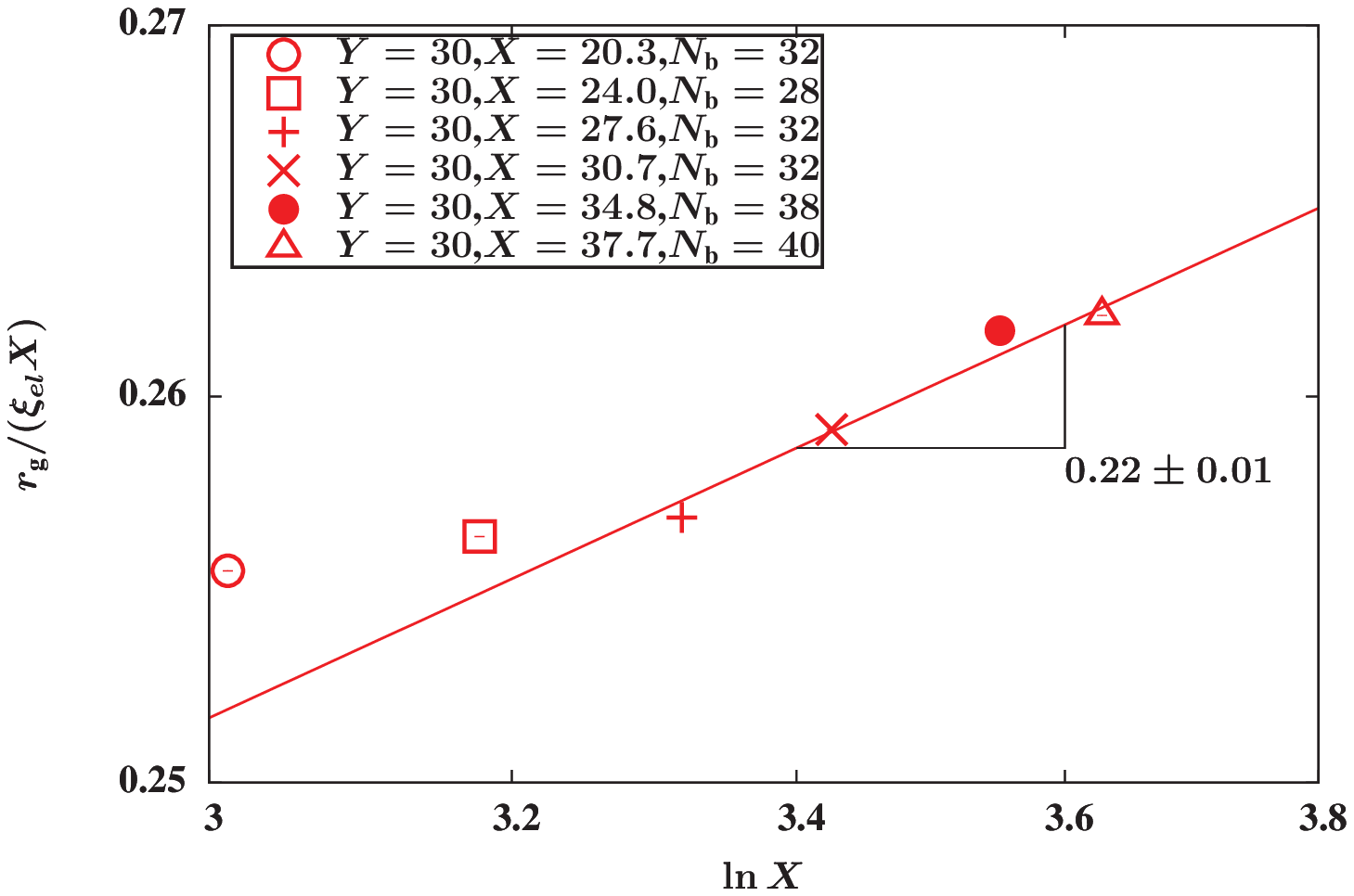} \\
(b)  \\
\end{tabular}
\end{center}
\caption{(Color online) Demonstration of the existence of weak logarithmic corrections to scaling at $Y=30$ (in the blob-pole regime) for the scaled radius of gyration. (a) The dotted line shows the onset of departure from $X^2$ scaling for $\langle r_\text{g}^2\rangle/\xi_\text{el}^2$ at large values of $X$.  (b) A power law fit through the last four data points for ${r_\text{g}}/{(\xi_\text{el}X)}$ versus $\ln X$ (solid line) shows that the logarithmic corrections have an exponent smaller than $1/3$. The various combinations of bare parameters corresponding to the value of $N_\text{b}$ are shown in Table \ref{table3}. Error bars are of the order of symbol size.} 
\label{fig:rvsx2}
\end{figure}

Notably, as long as the value of $X$ is the same, the predicted values of both $\langle r_\text{e}^2\rangle/\xi_\text{el}^2$, and $\langle r_\text{g}^2\rangle/\xi_\text{el}^2$, are independent of the specific choice of bead-spring chain parameters, at all values of $Y$. For instance at $X=48$, both $N_\text{b}=52$ and $N_\text{b}=60$, lead to identical results. Such a parameter-free data collapse was previously shown to occur  for BD simulation predictions of $\langle r_\text{e}^2\rangle/\xi_\text{el}^2$ by ~\citet{ravip2008}  \!\!\!\!\!\!\!\!\!{,} \, who also noted that the only constraint to the simulations appeared to be to ensure that there were more beads than blobs in a chain, i.e., $N_\text{b} \ge X$.

The OSFKK scaling picture ignores the presence of logarithmic corrections in the blob-pole regime. However, as seen in Appendix~\ref{sec:App}, and as shown previously by several different analytical studies~\cite{degennes,Dobrynin2005}, such corrections are expected to arise in this regime, and previous molecular simulations have shown that such is indeed the case~\cite{Liao2003}. In the absence of logarithmic corrections, the scaled end-to-end vector, $\langle r_\text{e}^2\rangle/\xi_\text{el}^2$, is expected to scale as $X^{2}$ in regime~II (see Table~\ref{table1}). It is clear from Fig.~\ref{fig:rvsx}~(a) that while this scaling is obeyed for values of $X \lesssim 30$, there is a departure at larger values of $X$. A similar deviation from the expected scaling in regime~II was observed previously in the BD simulations reported by ~\citet{ravip2008} \!\!\!\!\!\!\!\!{.} \, However, they did not verify if the source of the deviation was due to logarithmic corrections. In terms of the blob scaling variable $X$, Eq.~\ref{eq:relogX} shows that the logarithmic correction term has an exponent of $1/3$. We expect therefore that a plot of ${r_\text{e}}/{(\xi_\text{el}X)}$ versus $\ln X$ would be a straight line with a slope of $1/3$. Fig.~\ref{fig:rvsx}~(b) shows that the logarithmic corrections to OSFKK scaling, picked up by the BD simulations, do in fact have the expected dependence on $X$.

While the radius of gyration is also a measure of mean chain size, Figs.~\ref{fig:rvsx2} reveal that the logarithmic corrections to scaling for $\langle r_\text{g}^2\rangle/\xi_\text{el}^2$ in the blob-pole regime are much weaker than for $\langle r_\text{e}^2\rangle/\xi_\text{el}^2$. This could be attributed to the fact that the radius of gyration not only depends on the stretched axis of the chain (which is expected to be in the direction of the end-to-end vector), but also depends on the dimensions of the chain perpendicular to the elongation axis. The scaling of the chain in these lateral dimensions is generally considered to obey random walk statistics. We examine the validity of this expectation below, when we consider the eigenvalues of the gyration tensor.

\begin{figure}[!ht]
\begin{center}
\begin{tabular}{c}
\includegraphics[width=8cm,height=!]{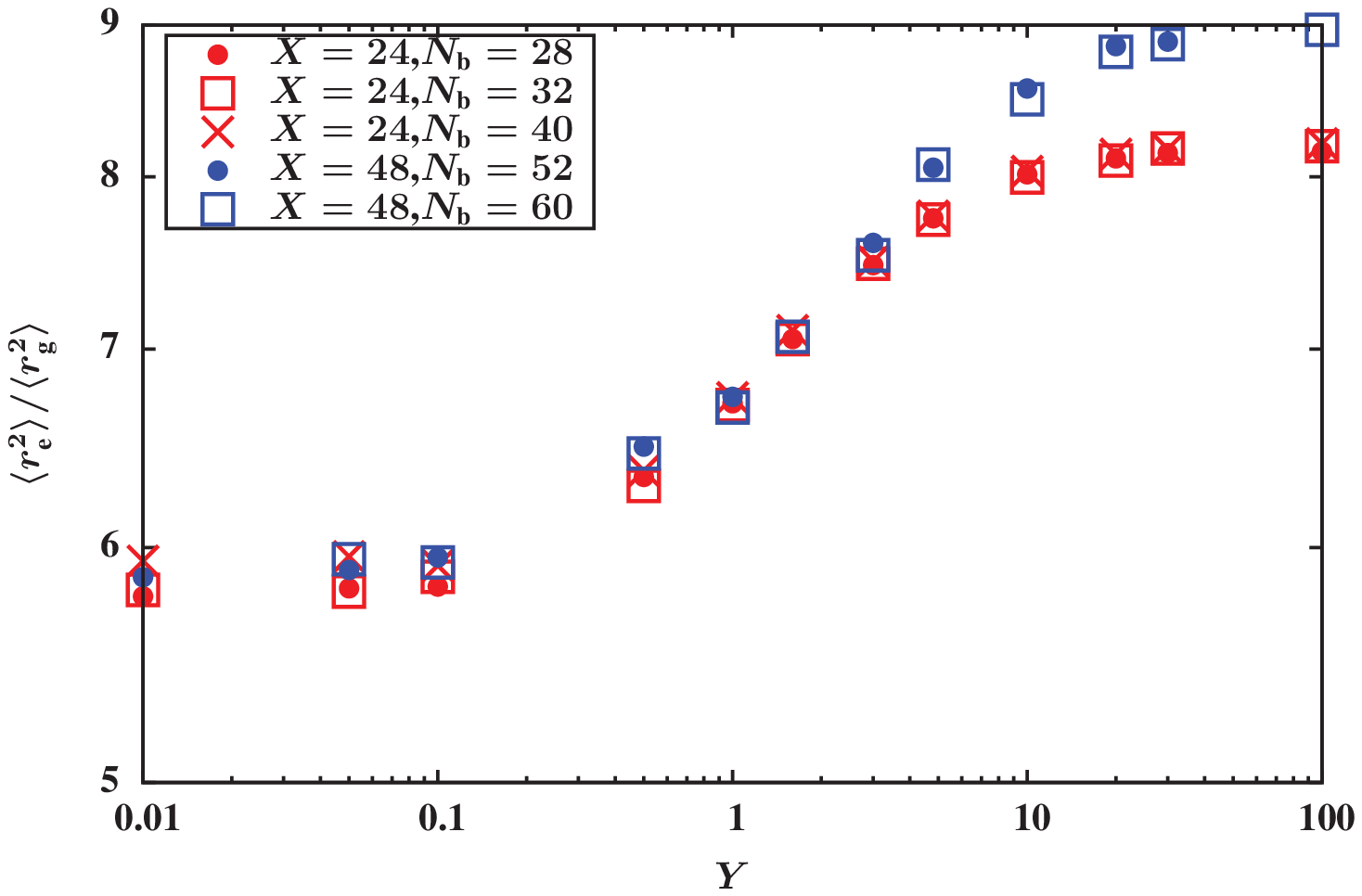} \\
(a)  \\
\includegraphics[width=8cm,height=!]{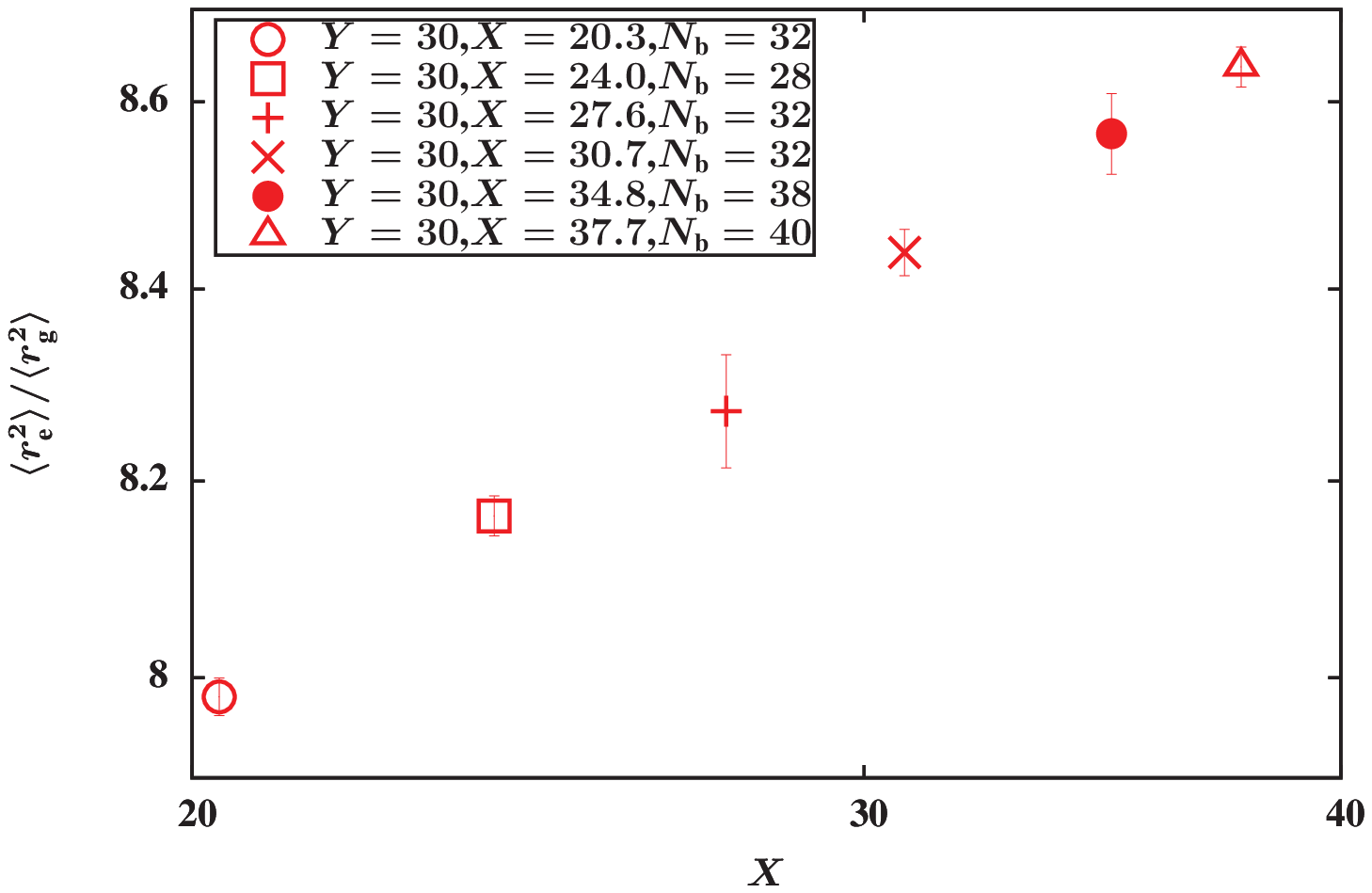} \\
(b)  \\
\end{tabular}
\end{center}
\caption{(Color online)  The variation of the ratio $\langle r_\text{e}^2\rangle/\langle r_\text{g}^2\rangle$ with: (a) the reduced screening length $Y$, for two different values of the number of blobs $X$, and (b) the number of electrostatic blobs $X$, at $Y=30$. Error bars are of the order of symbol size in~(a), and are explicitly displayed in~(b). }
\label{fig:ratio}
\end{figure}
In the limit of long chains, the ratio of the mean square end-to-end vector to the mean square radius of gyration, $\langle r_\text{e}^2\rangle/\langle r_\text{g}^2\rangle$, has a universal value of 6 for ideal chains, while it is equal to 12 for rigid rods~\cite{Doi1986,Rubinstein2003}. We anticipate, therefore, that for the polyelectrolyte solutions considered here, the value of the ratio will increase from  being close to 6 in the limit $Y \ll 1$ (where the chain behaves ideally), towards a value of 12, as $Y \to \infty$, where the polyelectrolyte chain adopts a stretched rodlike configuration. Fig.~\ref{fig:ratio}~(a) shows that the ratio does approach a constant value close to 6 as $Y$ approaches zero, and increases monotonically with increasing $Y$. However, though the ratio levels off as the chain enters the blob-pole regime in the limit of large $Y$, the asymptotic value is less than 12, for the values of $X$ considered here. The departure from the limiting value for rigid rods can be considered to reflect the degree of flexibility remaining in the chain in this regime. Interestingly, the ratio appears to be independent of $X$ for a range of values of $Y$, starting with the ideal regime, and progressing well into the crossover regime (i.e., the red and blue symbols overlap in Fig.~\ref{fig:ratio}~(a)). The ratio starts to become $X$ dependent for values of $Y$ approaching the blob-pole regime. As was seen in Figs.~\ref{fig:rvsx}~(b) and~\ref{fig:rvsx2}~(b) above, though both $\langle r_\text{e}^2\rangle$ and $\langle r_\text{g}^2\rangle$ depart from the OSFKK scaling theory predictions in this regime due to the presence of logarithmic corrections, the strength of these corrections is different in the two cases. This difference is responsible for a persistent dependence of $\langle r_\text{e}^2\rangle/\langle r_\text{g}^2\rangle$ on $X$ in the blob-pole regime, as can be seen in~Fig.~\ref{fig:ratio}~(b), where the value of the ratio is plotted as a function of $X$, at $Y=30$. We can anticipate, however, that with increasing values of $X$ (i.e., chain length), the appearance of logarithmic corrections, and the consequent onset of non-universal behaviour, is postponed to larger and larger values of $Y$, due to a delay in the inception of the blob-pole regime (see Fig.~\ref{fig:f1}). 

\subsection{\label{sec:eigen} Eigenvalues of the radius of gyration tensor}

\begin{figure}[!ht]
\begin{center}
\begin{tabular}{c}
\includegraphics[width=8cm,height=!]{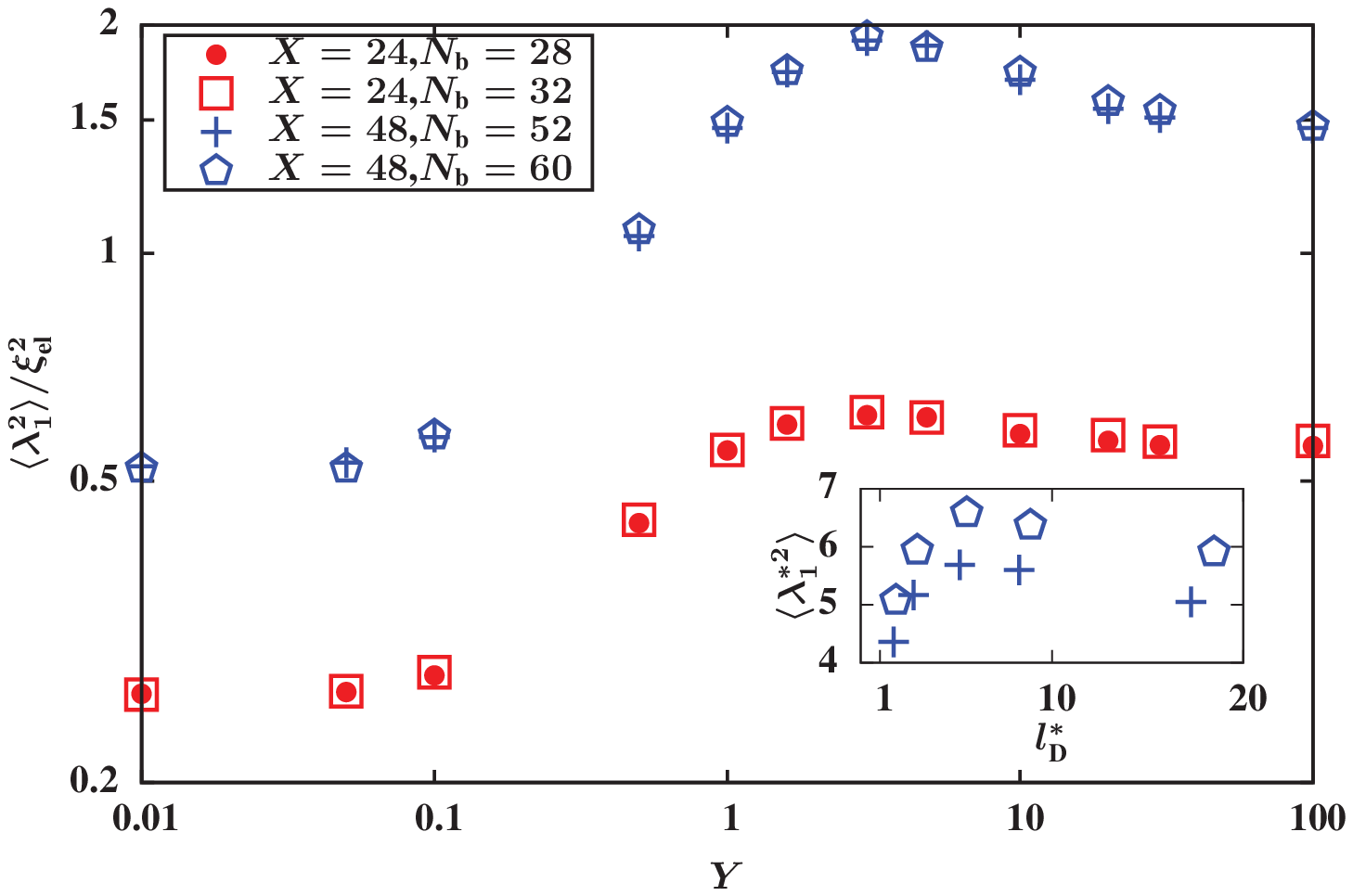} \\
(a) \\
\includegraphics[width=8cm,height=!]{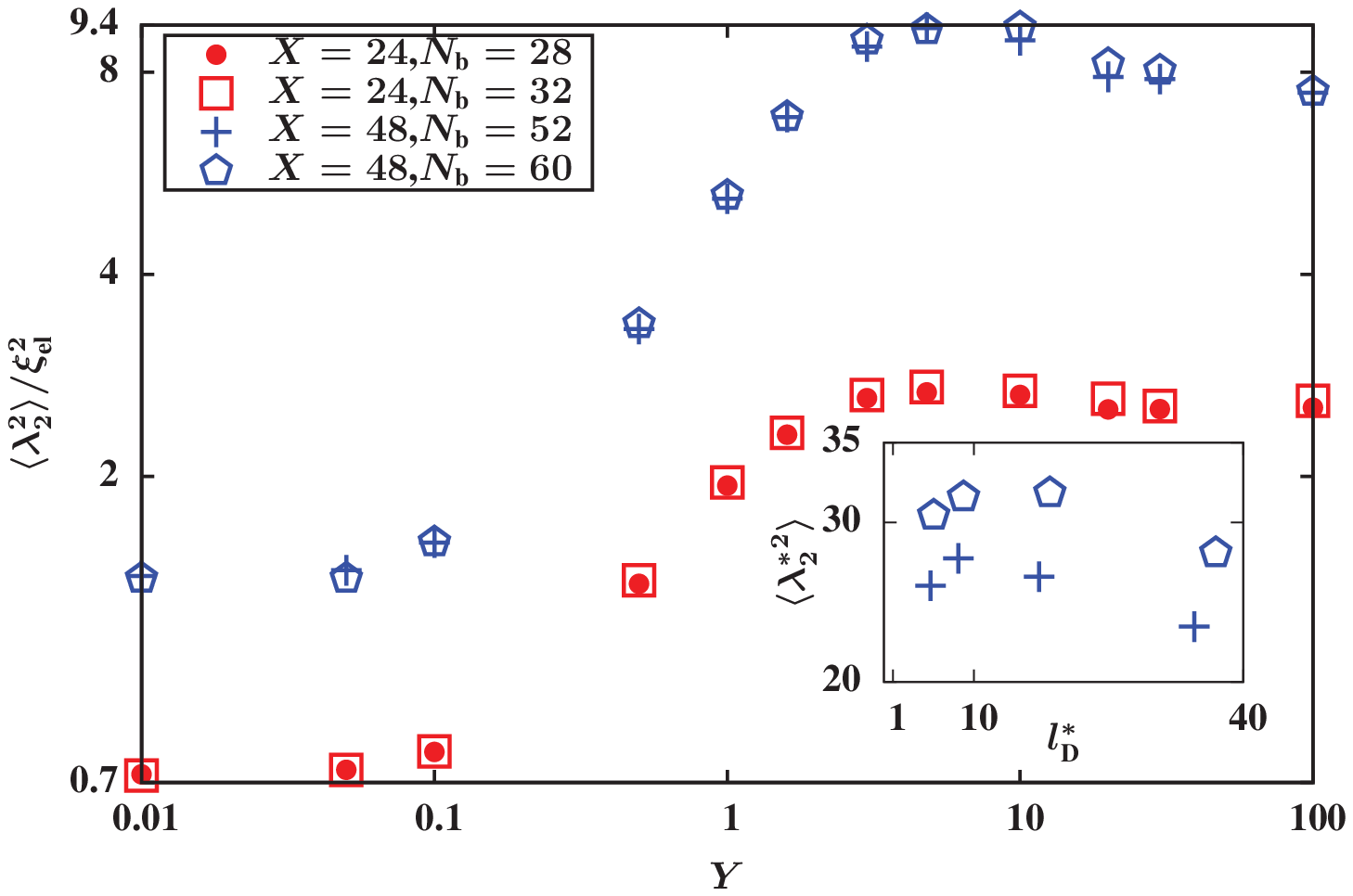} \\
(b)  \\
\includegraphics[width=8cm,height=!]{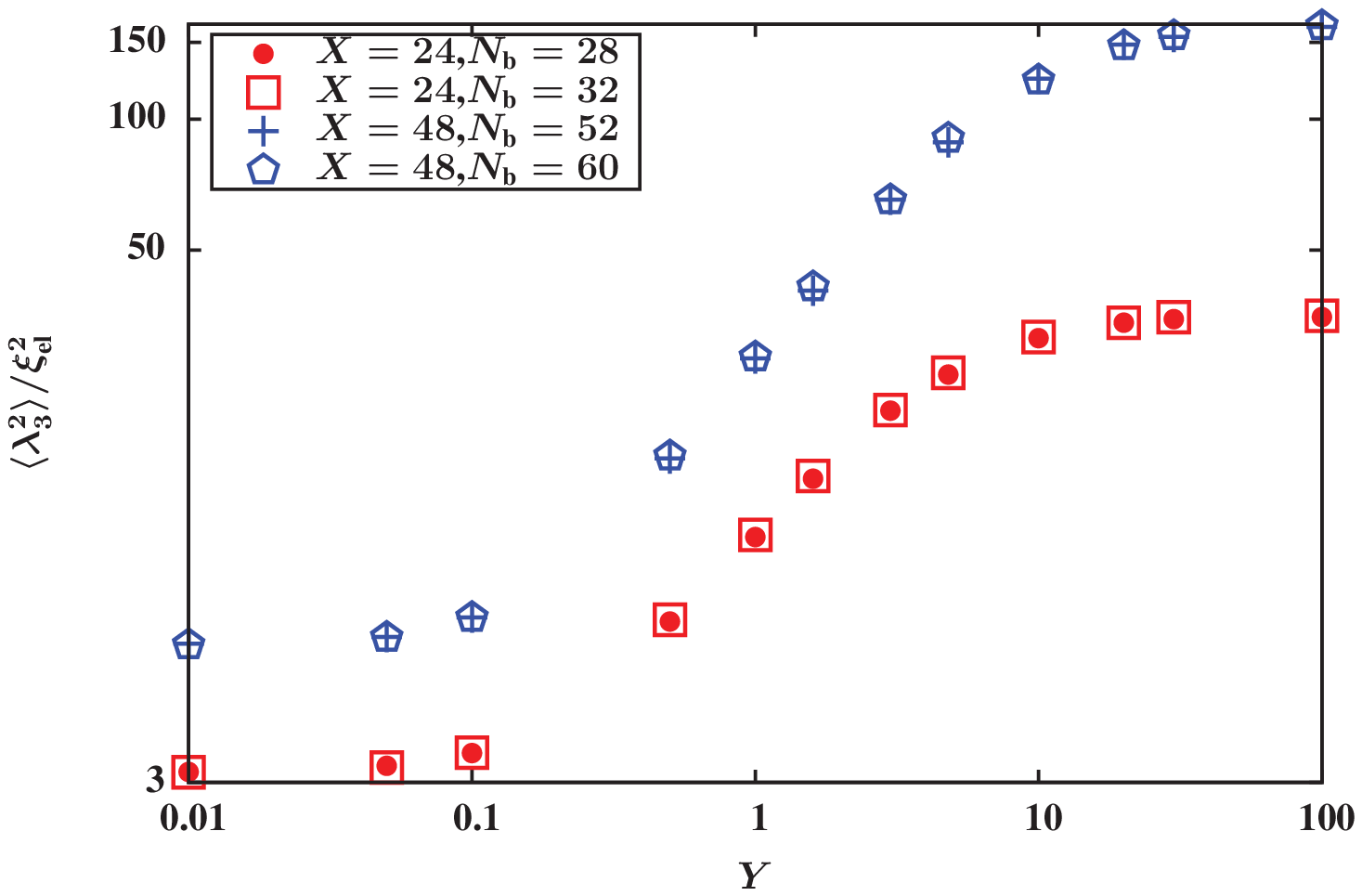} \\
(c)  \\
\end{tabular}
\end{center}
\vskip-20pt
\caption{(Color online) Scaled eigenvalues of the gyration tensor as a function of $Y$, at two  values of $X$: (a) the smallest eigenvalue $\langle \lambda^2_1 \rangle /\xi_\text{el}^2$, (b) the intermediate eigenvalue $\langle \lambda^2_2 \rangle /\xi_\text{el}^2$, and (c) the largest eigenvalue $\langle \lambda^2_3 \rangle /\xi_\text{el}^2$. The insets in~(a) and~(b) display the dependence of the unscaled eigenvalues on the non-dimensional Debye length $l_\text{D}^{*}$ close to the overshoot, for two different chain lengths, $N_\text{b}$, at the same value of $X$. Data obtained with the multi-chain algorithm, with $c/c^*=10^{-5}$. Error bars are of the order of symbol size.}
\label{fig:eigeny}
\end{figure}

\begin{figure}[t]
\begin{center}
\begin{tabular}{c}
\includegraphics[width=8cm,height=!]{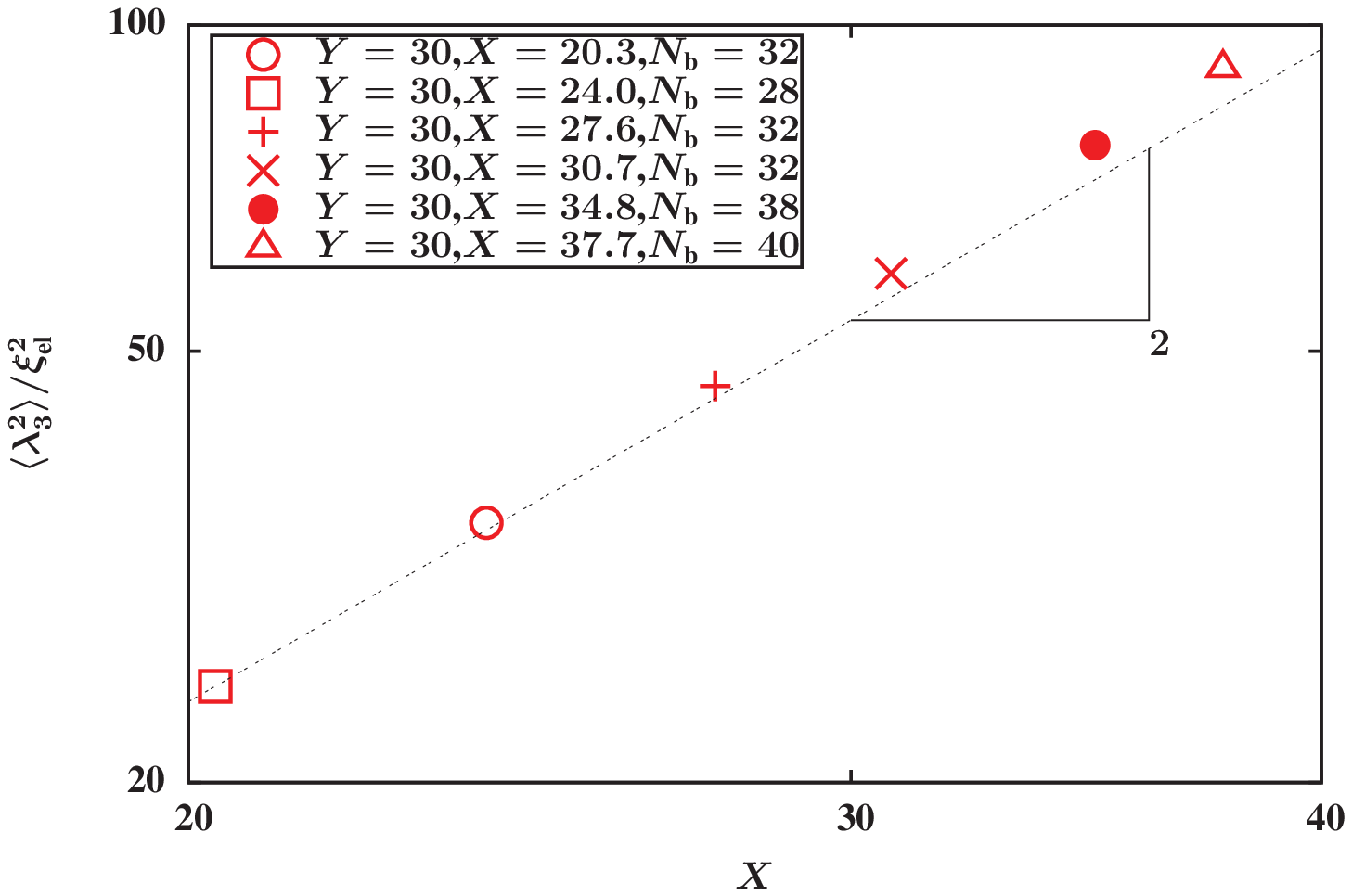} \\
(a) \\
\includegraphics[width=8cm,height=!]{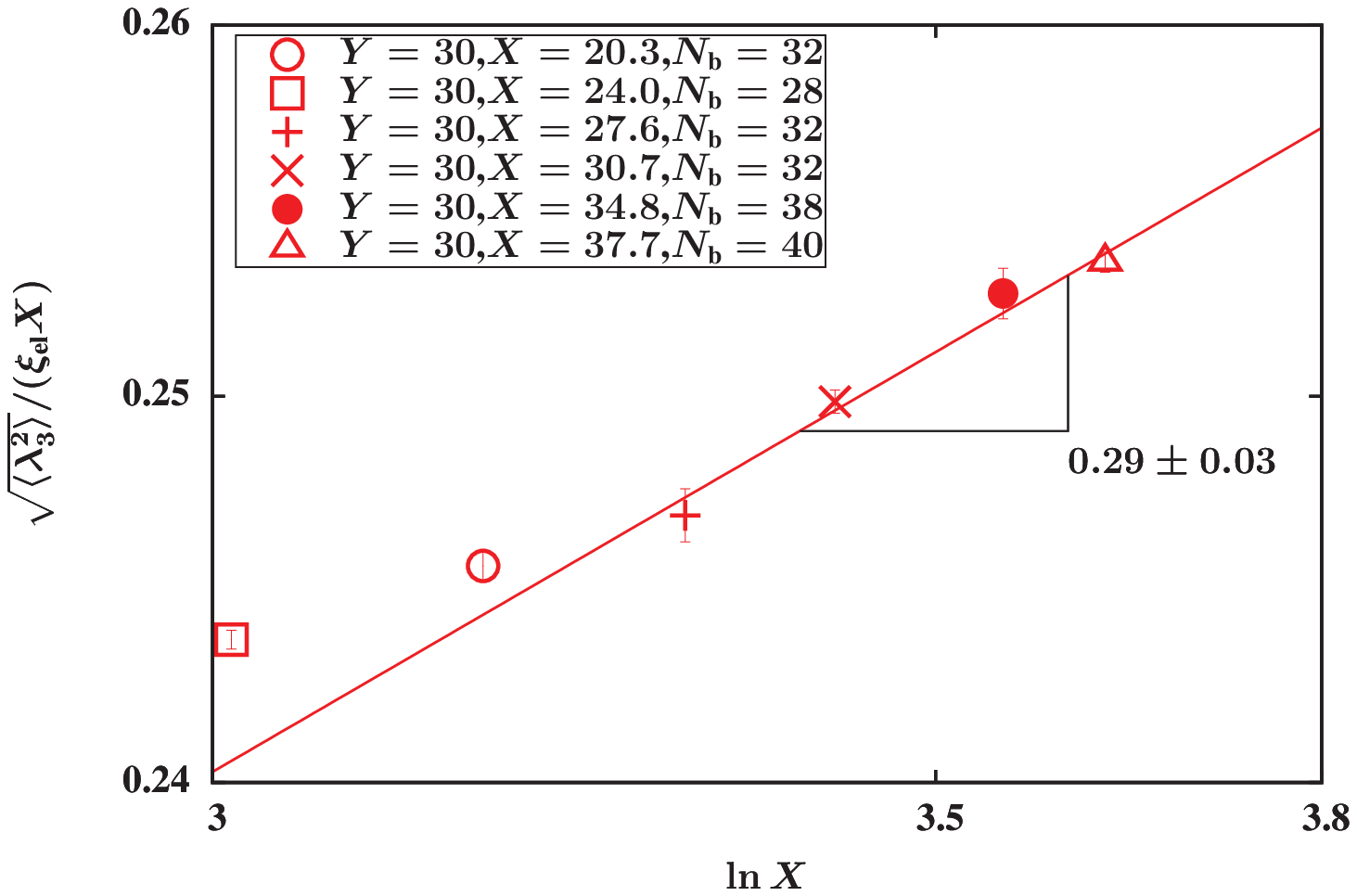} \\
(b)  \\
\end{tabular}
\end{center}
\vskip-15pt
\caption{(Color online) Demonstration of the existence of logarithmic corrections to scaling at $Y = 30$ (in the blob-pole regime) for the largest scaled eigenvalue of the gyration tensor. (a) The dotted line shows that $\langle \lambda^2_3 \rangle /\xi_\text{el}^2$ departs from the $X^2$ scaling predicted by OSFKK theory for $X \gtrsim 30$.  (b) A power law fit through the the last four data points for $\sqrt{\langle \lambda^2_3 \rangle}/\xi_{el}X$ versus $\ln X$ (solid line) shows that the logarithmic corrections have an exponent close to the expected value of $1/3$. Data was obtained using the multi-chain algorithm with $c/c^*=10^{-5}$. Error bars are of the order of symbol size in~(a), and are explicitly displayed in~(b). }
\label{fig:eigen3x}
\end{figure}

\begin{figure}[t]
\begin{center}
\begin{tabular}{c}
\includegraphics[width=8cm,height=!]{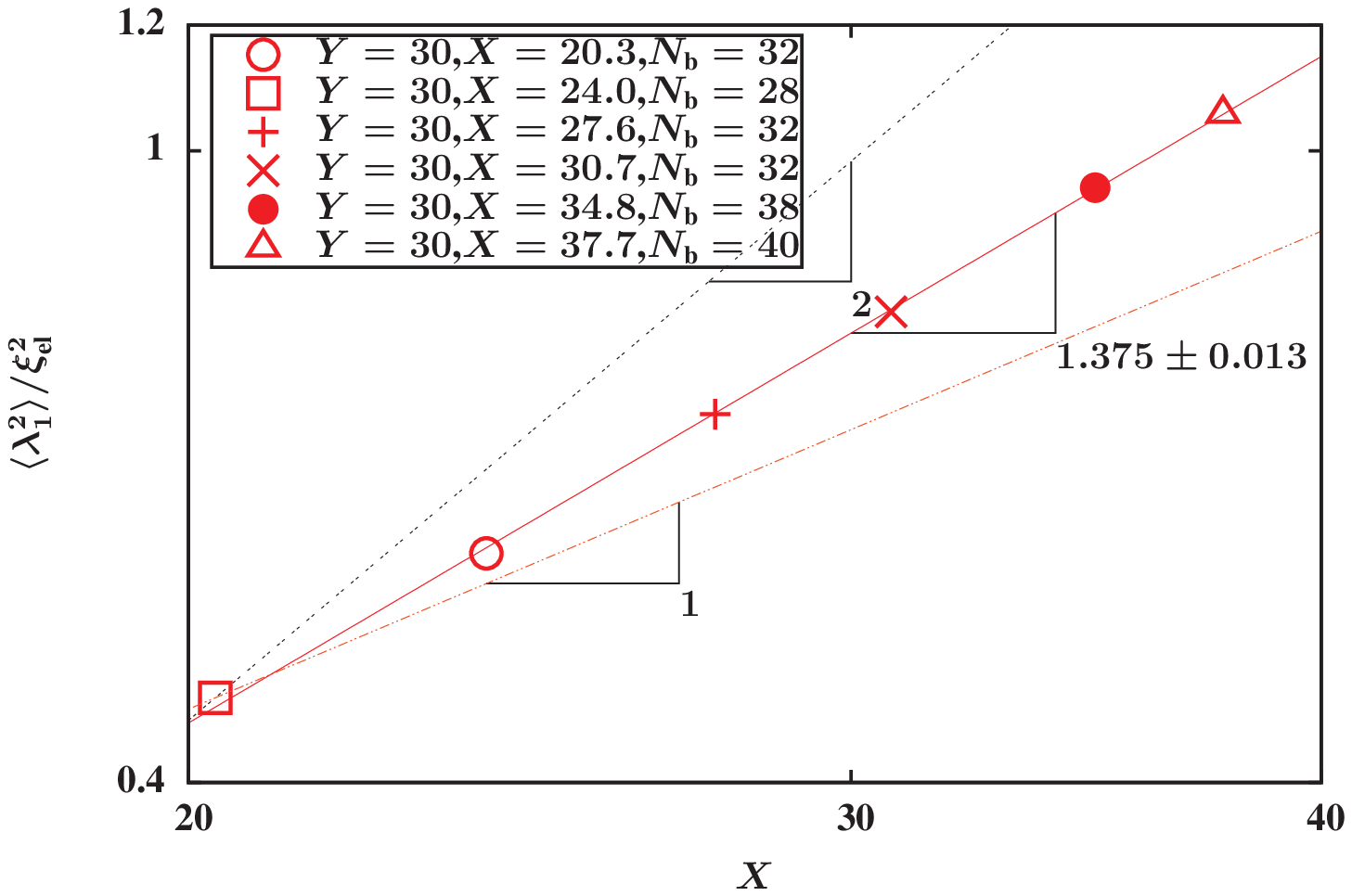} \\
(a) \\
\includegraphics[width=8cm,height=!]{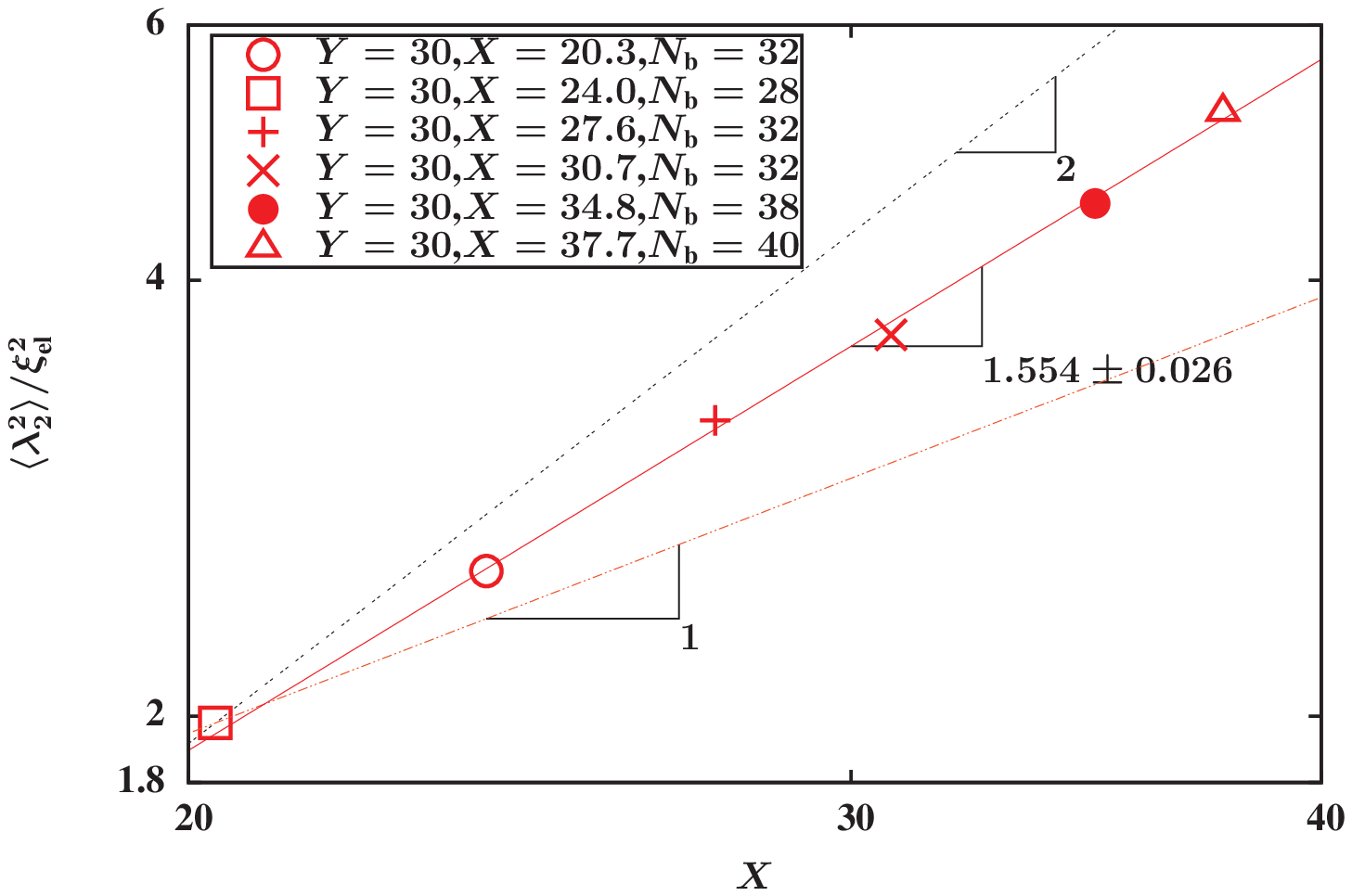} \\
(b)  \\
\end{tabular}
\end{center}
\caption{(Color online) Dependence on the number of blobs $X$, of  (a) the smallest scaled eigenvalue of the gyration tensor, $\langle \lambda^2_1 \rangle /\xi_\text{el}^2$, and (b) the intermediate scaled eigenvalue of the gyration tensor, $\langle \lambda^2_2 \rangle /\xi_\text{el}^2$,  at a reduced screening length, $Y=30$. Data was obtained using the multi-chain algorithm with $c/c^*=10^{-5}$. Error bars are of the order of symbol size. }
\label{fig:eigen12x}
\end{figure}

The dependence of the three scaled eigenvalues of the radius of gyration tensor, $\langle \lambda^2_1 \rangle/\xi_\text{el}^2$, $\langle \lambda^2_2 \rangle/\xi_\text{el}^2$, and $\langle \lambda^2_3 \rangle/\xi_\text{el}^2$, on the reduced screening length $Y$, is shown in Figs.~\ref{fig:eigeny}. As was observed for $\langle r_\text{e}^2 \rangle / \xi_\text{el}^2$ and $\langle r_\text{g}^2\rangle/\xi_\text{el}^2$, all the eigenvalues increase in the crossover regime from constant values at small values of $Y$, to constant values in the blob-pole regime at large values of $Y$. The collapse of data for different bead-spring chain model parameters, when represented in terms of blob scaling variables, is also observed for these properties. Clearly, the chain size increases with increasing electrostatic repulsion between blobs, not only in the direction of maximum chain stretching ($\langle \lambda^2_3\rangle$), but also in the lateral directions ($\langle \lambda^2_1\rangle$ and $\langle \lambda^2_2\rangle$). The significant differences in the magnitude of the chain dimensions in the three different principal directions clearly illustrates the highly anisotropic shape of the chain in all regions of the phase diagram. This is examined in more detail in terms of shape functions in the section below. A curious observation, for which we do not have an obvious explanation, is the presence of an overshoot in $\langle \lambda^2_1 \rangle /\xi_\text{el}^2$ and $\langle \lambda^2_2 \rangle /\xi_\text{el}^2$ for values of $Y$ at the threshold of the blob-pole regime, i.e. just before the chain enters the final screening-length-independent regime from the crossover regime. The continued stretching of $\langle \lambda^2_3\rangle$ with increasing $Y$ at the verge of the blob-pole regime appears to be accommodated by a shrinkage in $\langle \lambda^2_1\rangle$ and $\langle \lambda^2_2\rangle$. The overshoot also becomes more pronounced with increasing $X$. The insets in Figs.~\ref{fig:eigeny}~(a) and~(b) demonstrate that even the occurrence of an overshoot, which is distinct for different bead-spring chain parameters, collapses onto a single curve when represented in terms of scaling variables---signifying that blob variables accurately capture the essential physics, even for phenomena that go beyond the simple picture upon which they were originally based. The use of finitely extensible springs in the simulations is probably not  responsible for the overshoot, and it is more likely to be an electrostatic phenomenon, since the springs are weakly stretched even in the blob-pole regime. For instance, from the value of $\langle r_\text{e}^2\rangle/\xi_\text{el}^2$ in Fig.~\ref{fig:rvsy}~(a) ($\langle r_\text{e}^2\rangle/\xi_\text{el}^2 \approx 2000$, for $N_\text{b}= 60$ and $X=48$, in the limit of large $Y$), and the values of $\xi_\text{el}^{*}$ and $L^{*}$ corresponding to this value of $N_\text{b}$ and $X$ in Table~\ref{table3} ($\xi_\text{el}^{*} = 1.85$ and $L^{*} = 471$), we can see that the ratio of the end-to-end vector to the contour length is less than $16\%$. 

\begin{figure}[!t]
\begin{center}
\begin{tabular}{c}
\includegraphics[width=8cm,height=!]{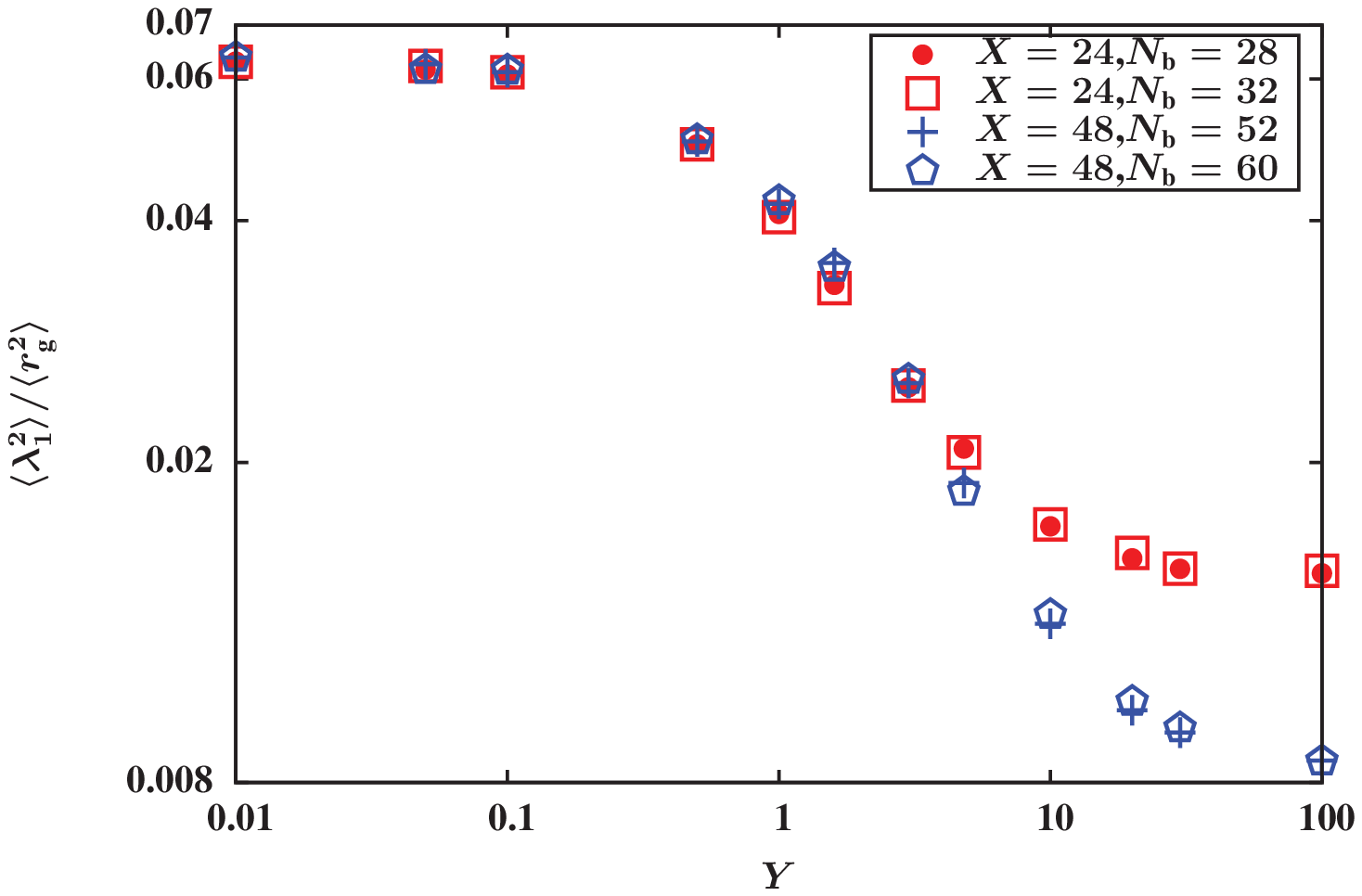} \\
(a) \\
\includegraphics[width=8cm,height=!]{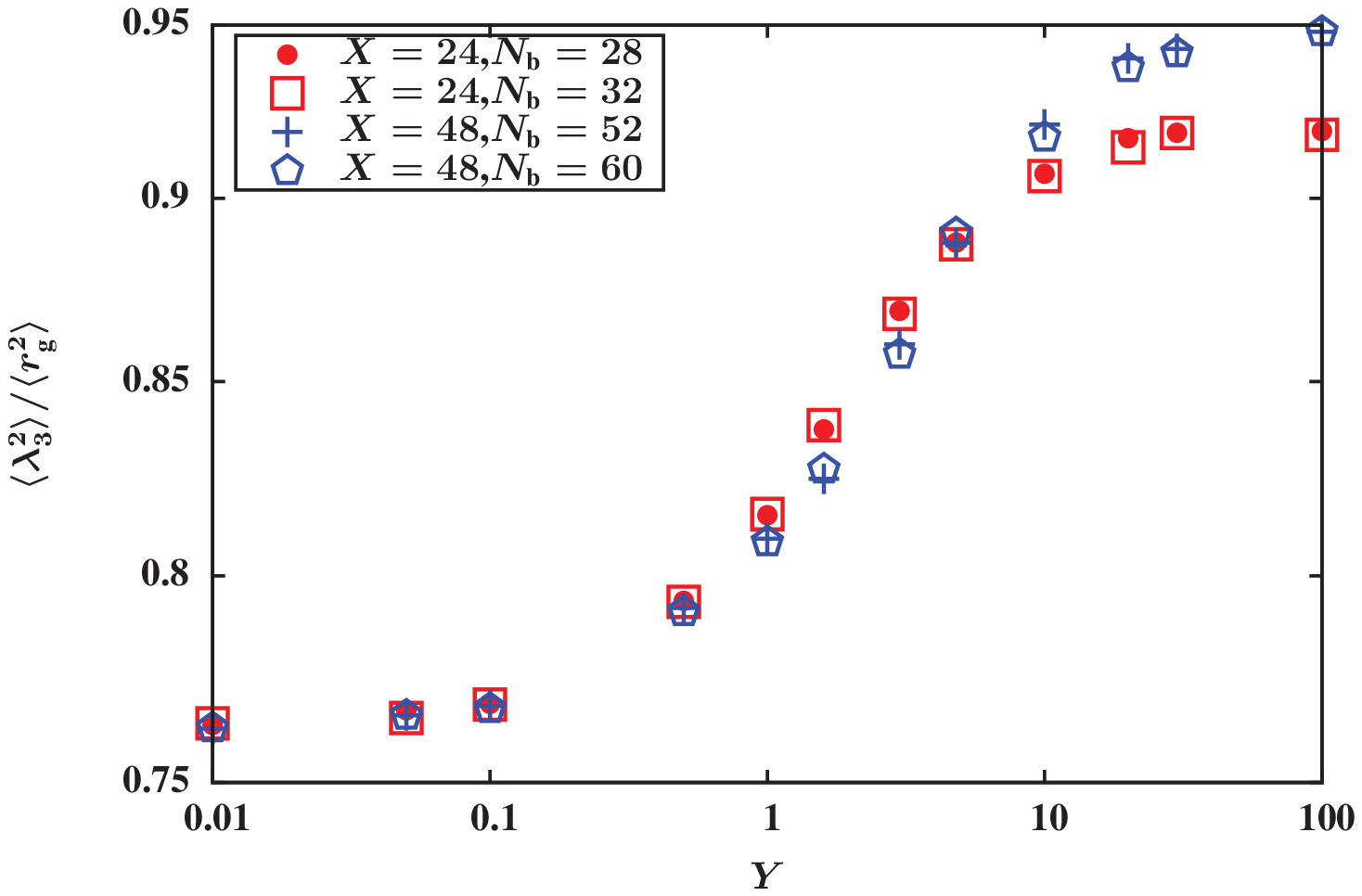} \\
(b)  \\
\end{tabular}
\end{center}
\vskip-15pt
\caption{(Color online) Dependence of the eigenvalues of the radius of gyration tensor, normalised by the radius of gyration, on the reduced screening length $Y$, at two values of the number of blobs, $X$: (a) the smallest eigenvalue $ \langle \lambda^2_1 \rangle /\langle r_\text{g}^2\rangle$,  
and (b) the largest eigenvalue $\langle \lambda^2_3 \rangle /\langle r_\text{g}^2\rangle$. Data was obtained using the multi-chain algorithm, with $c/c^*=10^{-5}$. Error bars are of the order of symbol size.}
\label{fig:eigennorm}
\end{figure}

\begin{table*}[t]
\caption{\label{table:eigen1} Normalized eigenvalues of the radius of gyration tensor for neutral random walk polymers compared with those for a polyelectrolyte chain (PE) in the ideal chain regime of the phase diagram. Predictions by ~\citet{Koyama1968} and ~\citet{Wei1997} were obtained from approximate analytical calculations for infinite Gaussian chains, while those by ~\citet{Solc1971,Solc1971a} and ~\citet{Kranbuehl1977} were obtained with lattice Monte Carlo simulations. Values reported from ~\citet{Theodorou1985} are for atactic polypropylene, obtained from Monte Carlo simulations based on a rotational isomeric state model, while those from ~\citet{Zifferer1999} are from lattice Monte Carlo simulations of ``nonreversal random walks'', which are considered to represent $\theta$-solvents. Data for the polyelectrolyte solution was obtained using the multi-chain algorithm, with $c/c^*= 10^{-5}$.} 
\vskip10pt
\begin{tabular}{l|c|c|c|c|c|c}
\hline  
\hline 
Solution & Chain length  & $\langle \lambda^2_1\rangle/\langle r_\text{g}^2\rangle$ & $\langle \lambda^2_2\rangle/\langle r_\text{g}^2\rangle$ & $\langle \lambda^2_3\rangle/\langle r_\text{g}^2\rangle$ &  $\langle \lambda^2_2\rangle/\langle \lambda^2_1\rangle$ & $\langle \lambda^2_3\rangle/\langle \lambda^2_1\rangle$  \\ 
\hline 
Neutral~\cite{Koyama1968} & $N_\text{b} \to \infty $ & $ 0.0646 $ & $ 0.175$ &$ 0.754 $ & $ 2.5$ & $ 10.6 $  \\
\hline  
Neutral~\cite{Wei1997} & $N_\text{b} \to \infty $ & $ 0.071 $ & $ 0.179$ &$ 0.750 $ & $ \cdots $ & $10.554  $  \\
\hline  
Neutral~\cite{Solc1971,Solc1971a} & $N_\text{b} = 100  $ & $ 0.065 $ & $ 0.175 $ & $ 0.75 $ & $ 2.7 $ & $ 11.7 $  \\
\hline  
Neutral~\cite{Kranbuehl1977} & $N_\text{b} = 63$ & $ 0.065 $&$ 0.176 $ & $ 0.76  $ & $ 2.70 $ & $ 11.7 $  \\
\hline 
Neutral~\cite{Theodorou1985} & $N_\text{b} = 999 $ & $ 0.060 \pm  0.004 $&$ 0.17 \pm 0.01 $ &$ 0.77 \pm 0.06 $ & $ 2.9 \pm 0.2 $& $ 12.7 \pm 0.9 $  \\
\hline 
Neutral~\cite{Zifferer1999} & $N_\text{b} \to \infty $ & $ 0.0633 $&$ 0.1721 $ &$ 0.7645 $ & $ \cdots  $& $ \cdots  $  \\
\hline 
PE ($Y = 0.01$) & $X = 24$  & $ 0.0631\pm 0.0003 $&$ 0.1734\pm 0.0009$&$ 0.7626\pm 0.0061 $ & $ 2.75\pm 0.02 $& $ 12.10 \pm 0.12$  \\
\hline  
\hline 
\end{tabular} 
\end{table*}

\begin{figure}[t]
\begin{center}
\begin{tabular}{c}
\includegraphics[width=8cm,height=!]{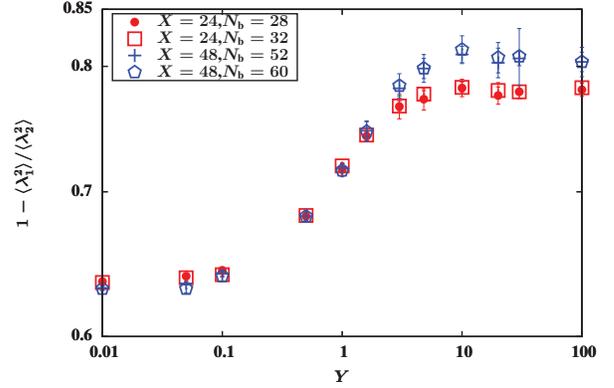} \\
\end{tabular}
\end{center}
\caption{(Color online) Asymmetry in chain shape in the plane perpendicular to the stretching axis as a function of the reduced screening length, $Y$. Data was obtained using the multi-chain algorithm, with $c/c^*=10^{-5}$. }
\label{fig:eigenratio}
\end{figure}

\begin{table*}[!t]
\caption{\label{table:eigen2} Normalized eigenvalues of the radius of gyration tensor for various values of the number of blobs $X$, at a reduced screening length, $Y=30$. Data was obtained using the multi-chain algorithm, with $c/c^*=10^{-5}$. } 
\vskip10pt
\begin{tabular}{c|c|c|c|c|c|c}
\hline  
\hline 
$X$ & $\langle \lambda^2_1\rangle/\langle r_\text{g}^2\rangle$ & $\langle \lambda^2_2\rangle/\langle r_\text{g}^2\rangle$ & $\langle \lambda^2_3\rangle/\langle r_\text{g}^2\rangle$  & $\langle \lambda^2_2\rangle/\langle \lambda^2_1\rangle$ & $\langle \lambda^2_3\rangle/\langle \lambda^2_1\rangle$ & $1 - \langle \lambda^2_1\rangle/\langle \lambda^2_2\rangle$  \\ 
\hline 
20.3 & $ 0.0168\pm 0.0001 $&$ 0.0734\pm 0.0003$&$ 0.9099\pm 0.0026 $ & $ 4.37\pm 0.02 $ & $ 54.2\pm 0.22$ &$ 0.771\pm 0.001$ \\
\hline  
24 & $0.0148\pm 0.0001$ & $0.0666\pm 0.0004$ & $0.9186\pm 0.0038$ & $4.52\pm 0.03$ & $62.3\pm 0.30$ &$ 0.779\pm 0.002$ \\
\hline 
27.6 &  $ 0.0136\pm 0.0001 $ & $ 0.0636\pm 0.0011$ & $ 0.9228\pm 0.0078 $  & $ 4.69\pm 0.09 $&$ 68.1\pm 0.88$ & $ 0.787\pm 0.004$ \\
\hline 
30.7 &  $ 0.0125\pm 0.0001 $&$ 0.0578\pm 0.0004$&$ 0.9297\pm 0.0034 $ & $ 4.63\pm 0.04 $&$ 74.5\pm 0.43$ &$ 0.784 \pm 0.002$ \\
\hline 
34.8 & $ 0.0114\pm 0.0001 $&$ 0.0545\pm 0.0011$&$ 0.9341\pm 0.0076 $  & $ 4.77\pm 0.11 $&$ 81.8\pm 1.12$ & $ 0.790 \pm 0.004$ \\
\hline 
37.7 &  $ 0.0108\pm 0.0001 $&$ 0.0536\pm 0.0004$&$0.9356\pm 0.0030 $ & $ 4.95\pm 0.04 $&$ 86.5\pm 0.40$ &$ 0.798 \pm 0.001$ \\ 
\hline 
48 &  $0.0094\pm 0.0002$ & $0.0488\pm 0.0013$ & $0.9418\pm 0.0074$ & $5.22\pm 0.16$ & $100.7\pm 1.7$ & $0.808 \pm 0.006$ \\ 
\hline 
\hline 
\end{tabular} 
\end{table*}

The scaling of $ r_\text{e}$ with $X$ in the blob-pole regime was seen in Fig.~\ref{fig:rvsx}~(b) to exhibit logarithmic corrections in line with the prediction of Eq.~(\ref{eq:relogX}).   Since the largest eigenvalue of the radius of gyration tensor is expected to be in the direction of maximum chain stretching, we expect $\sqrt{\langle \lambda^2_3 \rangle} /\xi_\text{el}$ to also exhibit logarithmic corrections according to Eq.~(\ref{eq:relogX}). Figure~\ref{fig:eigen3x}~(a) displays the departure from the OSFKK scaling prediction in the blob-pole regime, while Fig.~\ref{fig:eigen3x}~(b) shows that the logarithmic corrections have an exponent close to the expected value of $1/3$. 

According to OSFKK theory, the scaling of chain dimensions lateral to the elongation axis is expected to remain unperturbed in the blob-pole regime, i.e., to obey random walk statistics. To our knowledge, there are no theories that describe modifications to this scaling behaviour due to the nonuniform stretching of the chain (unlike the refined theories developed for scaling in the direction of maximum stretching). The dependence of chain dimensions on the number of blobs, in the directions perpendicular to the stretching direction, is examined in Figs.~\ref{fig:eigen12x}~(a) and~(b), where, $\langle \lambda^2_1 \rangle /\xi_\text{el}^2$ and $\langle \lambda^2_2 \rangle /\xi_\text{el}^2$ are plotted as functions of $X$. For the limited range of values of $X$ examined here, the smallest and intermediate eigenvalues appear to obey power law scaling with exponents that lie somewhere between ideal chain and blob-pole scaling.

Eigenvalues of the radius of gyration tensor for neutral polymer chains are usually reported in terms of ratios, either between individual eigenvalues, or with the mean square radius of gyration, since they are expected to attain universal values in the limit of long chains. For a chain with a spherically symmetric shape about the centre of mass, we expect $\langle \lambda^2_i \rangle /\langle r_\text{g}^2\rangle = 1/3$, for $i=1, 2, 3$, and $ \langle \lambda^2_i \rangle /  \langle \lambda^2_j \rangle = 1$ for all combinations  $i$ and $j$. Predictions reported previously in the literature for flexible neutral chains in $\theta$-solutions, from a variety of different approaches, are displayed in Table~\ref{table:eigen1}. The strong asymmetry in the shapes of chains is clearly apparent from the extent of departure from the expected values. ~\citet{Steinhauser2005}  has carried out molecular dynamics simulations to examine the influence of solvent quality on the ratios of eigenvalues, and noted that while their values are relatively similar in good and theta solvents, they change significantly in poor solvents, with values that imply that chains become much more spherical as they collapse into globules. From OSFKK theory, we expect polyelectrolyte chains to behave like  ideal chains for $Y \ll 1$. As can be seen from the last row of Table~\ref{table:eigen1}, this is indeed the case for the results obtained with the current simulations, with all the different ratios being in good agreement with results for neutral chains. 

The dependence of the eigenvalues scaled with the mean square radius of gyration, on the reduced screening length $Y$, is displayed in Figs.~\ref{fig:eigennorm}. Starting with the limiting values reported in Table~\ref{table:eigen1} for polyelectrolyte chains in the ideal chain regime, both $ \langle \lambda^2_1 \rangle /\langle r_\text{g}^2\rangle$ and $\langle \lambda^2_2 \rangle /\langle r_\text{g}^2\rangle$ (not shown here) decrease with increasing $Y$ in the crossover regime until they reach their asymptotic values in the blob-pole regime. The reason for the decrease in the magnitude of these ratios is because $\langle r_\text{g}^2\rangle = \langle \lambda^2_1 \rangle + \langle \lambda^2_2 \rangle + \langle \lambda^2_3 \rangle$ is dominated by the behaviour of $\langle \lambda^2_3 \rangle$, which can be seen from Figs.~\ref{fig:eigeny} to increase much more rapidly with $Y$ than both $\langle \lambda^2_1 \rangle$ and $\langle \lambda^2_2 \rangle$. For the same reason, the ratio $\langle \lambda^2_3 \rangle /\langle r_\text{g}^2\rangle$ increases with $Y$, nearly approaching a value of 1 in the limit of large $Y$. The overshoot in $\langle \lambda^2_1 \rangle$ and $\langle \lambda^2_2 \rangle$ that was observed in Figs.~\ref{fig:eigeny}~(a) and~(b) at the threshold of the blob-pole regime, is not noticeable when the eigenvalues are scaled with $\langle r_\text{g}^2\rangle$ in place of $\xi_\text{el}^2$ (as displayed in Fig.~\ref{fig:eigennorm}~(a) for $\langle \lambda^2_1 \rangle/\langle r_\text{g}^2\rangle$).

The extent of asymmetry in chain shape in the plane perpendicular to the elongation axis can be gauged by the quantity, $1-\langle \lambda_1^2 \rangle/\langle \lambda_2^2 \rangle$, which would be zero if the shape was symmetric. For neutral chains, and for polyelectrolyte chains in the ideal chain regime, using the values of  $\langle \lambda^2_1 \rangle /\langle r_\text{g}^2\rangle$ and $\langle \lambda^2_2 \rangle /\langle r_\text{g}^2\rangle$ reported in Table~\ref{table:eigen1}, we can calculate $1-\langle \lambda_1^2 \rangle/\langle \lambda_2^2 \rangle \approx 0.63$. Fig.~\ref{fig:eigenratio} shows that with increasing electrostatic repulsion between blobs, the already highly asymmetric shape becomes even more so, levelling off to a constant value in the limit of large $Y$. 

A notable aspect of the normalised eigenvalue ratios displayed in Figs.~\ref{fig:eigennorm} and~\ref{fig:eigenratio} is that their values are independent of the number of blobs $X$ in the chain, for a wide variety of values of $Y$, ranging from the ideal chain regime and well into the crossover regime. The ratios become $X$ dependent only as the chains approach the blob-pole regime with increasing screening length. This suggests that for sufficiently long chains, as in the case of neutral polymer chains,  the normalised eigenvalues of polyelectrolyte chains attain universal values in those parts of the phase diagram that lie outside the blob-pole regime. Within the blob-pole regime, we expect that ratios of quantities that scale nearly identically with $X$ will attain roughly constant values, such as $\langle \lambda^2_3\rangle/\langle r_\text{e}^2\rangle$ and $\langle \lambda^2_3\rangle/\langle r_\text{g}^2\rangle$ (which follows from the dominance of the growth of $\langle \lambda^2_3\rangle$ with $X$, compared to $\langle \lambda^2_1\rangle$ and $\langle \lambda^2_2\rangle$), while ratios of quantities that do not scale identically, would retain a dependence on the number of blobs. This is examined in Table~\ref{table:eigen2}, where the dependence of a variety of ratios on the number of blobs $X$, at the particular value, $Y=30$, which is in the blob-pole regime, is tabulated. At any fixed value of $Y$, however, we expect chains to move out of the blob-pole regime as $X \to \infty$, and for all the ratios to become universal.

\begin{figure*}[t]
\begin{center}
\begin{tabular}{cc}
\resizebox{8.25cm}{!} {\includegraphics*[width=8cm]{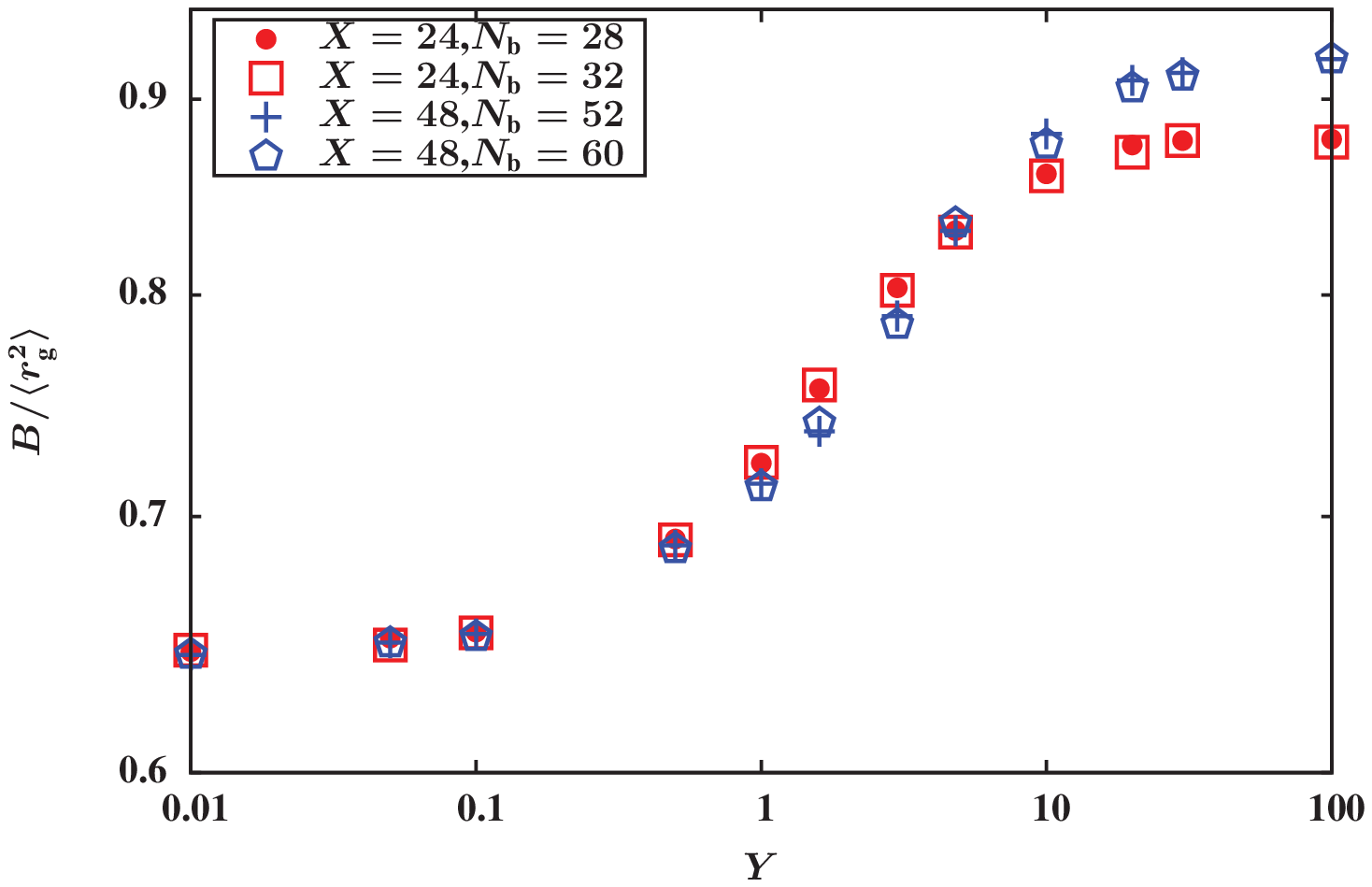}} &
\resizebox{8.25cm}{!} {\includegraphics*[width=8cm]{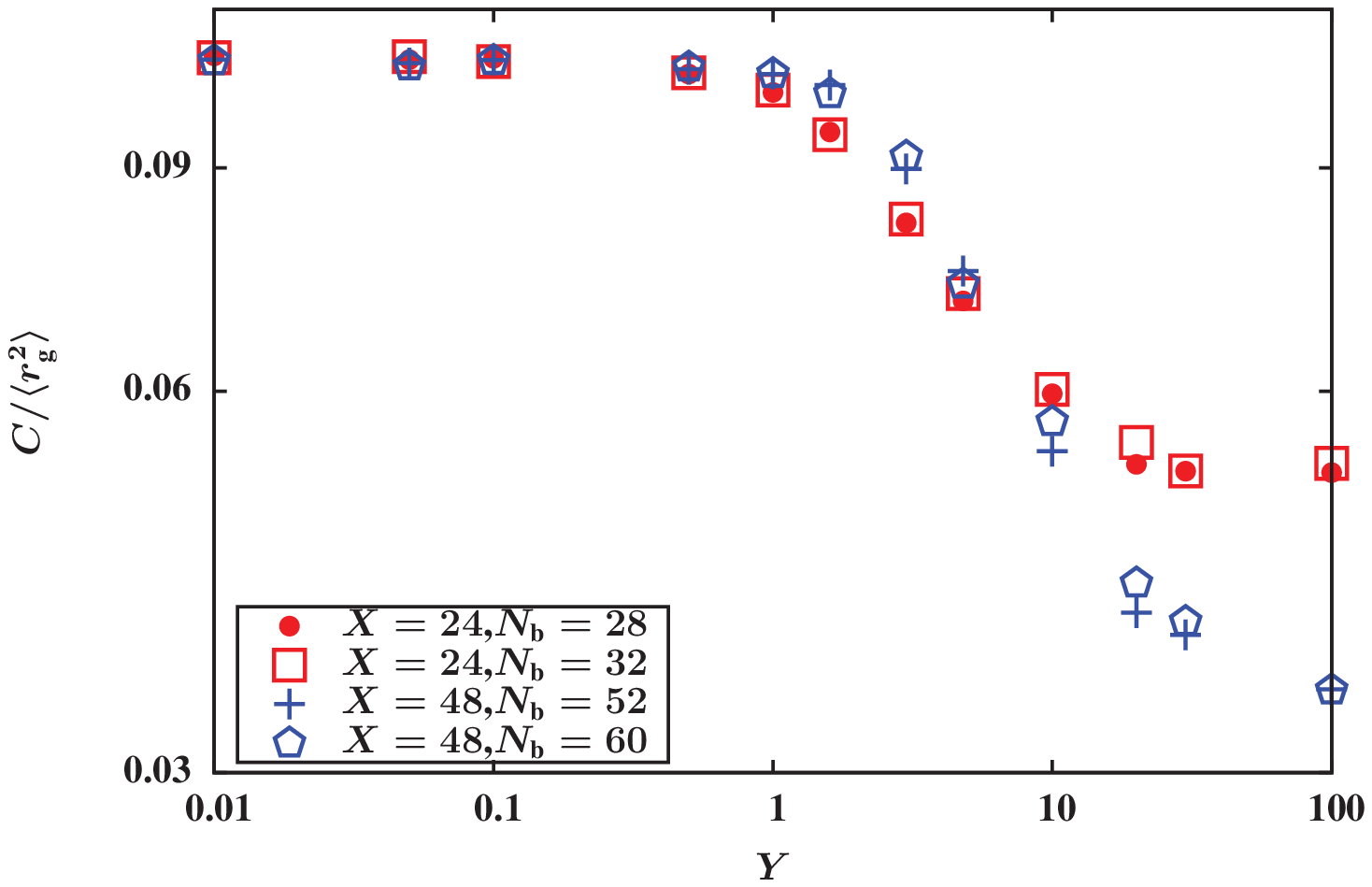}}\\
(a) & (b)  \\
\resizebox{8.25cm}{!} {\includegraphics*[width=8cm]{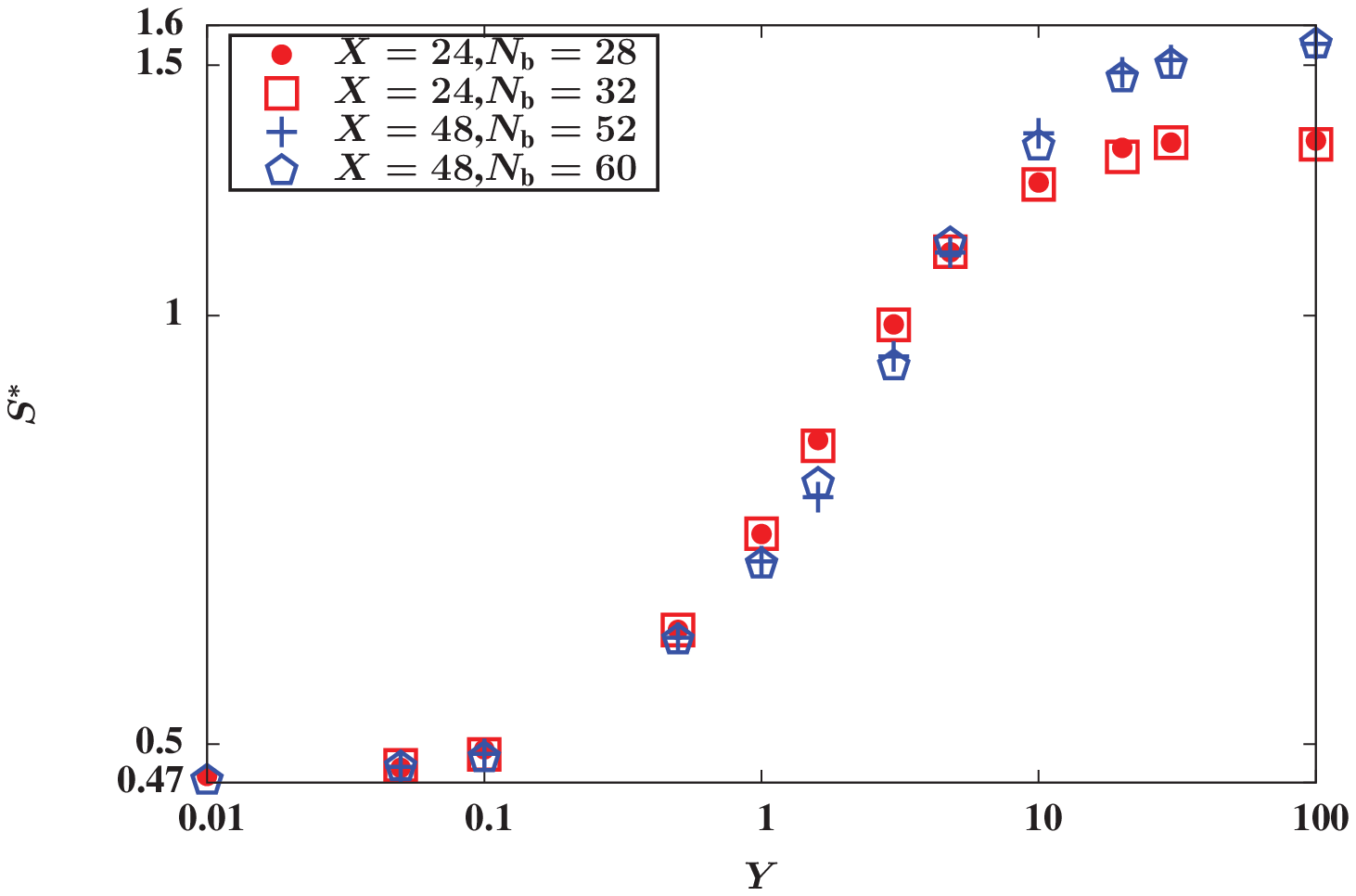}} &
\resizebox{8.25cm}{!} {\includegraphics*[width=8cm]{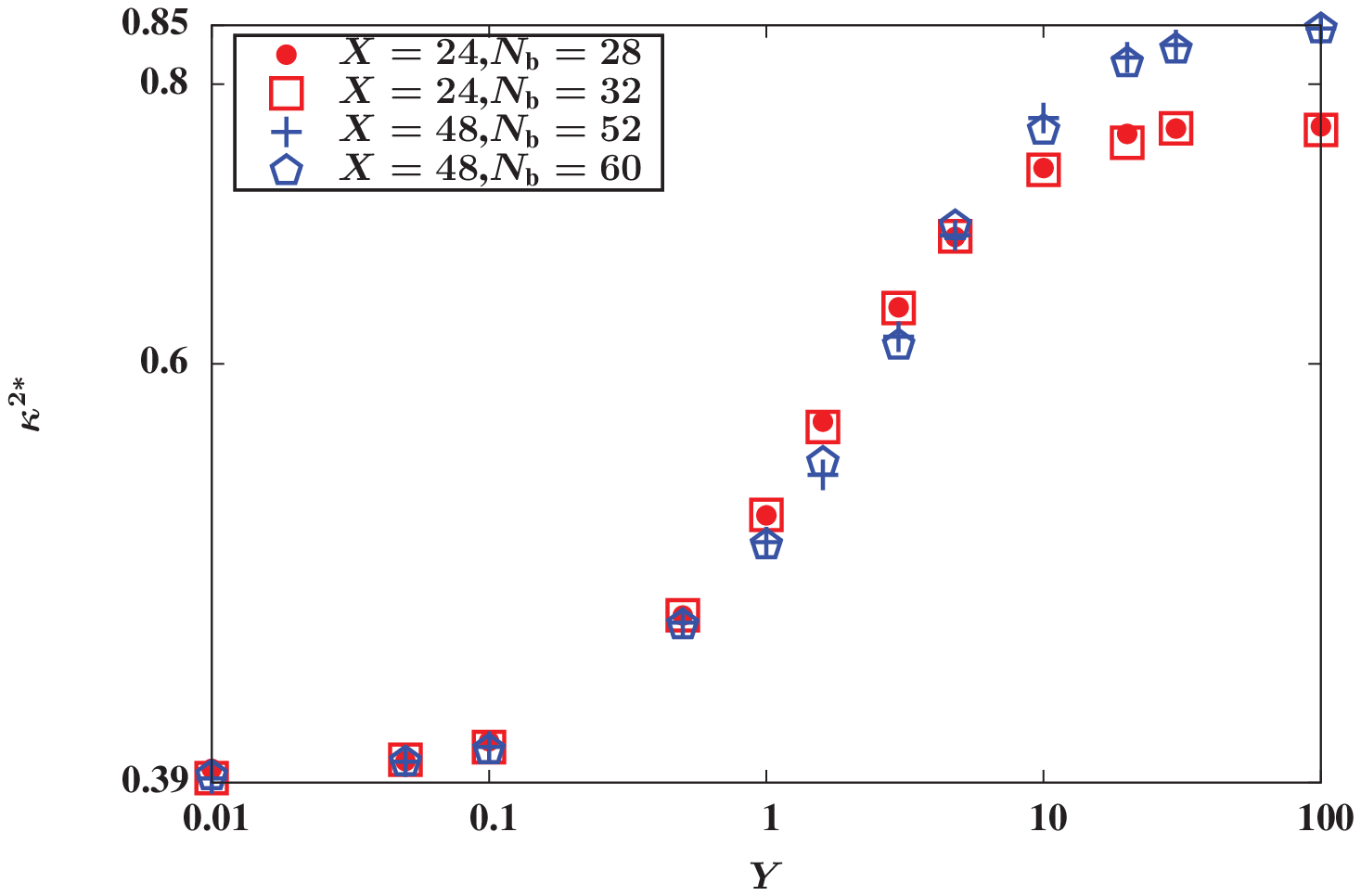}}\\
(c) & (d)  \\
\end{tabular}
\end{center}
\caption{(Color online) The dependence of shape functions on the reduced screen length $Y$, at two values of $X$: (a) the normalised asphericity $B/\langle r_\text{g}^2\rangle$ (see Eq.~(\ref{eq:sph})), (b) the normalised acylindricity $C/\langle r_\text{g}^2\rangle$ (see Eq.~(\ref{eq:cyl})), (c) the degree of prolateness, $S^{*}$ (see Eq.~(\ref{eq:pro2})), and, (d) the shape anisotropy, ${\kappa^{2}}^{*}$ (see Eq.~(\ref{eq:kappa2})). When not displayed explicitly, error bars are of the order of symbol size. Data obtained with the multi-chain code, with $c/c^*=10^{-5}$.}
\label{fig:acy}
\end{figure*}

\begin{figure*}[t]
\begin{center}
\begin{tabular}{cc}
\resizebox{8 cm}{!}{\includegraphics*[width=7.5cm]{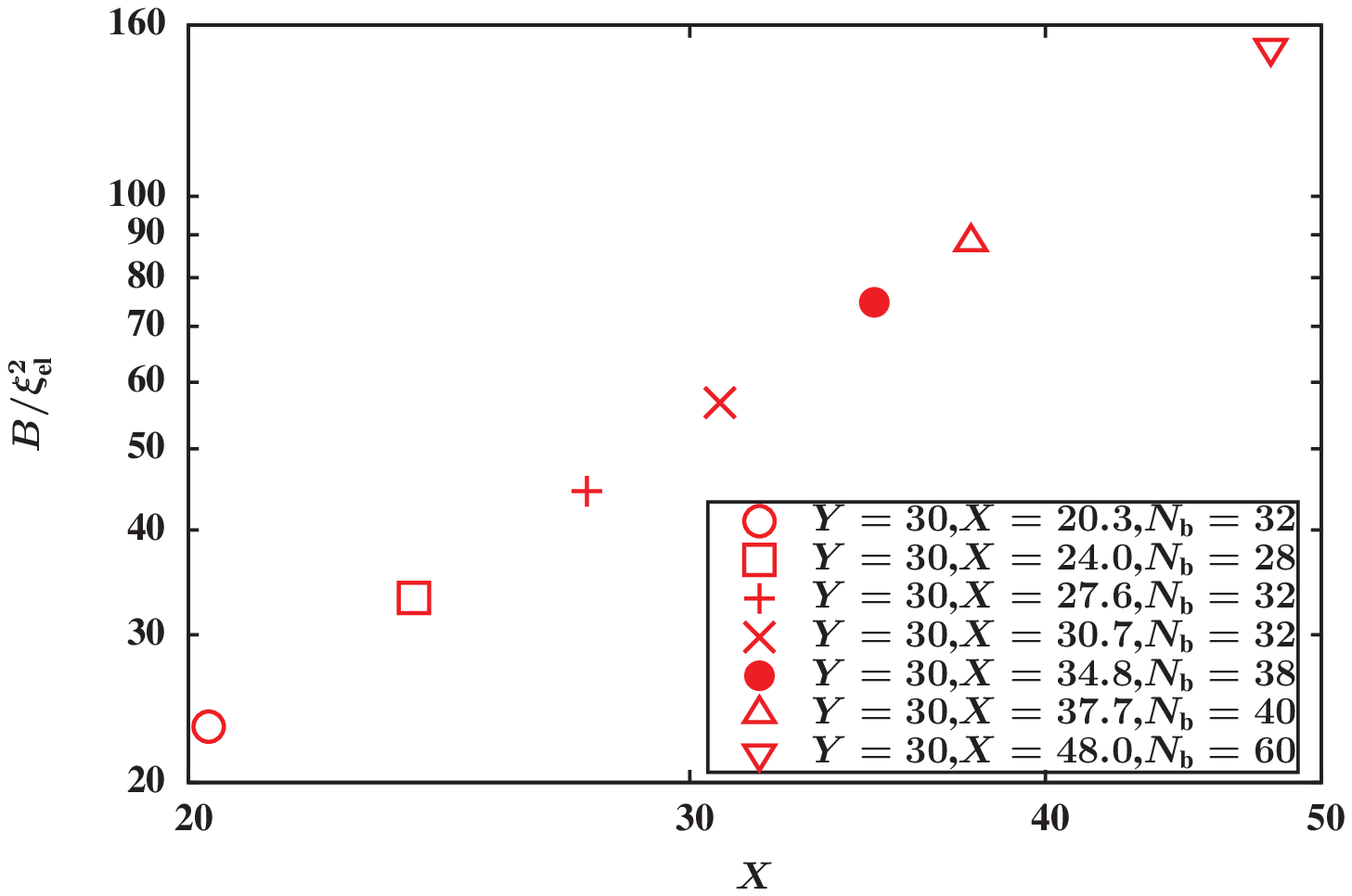}}  &
\resizebox{8 cm}{!}{\includegraphics*[width=7.5cm]{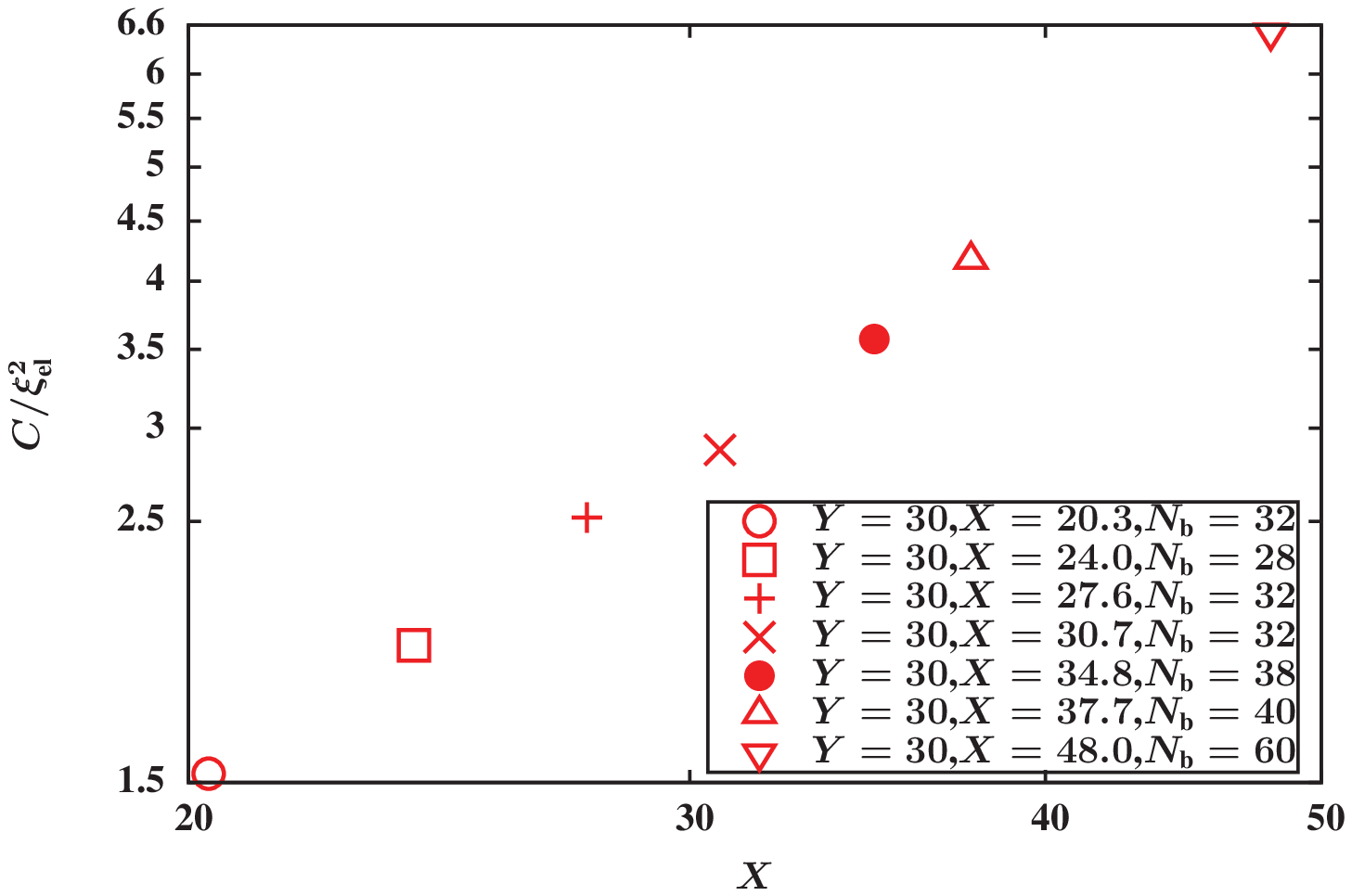}} \\
(a) & (b) \\
\resizebox{8 cm}{!}{\includegraphics*[width=7.5cm]{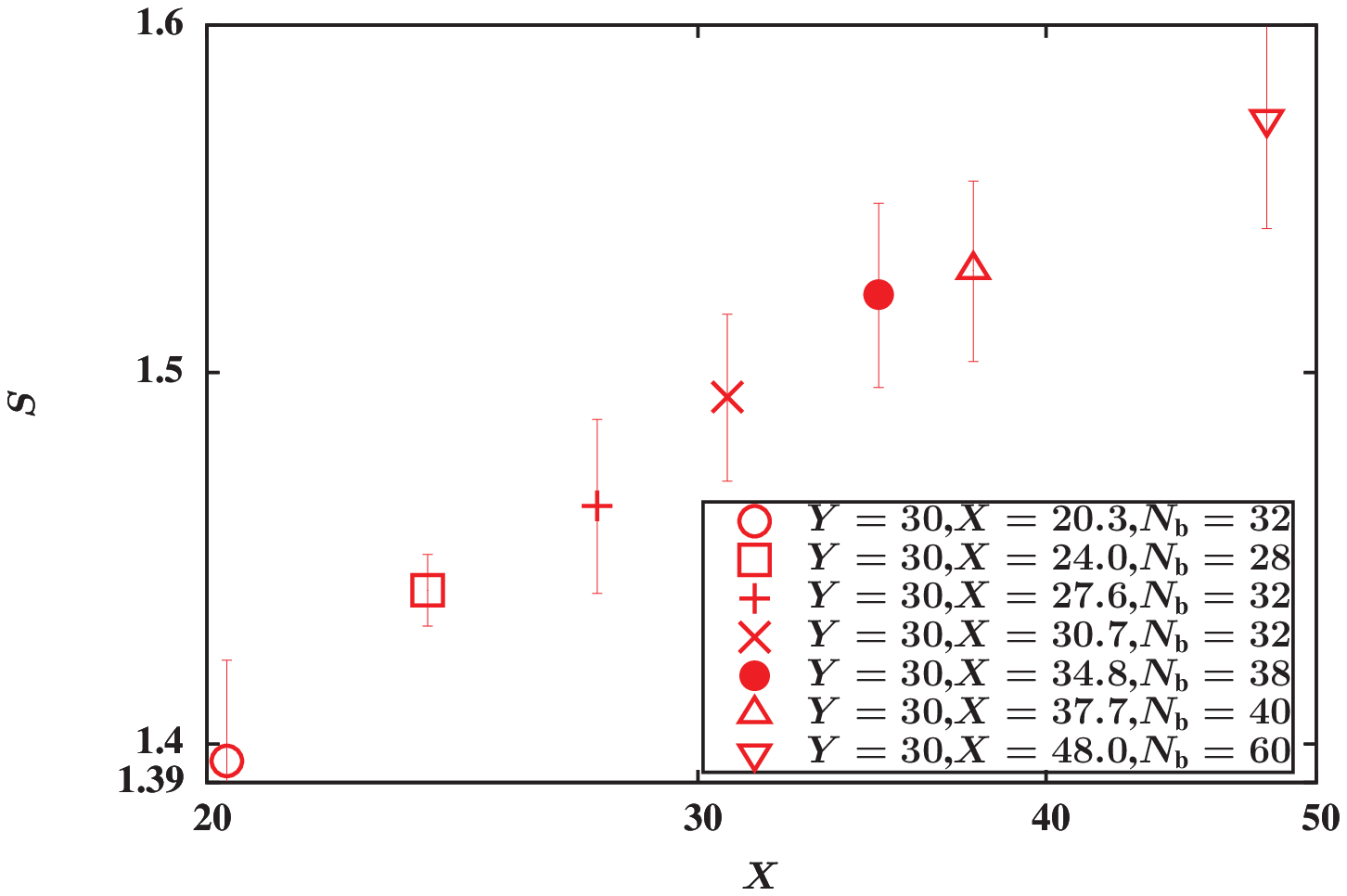}}  &
\resizebox{8 cm}{!}{\includegraphics*[width=7.5cm]{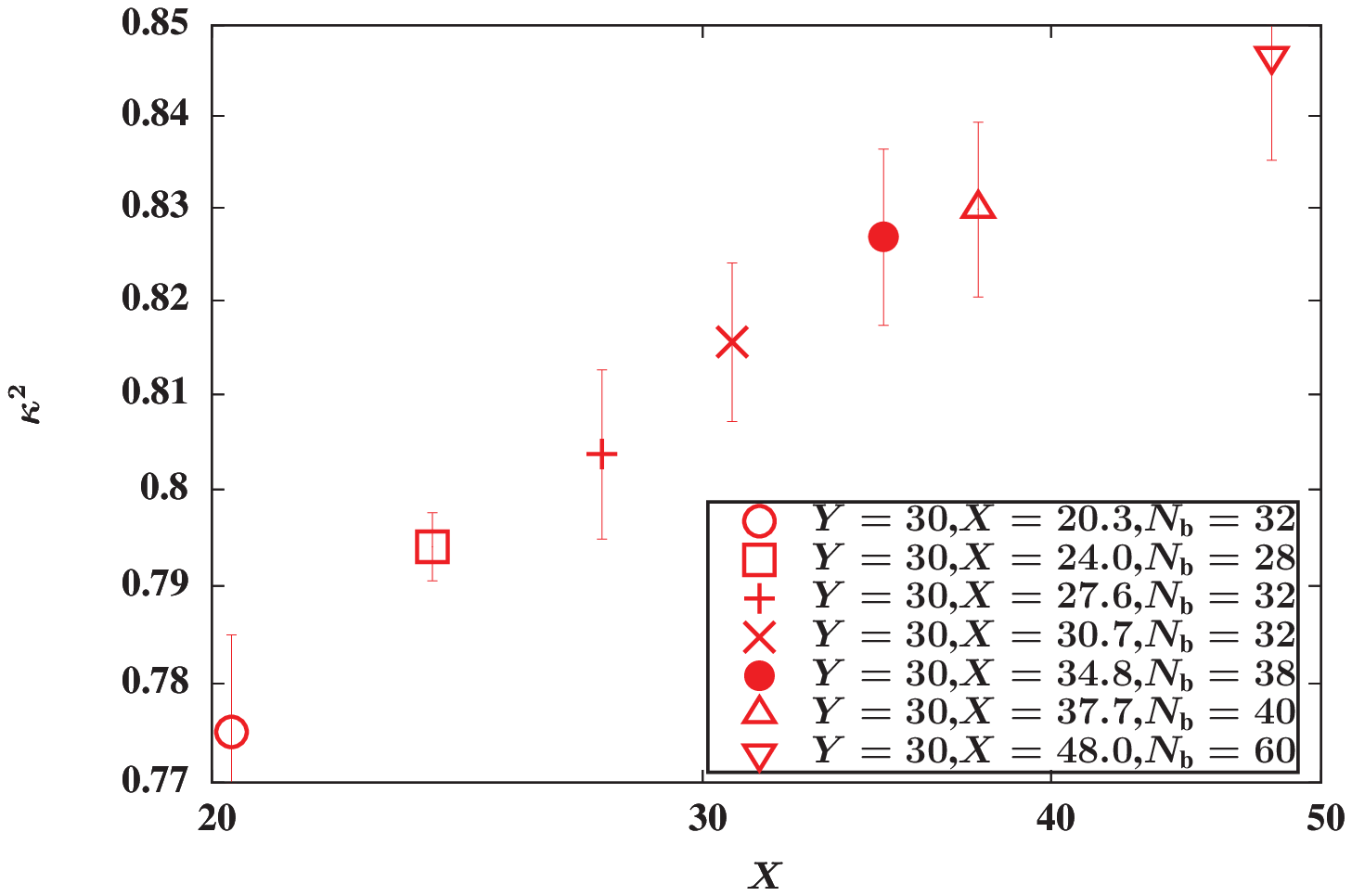}} \\
(c) & (d) \\
\end{tabular}
\end{center}
\caption{(Color online) The dependence of shape functions on the number of blobs $X$, in the blob-pole regime, at a particular value of the reduced screening length $Y = 30$: (a) the scaled asphericity $B/\xi_\text{el}^{2}$, (b) the scaled acylindricity $C/\xi_\text{el}^{2}$, (c) the degree of prolateness $S$, and, (d) the shape anisotropy $\kappa^{2}$. Error bars are of the order of symbol size in~(a) and~(b). Data obtained with the multi-chain code, with $c/c^*=10^{-5}$.}
\label{fig:pro}
\end{figure*}

\begingroup
\squeezetable
\begin{table*}[t]
\caption{\label{table:sf} Shape functions for neutral random walk polymers compared with those for a polyelectrolyte chain (PE) in the ideal chain regime of the phase diagram. Predictions by ~\citet{Steinhauser2005} were obtained with molecular dynamics simulations, while those by ~\citet{Bishop1986} and ~\citet{Bishop1991} were obtained with Brownian dynamics, and off-lattice Monte Carlo simulations, respectively. Methods used to evaluate shape functions in  Refs.~\citenum{Theodorou1985}, \citenum{Wei1997} and~\citenum{Zifferer1999} are given in the caption to Table~\ref{table:eigen1}. Data for the polyelectrolyte solution was obtained using the multi-chain algorithm, with $c/c^*= 10^{-5}$.} 
\vskip10pt
\begin{tabular}{l|c|c|c|c|c|c|c}
\hline  
\hline 
Solution & Chain length  & $B/\langle r_\text{g}^2\rangle$ & $C/\langle r_\text{g}^2\rangle$ & $S^{*}$ &  $S$ & ${\kappa^{2}}^{*}$ & $\kappa^{2}$ \\ 
\hline  
Neutral~\cite{Wei1997} & $N_\text{b} \to \infty $ & $ \cdots   $ & $ \cdots  $ & $0.4750 $ &$ 0.8875 $  & $0.3943$  & $0.5263  $\\
\hline  
Neutral~\cite{Theodorou1985} & $N_\text{b} = 999 $ & $ 0.66  \pm 0.04$ & $ 0.106  \pm 0.007  $ & $   \cdots $ & $  \cdots  $ & $0.41 \pm 0.01$ & $ \cdots    $ \\
\hline 
Neutral~\cite{Bishop1986} & $N_\text{b} =  48$ & $\cdots   $ & $ \cdots   $ & $ 0.460\pm 0.021 $ & $  0.859\pm 0.245 $  & $ 0.384 \pm 0.011 $  &  $ 0.518\pm 0.095 $\\
\hline  
Neutral~\cite{Bishop1991} & $N_\text{b} \to \infty $ & $ \cdots    $ & $ \cdots   $ & $\cdots  $ & $ \cdots $ & $ 0.397 \pm 0.001 $  &  $ 0.529\pm 0.001 $\\
\hline 
Neutral~\cite{Steinhauser2005} & $N_\text{b} \to \infty $ & $  0.6255 \pm 0.0005  $ & $ \cdots    $ & $ \cdots  $ & $ \cdots $ &  $ 0.394 \pm 0.003 $ &  $ 0.5221\pm 0.0013 $\\
\hline 
Neutral~\cite{Zifferer1999} & $N_\text{b} \to \infty $ &$ \cdots   $ & $ \cdots   $ & $0.4751 $ & $ 0.8875 $ & $0.3943$  &  $0.5264  $\\
\hline 
PE ($Y = 0.01$) & $X = 24$  & $0.6453\pm 0.0056$ & $0.1103\pm 0.0006$ & $0.4751\pm 0.0024$ & $0.8810\pm 0.0205$  & $0.3955\pm 0.0011$  &  $0.5253\pm 0.0074$\\
\hline  
\hline 
\end{tabular} 
\end{table*}
\endgroup

\subsection{\label{sec:shape} Shape functions}

For chain shapes with tetrahedral or greater symmetry, the asphericity $B = 0$, otherwise $B > 0$. For chain shapes with cylindrical symmetry, the acylindricity $C = 0$, otherwise $C > 0$. With regard to the degree of prolateness, its sign determines whether chain shapes are preponderantly oblate ($S, S^{*} \in \left[-0.25,0 \right]$) or prolate ($S, S^{*} \in \left[0,2 \right]$). The relative anisotropy ($\kappa^{2}$ and ${\kappa^{2}}^{*}$), on the other hand, lies between 0 (for spheres) and 1 (for rods). In this context, the values of the normalised  shape functions $B/\langle r_\text{g}^2\rangle$ and $C/\langle r_\text{g}^2\rangle$, $S$ and $S^{*}$, and $\kappa^{2}$ and ${\kappa^{2}}^{*}$, predicted by various techniques for neutral polymer chains are compared with the predictions of the current simulations for polyelectrolyte chains in the ideal chain regime, in Table~\ref{table:sf}. Clearly, in this regime, polyelectrolyte chains behave identically to neutral random walk chains, and share the same high degree of anisotropy. 

The dependence of the shape functions $B/\langle r_\text{g}^2\rangle$, $C/\langle r_\text{g}^2\rangle$, $S^{*}$ and ${\kappa^{2}}^{*}$ on $Y$, for two different values of $X$, is displayed in Figs.~\ref{fig:acy}. As we might expect, chains appear to become more aspherical, more cylindrical, more prolate, and more rodlike, with increasing electrostatic repulsion between the blobs. The functions $S$ and ${\kappa^{2}}$ behave similarly to $S^{*}$ and ${\kappa^{2}}^{*}$.

The independence of the values of the shape functions from the number of blobs $X$, in all the regimes except the blob-pole regime, suggests that for sufficiently long chains, at any given value of the reduced screening length $Y$, the shapes of polyelectrolyte chains are universal. In the blob-pole regime however, as noted earlier, the appearance of corrections to scaling (to varying extents in the three principal directions), leads to a dependence of chain shapes on the chain length. The nature of this dependence on $X$, at a particular value of $Y=30$, is displayed in Figs.~\ref{fig:pro} for $B/\xi_\text{el}^{2}$, $C/\xi_\text{el}^{2}$, $S$ and ${\kappa^{2}}$. The functions $S^{*}$ and ${\kappa^{2}}^{*}$ behave similarly to $S$ and ${\kappa^{2}}$.

\subsection{\label{sec:diff} Translational diffusivity interpreted with the OSFKK scaling picture}

The dependence on the reduced screening length $Y$, of all the static properties examined here so far, has followed a similar pattern: they exhibit  $Y$ independent behaviour in the ideal chain and blob-pole regimes, while they are functions of $Y$ in the crossover regime between these two limits. By defining the Zimm diffusivity of an electrostatic blob through the expression,
\begin{equation}
\label{eq:theta}
D_{\xi}= \frac{k_\text{B} T}{6 \pi \eta_\text{s} (\xi_\text{el}/2)}
\end{equation}
we find that a similar pattern is obeyed by the ratio $D/D_{\xi}$, where $D$ is the translational diffusivity of the polyelectrolyte chain as a whole. 

\begin{figure}[t]
\begin{center}
\begin{tabular}{c}
\includegraphics[width=8cm,height=!]{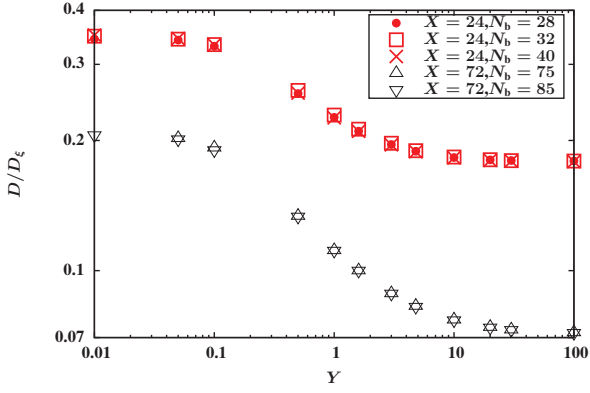} \\
\end{tabular}
\end{center}
\caption{(Color online) Dependence of the ratio of the translational diffusivity of the chain $D$, to the diffusivity of the electrostatic blob $D_{\xi}$, on the reduced screening length $Y$, at two values of the number of blobs, $X$. Error bars are of the order of symbol size.}
\label{fig:DbyDxi}
\end{figure}

\begin{figure}[!ht]
\begin{center}
\begin{tabular}{c}
\includegraphics[width=8cm,height=!]{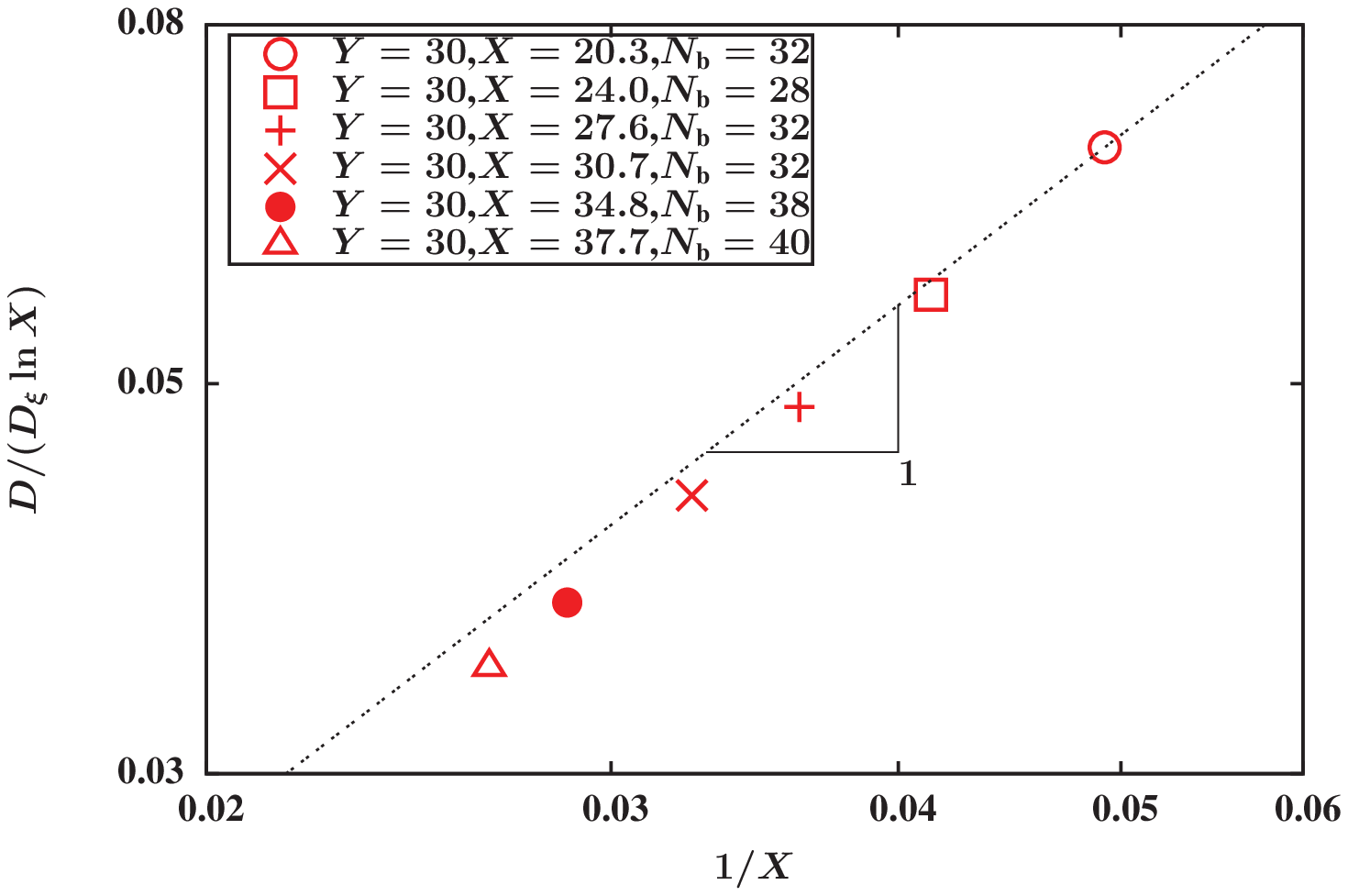}  \\ 
(a) \\
\includegraphics[width=8cm,height=!]{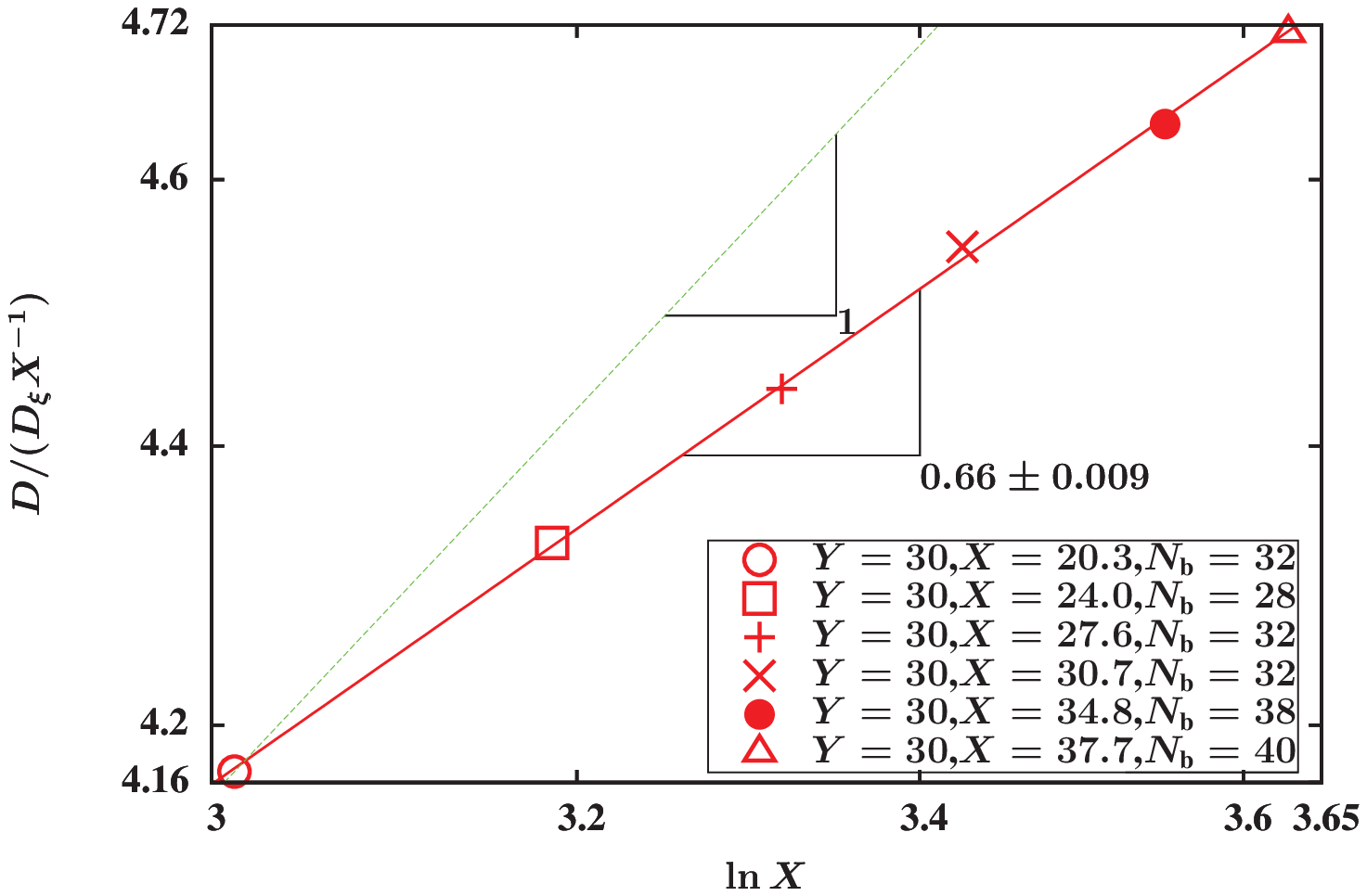} \\
(b)  \\
\end{tabular}
\end{center}
\caption{(Color online) Demonstration of the importance of accounting for logarithmic corrections to the scaling of $r_\text{e}$ with $X$ in Eq.~(\ref{eq:diffg}) for the translational diffusivity $D$ in the blob-pole regime: (a) Departure from OSFKK scaling at large values of $X$ (see Eq.~(\ref{eq:polelaw})) (b) Scaling behaviour in agreement with Eq.~(\ref{eq:polelawc}), with the expected exponent of $\ln X \approx 2/3$. Error bars are of the order of symbol size.} 
\label{fig:diffvsx}
\end{figure}

In the ideal chain regime, since $D \sim k_\text{B} T/(6 \pi \eta_\text{s}r_\text{e})$, we expect $D/D_{\xi} \sim X^{-1/2}$. In the blob-pole regime, an expression for the diffusivity can be derived by drawing an analogy with the shish-kebab model for rodlike polymers. In the latter, a rodlike  polymer of length $L$ is modelled as $N=(L/d)$ beads of diameter $d$, placed along a straight line. The translational diffusivity of such a shish-kebab can be shown to be~\cite{Doi1986},
\begin{equation}
D=\dfrac{\ln(L/d)}{3\pi \eta_\text{s} L}\, k_\text{B} T 
\end{equation}
By mapping $d \to \xi_\text{el}$, and $L \to r_\text{e}$, it follows that the diffusivity of a blob-pole is given by,
\begin{equation}
\label{eq:diffg}
D=\dfrac{\ln(X)}{3\pi \eta_\text{s} r_\text{e}}\, k_\text{B} T 
\end{equation}
Within the OSFKK scaling ansatz, since $(r_\text{e}/\xi_\text{el}) \sim X$ in the blob-pole regime, we see that,
\begin{equation}
\label{eq:polelaw}
\frac{D}{D_{\xi}} = \frac{\ln(X)}{X} 
\end{equation}
If logarithmic corrections to the scaling of $r_\text{e}$ (according to Eq.~(\ref{eq:relogX})) are taken into account, we expect,
\begin{equation}
\label{eq:polelawc}
\frac{D}{D_{\xi}}= \frac{\left[ \ln(X)\right]^{{2}/{3}}}{X} 
\end{equation}
In either case, we see that the ratio $D/D_{\xi}$ is independent of $Y$ in the both the limiting regimes of small and large $Y$. In between these two limits, we expect $D/D_{\xi}$ to depend on $Y$. 

These arguments are substantiated in the plot of $D/D_{\xi}$ versus $Y$, at two values of $X$,  displayed in Fig.~\ref{fig:DbyDxi}. We see that for each value of $X$, the ratio $D/D_{\xi}$ crosses over from a constant value in the ideal chain regime to a constant value in the blob-pole regime. As has been observed in the case of all the properties examined so far, the use of blob scaling variables to interpret the results of BD simulations leads to behaviour that is independent of the specific choice of bead-spring chain parameters. 

The importance of accounting for logarithmic corrections to the scaling of $r_\text{e}$ with $X$ in Eq.~(\ref{eq:diffg}), is studied in Figs.~\ref{fig:diffvsx}. The plot of $D/(D_{\xi} \ln X)$ versus $1/X$ in Fig.~\ref{fig:diffvsx}~(a) shows that while the scaling described by Eq.~(\ref{eq:polelaw})  is obeyed at relatively small values of $X$, there is a departure from OSFKK scaling for $X \gtrsim 30$. On the other hand, the plot of $D/(D_{\xi} X^{-1})$ as a function of $\ln X$ displayed in Fig.~\ref{fig:diffvsx}~(b) shows that Eq.~(\ref{eq:polelawc}) is indeed obeyed, with the exponent of $\ln X$ close to the expected value of $2/3$.

\begin{figure}[t]
\begin{center}
\includegraphics[width=8cm,height=!]{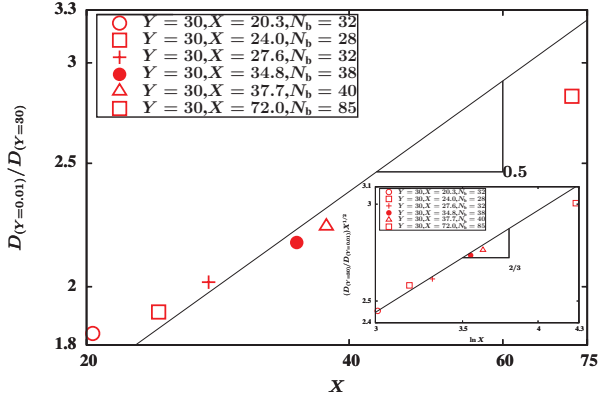}  
\end{center}
\caption{(Color online) Dependence on the number of blobs $X$, of the ratio of the translational diffusivity of the chain in the ideal chain regime ($Y=0.01$) to the diffusivity of a chain in the blob-pole regime ($Y=30$). Inset shows the importance of accounting for logarithmic corrections to scaling. Error bars are of the order of symbol size.} 
\label{fig:DvsX}
\end{figure}

A comparison of the magnitude of the diffusivity in the ideal chain and blob-pole regimes in Fig.~\ref{fig:DbyDxi}, suggests that strongly screened chains ($Y \to 0$) diffuse more than twice as fast as nearly unscreened chains ($Y \gg 1$), and that this ratio increases with $X$. This behaviour is examined in more detail in Fig.~\ref{fig:DvsX}, which displays the ratio of the diffusivity of a chain with $Y=0.01$ to that of a chain with $Y=30$, for various values of $X$. When logarithmic scaling corrections are ignored, we expect this ratio to scale as $X^{\frac{1}{2}}$. As can be seen from Fig.~\ref{fig:DvsX}, this scaling is approximately satisfied. As in other examples considered here, the inclusion of logarithmic corrections in the blob-pole regime leads to a better fit of the data, as shown in the inset to Fig.~\ref{fig:DvsX}. 

\begin{figure}[!ht]
\begin{center}
\begin{tabular}{c}
\includegraphics[width=8cm,height=!]{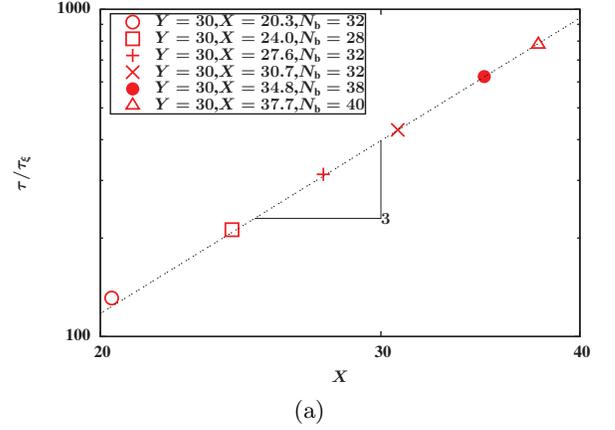}  \\ 
(a) \\
\includegraphics[width=8cm,height=!]{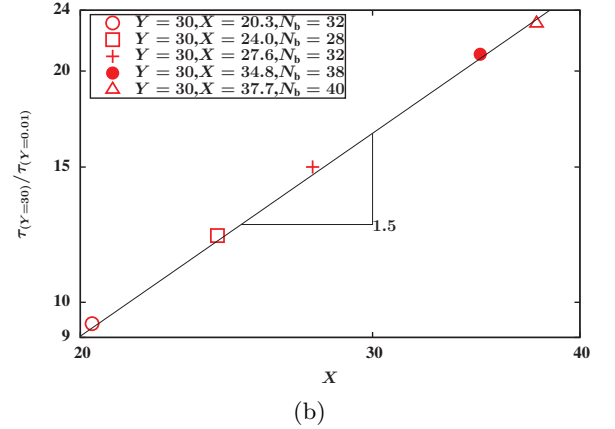} \\
(b)  \\
\end{tabular}
\end{center}
\caption{(Color online) Dependence on the number of blobs $X$, of (a) the ratio of the relaxation time of the chain in the blob-pole regime to the relaxation time of an electrostatic blob, and (b) the ratio of the relaxation time of the chain in the blob-pole regime (at $Y=30$) to the relaxation time in the ideal chain regime (at $Y=0.01$). Error bars are of the order of symbol size.} 
\label{fig:tauvsx}
\end{figure}

From the scaling of ${r_\text{e}}/{\xi_\text{el}}$ (see Eq.~(\ref{eq:relogX})) and ${D}/{D_{\xi}}$  (see Eq.~(\ref{eq:polelawc})) with $X$ in the blob-pole regime, it can be seen that the ratio, 
\[ \frac{\langle {r_\text{e}^{2}}\rangle/{\xi_\text{el}^{2}}}{D/D_{\xi}}    = \frac{\langle r_\text{e}^{2}\rangle/D}{\tau_{\xi}}\sim X^{3} \]
where, $\tau_{\xi}$, the blob relaxation time, is defined by, $\tau_{\xi} = {\xi_\text{el}^{2}}/{D_{\xi}}$. Notably, this  ratio is free from logarithmic corrections even in the blob-pole regime. Figure~\ref{fig:tauvsx}~(a) displays the dependence   on the number of blobs $X$, of the ratio of the relaxation time in the blob-pole regime to the relaxation time of an electrostatic blob, $\tau/\tau_{\xi}$. Though the relaxation time $\tau$ has been defined in terms of $\langle r_\text{g}^2\rangle$ instead of $\langle {r_\text{e}^{2}}\rangle$ (see Eq.~(\ref{eq:relax})), we see from Fig.~\ref{fig:tauvsx}~(a) that (in the range of $X$ values that have been considered), any residual logarithmic corrections are extremely weak, and that the ratio scales with $X$ with the expected power law exponent. 

The difference in the conformations of the chain in the two regimes, leads to the long-time relaxation in the blob-pole regime being nearly an order of magnitude slower than in the ideal chain regime, as displayed in Fig.~\ref{fig:tauvsx}~(b). As can be seen from the figure, the ratio of the two limiting relaxation times, $\tau_{(Y=30)}/\tau_{(Y=0.01)}$, increases with the number of blobs $X$ with the expected power law, $X^{\frac{3}{2}}$.

The ratio $U_\text{RD}$, defined by,
\begin{equation}
\label{eq:uniratio}
U_\text{RD} = \frac{r_\text{g}}{r_\text{h}} 
\end{equation}
where, ${r_\text{h}}$ is the hydrodynamic radius, ${r_\text{h}} ={k_\text{B} T}/(6\pi \eta_\text{s} D)$, has an universal value for neutral polymer solutions in the long chain limit, since both ${r_\text{g}}$ and ${r_\text{h}}$ scale identically with chain length~\cite{Zimm1956,Ottinger1996,Kroger2000}. In particular, by extrapolating finite chain data  acquired from highly accurate BD simulations, to the long chain limit, ~\citet{Sunthar2006}  have shown that for $\theta$-solutions, $U_\text{RD}^{\theta} =1.38 \pm 0.01$. We anticipate that for the polyelectrolyte solutions considered here, $U_\text{RD}$ will have a similar value in the ideal chain regime. 

\begin{figure}[t]
\begin{center}
\begin{tabular}{c}
\includegraphics[width=8cm,height=!]{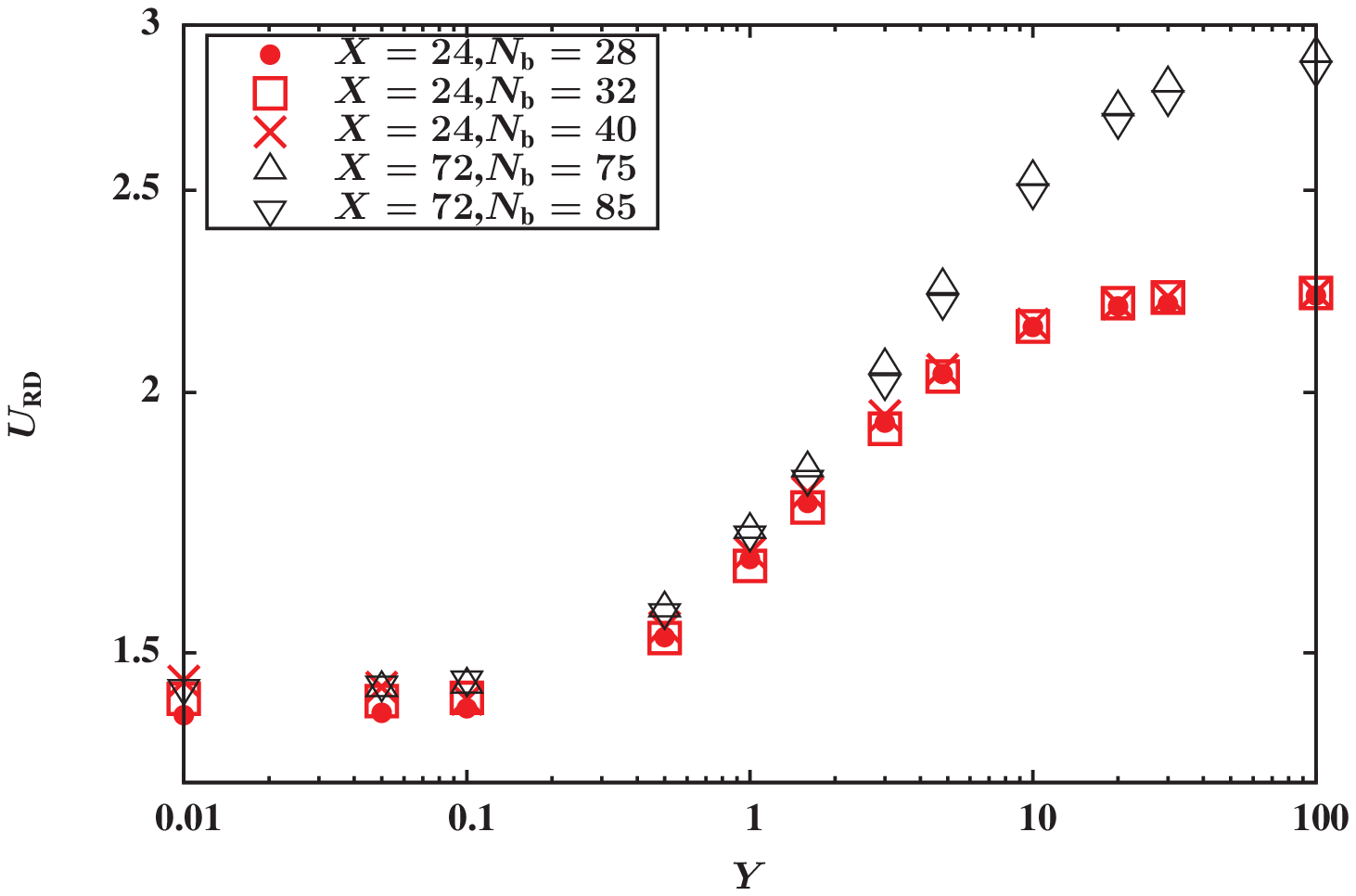}  \\ 
(a) \\
\includegraphics[width=8cm,height=!]{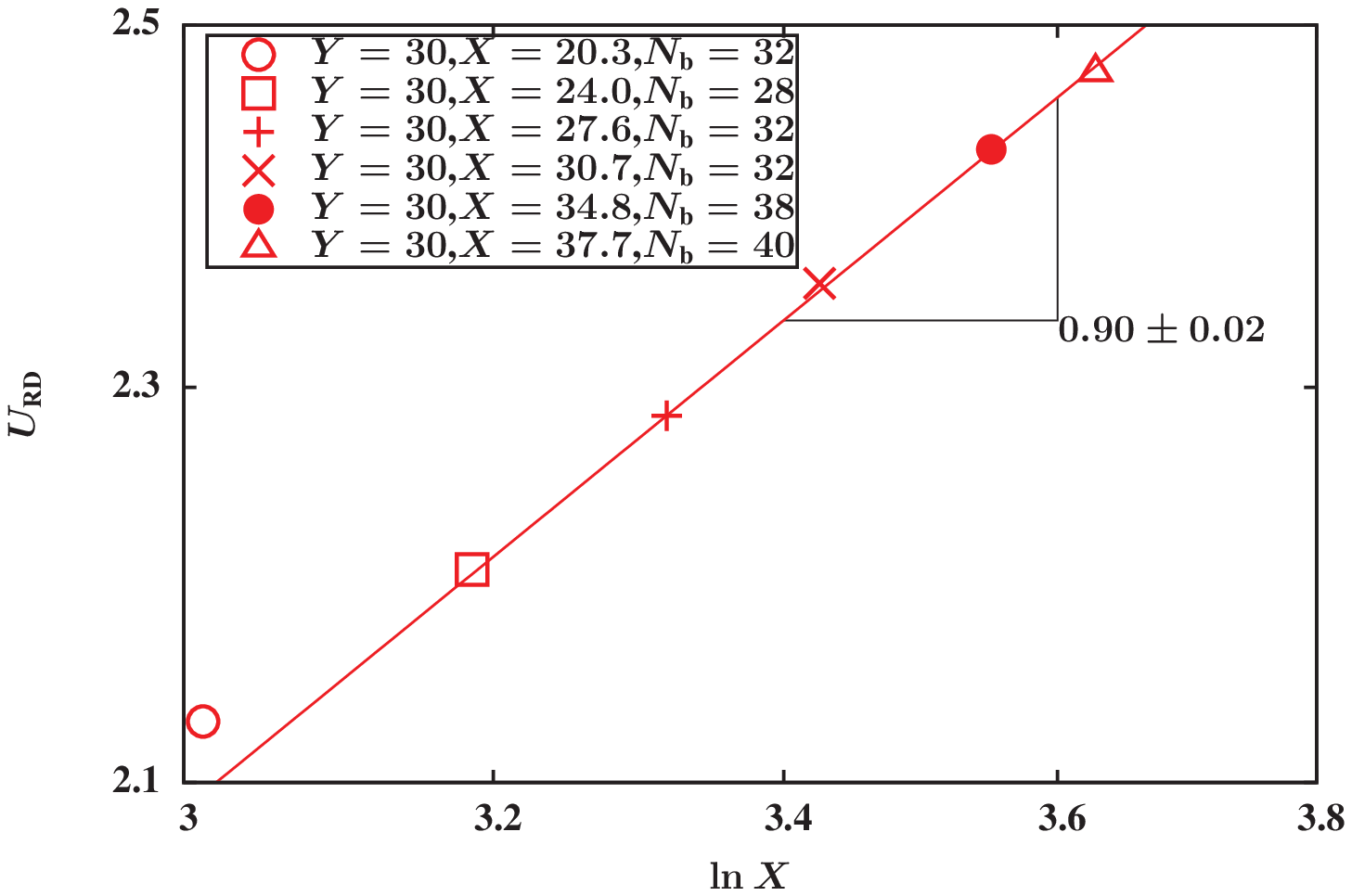} \\
(b)  \\
\end{tabular}
\end{center}
\caption{(Color online) Ratio of the radius of gyration to the hydrodynamic radius, $U_\text{RD}$ (see Eq.~(\ref{eq:uniratio})), as a function of (a) the reduced screening length $Y$, at two different values of the number of blobs $X$, and (b) demonstration of logarithmic corrections to scaling in the blob-pole regime, at a constant value of the reduced screening length, $Y=30$. Error bars are of the order of symbol size.} 
\label{fig:uniratio}
\end{figure}

In the blob-pole regime, by substituting for $D$ from Eq.~(\ref{eq:diffg}) into Eq.~(\ref{eq:uniratio}), we see that,
\begin{equation}
\label{eq:uniratiobp}
U_\text{RD}  \sim \left(\frac{r_\text{g}}{r_\text{e}} \right) \ln X
\end{equation}
As a result,  $U_\text{RD}$ does not have a universal value in this regime, but rather depends on the number of blobs $X$. In the context of the current simulations, Figs.~\ref{fig:rvsx}~(b) and~\ref{fig:rvsx2}~(b) show that the scaling expressions, $r_\text{e}/(\xi_\text{el} X) \sim (\ln X)^{0.36}$, and $r_\text{g}/(\xi_\text{el} X) \sim (\ln X)^{0.22}$, respectively, provide good fits to the simulation data. As a result, we expect from~Eq.~(\ref{eq:uniratiobp}) that,~$U_\text{RD} \sim  (\ln X)^{0.86}$.

Figure~\ref{fig:uniratio}~(a) displays the dependence of $U_\text{RD}$ on the reduced screening length $Y$, for two different values of $X$. We see that for small values of $Y$, the ratio has a constant universal value which is close to that for neutral chains. The ratio increases, while remaining universal (i.e., independent of $X$) for increasing values of $Y$ in the crossover regime, before levelling off to a non-universal value in the blob-pole regime. The dependence of $U_\text{RD}$ on $\ln X$ in the blob-pole regime, at the particular value $Y=30$, is displayed in Fig.~\ref{fig:uniratio}~(b). We see that the exponent of $\ln X$ is indeed close to the expected value of $0.86$. 

\section{\label{sec:conc}Conclusions}

The size, shape and diffusivity of a weakly-charged polyelectrolyte chain in solution have been examined in the limit of low polymer concentration using Brownian dynamics simulations of a coarse-grained bead-spring chain model, with Debye--H\"{u}ckel electrostatic interactions between the beads. Simulation results have been recast in terms of the scaling variables $X$ (the number of electrostatic blobs), and $Y$ (the reduced screening length), which are defined within the framework of the OSFKK blob scaling theory. While the root-mean-square end-to-end vector and radius of gyration have been used to represent the mean size of the chain, various functions defined in terms of the eigenvalues of the radius of gyration tensor have been used to characterise chain shape in all parts of the $\{X, Y\}$ phase-space. The translational diffusion coefficient of the chain, which is a dynamic property, has been determined accurately from the displacement of the chain centre of mass, by taking hydrodynamic interactions into account through incorporation of the Rotne-Prager-Yamakawa tensor into the BD algorithm. 

The key results of the present work are summarised in the list below:
\begin{enumerate}
\item Interpretation of simulation results in terms of blob scaling variables $X$ and $Y$ leads to a description of solution behaviour independent of the level of coarse-graining, i.e., of the choice of the number of beads $N_\text{b}$, and other parameters in the  bead-spring chain model. This should prove extremely useful for comparing simulation results with experiments. The procedure for mapping from experimental variables to blob scaling variables, in order to facilitate this comparison, has been provided in Appendix~\ref{sec:AppB}.
\item Three broad domains of behaviour can be clearly identified: (i) the ideal chain regime corresponding to small values of $Y$, (ii) the crossover regime corresponding to intermediate values of $Y$, and (iii) the blob-pole regime corresponding to large values of $Y$.
\item In the blob-pole regime, BD simulations appear to pick up the existence of logarithmic corrections for fairly short chains, which are in good agreement with predictions of refined scaling theories. 
\item Various universal ratios of eigenvalues, the asphericity, the acylindricity, the degree of prolateness and the shape anisotropy of polyelectrolyte chains in the ideal chain regime, have been compared with published results in the literature for flexible neutral random walk polymers. 
\item When suitably scaled, the size, shape and diffusivity of a chain are independent of the reduced screening length $Y$ in the ideal chain and blob-pole regimes, and depend on $Y$ only in the crossover regime.
\item The translational diffusivity of the chain in the blob-pole regime can be described by drawing an analogy with the translational diffusivity of a rodlike polymer modelled as a shish-kebab. The normalisation parameter that enables the collapse of BD data for the translational diffusivity is the Zimm diffusivity of the electrostatic blob. 
\item All properties of polyelectrolyte solutions, when suitably normalised, appear to exhibit universal behaviour, independent of the number of electrostatic blobs $X$ in the chain, in all regimes of the $\{X, Y\}$ phase-space, except in the blob-pole regime, where the occurrence of logarithmic corrections to scaling leads to non-universal behaviour.
\end{enumerate}
\section{\label{sec:Ak}Acknowledgements}

The authors gratefully acknowledge CPU time grants from the National Computational Infrastructure (NCI) facility hosted by the Australian National University, and Victorian Life Sciences Computation Initiative (VLSCI) hosted by the University of Melbourne.

\appendix 

\section{\label{sec:App} Scaling of the end-to-end vector in the blob-pole regime}

The scaling of the end-to-end vector in the blob-pole regime can be derived from a Flory type energy minimisation argument. The essential assumption is that chains adopt an ellipsoidal shape, with electrostatic interactions causing stretching in the direction of the major-axis, while leaving chain conformations in the direction of the minor-axes unaffected. As a result, the long half-axis of the ellipsoid scales as $r_\text{e}/2$, while the aspect ratio scales as $r_\text{e}/(b_\text{K}N_\text{K}^{1/2})$. The total energy of such a chain can be obtained by combining the electrostatic energy of an ellipsoid with the elastic energy of a stretched chain. Ignoring prefactors of order unity, the electrostatic energy of an ellipsoid is~\cite{Dobrynin2005},
\[ k_\text{B}T \, \frac{l_\text{B} \left( f N_\text{K} \right)^2}{r_\text{e}}\ln \left(\frac{r_\text{e}}{b_\text{K}N_\text{K}^\frac{1}{2}} \right) \]
while the elastic energy of a stretched chain is~\cite{degennes},
\[ k_\text{B}T \,   \left( \frac{r_\text{e}^2}{b_\text{K}^2N_\text{K}} \right) \]
From Flory theory, it follows that the equilibrium end-to-end vector can be derived by minimising the total energy,
\begin{equation}
\label{eq:et}
\frac{U}{ k_\text{B}T}   \sim  \left( \frac{r_\text{e}^2}{b_\text{K}^2N_\text{K}} \right) +  \, \frac{l_\text{B} \left( f N_\text{K} \right)^2}{r_\text{e}}\ln \left(\frac{r_\text{e}}{b_\text{K}N_\text{K}^\frac{1}{2}} \right) 
\end{equation}
It is convenient to proceed in two steps---first deriving the equilibrium end-to-end vector by neglecting the logarithmic term in the electrostatic energy, and then accounting for it in the next step. Minimising the sum,
\[  \left( \frac{r_\text{e}^2}{b_\text{K}^2N_\text{K}} \right) +  \frac{l_\text{B} \left( f N_\text{K} \right)^2}{r_\text{e}} \]
with respect to ${r_\text{e}}$, leads to,
\begin{equation}
\label{eq:rea}
r_\text{e}=f^\frac{2}{3} \, {\hat l}_\text{B}^\frac{1}{3}\, N_\text{K}   \, b_\text{K} 
\end{equation}
This is identical to the expression derived by OSFKK theory for the end-to-end vector in the blob-pole regime (see Table~\ref{table1}). We now assume that including the logarithmic term in the electrostatic energy leads to a modification of the equilibrium end-to-end vector,
\begin{equation}
\label{eq:reax}
r_\text{e}=f^\frac{2}{3} \, {\hat l}_\text{B}^\frac{1}{3}\, N_\text{K}   \, b_\text{K}   \, x
\end{equation}
where, the factor $x$ remains to be determined. Substituting the modified expression for $r_\text{e}$ from Eq.~(\ref{eq:reax}) into Eq.~(\ref{eq:et}) (and absorbing the common factor $N_\text{K} (f^2 \, {\hat l}_\text{B})^{\frac{2}{3}}$ into the energy), leads to the following sum,
\[ x^{2} + \frac{1}{x} \, \ln \left[ N_\text{K}^\frac{1}{2} \,  f^\frac{2}{3} \, {\hat l}_\text{B}^\frac{1}{3} \, x \right] \]
which must be minimised with respect to $x$ in order to determine $x$. This leads to,
\begin{equation}
\label{eq:xlogaln}
x^3 = \ln \left[ N_\text{K}^\frac{1}{2} \,  f^\frac{2}{3} \, {\hat l}_\text{B}^\frac{1}{3}  \right]+\ln x - 1  
\end{equation}
Assuming $x \gg 1$, implies,
\begin{equation}
\label{eq:xloga}
x =   \left\lbrace \frac{1}{2} \ln \left[ N_\text{K} \,  f^\frac{4}{3} \,  {\hat l}_\text{B}^\frac{2}{3} \right] \right\rbrace^{\frac{1}{3}}
\end{equation}
where, the factor $(1/2)$ in front of the logarithmic term has been introduced for future convenience. The assumption that $x$ is large can be seen to be justified for $N_\text{K} \gg 1$.

Substituting the expression for $x$ from Eq.~(\ref{eq:xloga}) into Eq.~(\ref{eq:reax}), shows that this two-step procedure leads to a prediction of the equilibrium end-to-end vector scaling with chain size according to,
\begin{equation}
\label{eq:reloga}
\frac{r_\text{e}}{b_\text{K}} \sim f^\frac{2}{3} \, {\hat l}_\text{B}^\frac{1}{3} \, N_\text{K} \left\lbrace  \ln \left[ N_\text{K} \,  f^\frac{4}{3} \,  {\hat l}_\text{B}^\frac{2}{3} \right] \right\rbrace^{\frac{1}{3}}
\end{equation}
Clearly, the logarithmic correction to the OSFKK scaling expression in the blob-pole regime  comes from taking the logarithmic term in the expression for the electrostatic energy of an ellipsoid into account. Equation~(\ref{eq:reloga}) is similar to the expression derived previously by\ ~\citet{Dobrynin2005} \!\!\!\!\!\!\!\!{.} \, This expression takes a particularly simple form in terms of blob scaling variables (see Eqs.~(\ref{eq:xi}) and~(\ref{eq:x})),
\begin{equation}
\frac{r_\text{e}}{\xi_\text{el}} \sim X \left[  \ln X \right]^{\frac{1}{3}}
\end{equation}

\begin{table*}[t]
\caption{\label{tableB}  Mapping experimental data for sodium poly(styrene sulfonate) to the blob model}\vskip10pt
\centering       
{\footnotesize
\begin{tabular}{c | c | c | c | c | c | c | c | c }    
\hline
\hline        
\multicolumn{9}{c}{$M=20600$ ($N_\text{m} = 100, N_\text{K} = 10$) } \\
\hline  
  \multicolumn{6}{c}{Solution properties}  \vline&  \multicolumn{3}{c}{Blob variables} \\
\hline  
$\alpha \, (f)$ & $c_\text{s}^\text{m}$  \,  (I) (mol/ltr) & $T \, ^{\circ}$C  & $l_\text{B}$ (nm) & $l_\text{D}$  (nm)   &  $\gamma_{0}$ &  $\xi_\text{el}$  (nm) & $X$ & $Y$ \\ 
\hline          
 \multirow{9}{*}{0.01 (0.1)} &   \multirow{3}{*}{0 ($5 \times 10^{-9}$) }  
 & 15 &  0.7077 & 4321.0  & 0.0283 & 17.67 & 0.2001 & 244.5084 \\
 \cline{3-9}
 & & 25 &   0.7158  & 4297.0  & 0.0286 & 17.61 & 0.2016 &  244.0463 \\
  \cline{3-9}
  & & 35 &  0.7247  & 4270.0  & 0.0290 & 17.53 & 0.2033 &  243.5422 \\
  \cline{2-9}
   &  \multirow{3}{*}{0.01 (0.01)}  
 & 15 &  0.7077 & 3.056  & 0.0283 & 17.67 & 0.2001 & 0.1729 \\
 \cline{3-9}
 & & 25 &   0.7158  & 3.038  & 0.0286 & 17.61 & 0.2016 & 0.1726 \\
  \cline{3-9}
  & & 35 &  0.7247  & 3.019  & 0.0290 & 17.53 & 0.2033 & 0.1722 \\
    \cline{2-9}
  &  \multirow{3}{*}{0.5 (0.5)}  
 & 15 &  0.7077 & 0.4321  & 0.0283 & 17.67 & 0.2001 & 0.0245 \\
 \cline{3-9}
 & & 25 &   0.7158  & 0.4297  & 0.0286 & 17.61 & 0.2016 & 0.0244 \\
  \cline{3-9}
  & & 35 &  0.7247  & 0.4270  & 0.0290 & 17.53 & 0.2033 & 0.0244 \\
\hline     
 \multirow{9}{*}{0.3 (3.0) } &   \multirow{3}{*}{0 ($1.5 \times 10^{-7}$)}  
 & 15 &  0.7077 & 788.9  & 0.8492 & 1.830 & 18.6534 & 431.0033 \\
 \cline{3-9}
 & & 25 &   0.7158  & 784.5  & 0.8589 & 1.824  & 18.7951 & 430.1889 \\
  \cline{3-9}
  & & 35 &  0.7247  & 779.6  & 0.8696 & 1.816  & 18.9512 &  429.3002 \\
  \cline{2-9}
   &  \multirow{3}{*}{0.01 (0.01)}  
 & 15 &  0.7077 & 3.056  & 0.8492 & 1.830 & 18.6534 & 1.6693 \\
 \cline{3-9}
 & & 25 &   0.7158  & 3.038  & 0.8589 & 1.824  & 18.7951 & 1.6661 \\
  \cline{3-9}
  & & 35 &  0.7247  & 3.019  & 0.8696 & 1.816  & 18.9512 &  1.6627 \\
  \cline{2-9}
  &  \multirow{3}{*}{0.5 (0.5)}  
 & 15 &  0.7077 & 0.4321 & 0.8492 & 1.830 & 18.6534 & 0.2361 \\
 \cline{3-9}
 & & 25 &   0.7158  & 0.4297  & 0.8589 & 1.824  & 18.7951 & 0.2356 \\
  \cline{3-9}
  & & 35 &  0.7247  & 0.4270  & 0.8696 & 1.816  & 18.9512 &  0.2351 \\
\hline       
\hline         
\multicolumn{9}{c}{$M=206000$ ($N_\text{m} = 1000, N_\text{K} = 100$) } \\
\hline  
$\alpha \, (f)$ & $c_\text{s}^\text{m}$  \,  (I) (mol/ltr) & $T \, ^{\circ}$C  & $l_\text{B}$ (nm) & $l_\text{D}$  (nm)   &  $\gamma_{0}$ &  $\xi_\text{el}$  (nm) & $X$ & $Y$ \\ 
\hline     
 \multirow{9}{*}{0.1 (1.0)} &   \multirow{3}{*}{$4 \times 10^{-6}$ ($4.05 \times 10^{-6}$) }  
 & 15 &  0.7077 & 151.8  & 0.2831 & 3.808 & 43.1119 & 39.8766 \\
 \cline{3-9}
 & & 25 &   0.7158  & 151.0  & 0.2863 & 3.793 & 43.4393 &  39.8012 \\
  \cline{3-9}
  & & 35 &  0.7247  & 150.0  & 0.2899 & 3.777 & 43.8001 &  39.7190 \\
  \cline{2-9}
   &  \multirow{3}{*}{0.1 (0.1)}  
& 15 &  0.7077 & 0.9662  & 0.2831 & 3.808 & 43.1119 & 0.2538 \\
 \cline{3-9}
 & & 25 &   0.7158  & 0.9608  & 0.2863 & 3.793 & 43.4393 &  0.2533 \\
  \cline{3-9}
  & & 35 &  0.7247  & 0.9548  & 0.2899 & 3.777 & 43.8001 &   0.2528 \\
    \cline{2-9}
  &  \multirow{3}{*}{0.8 (0.8)}  
& 15 &  0.7077 & 0.3416  & 0.2831 & 3.808 & 43.1119 & 0.0897 \\
 \cline{3-9}
 & & 25 &   0.7158  & 0.3397  & 0.2863 & 3.793 & 43.4393 &  0.0896 \\
  \cline{3-9}
  & & 35 &  0.7247  & 0.3376  & 0.2899 & 3.777 & 43.8001 &  0.0894 \\
\hline     
 \multirow{9}{*}{0.35 (3.5) } &   \multirow{3}{*}{$4 \times 10^{-6}$ ($4.175 \times 10^{-6}$) }  
 & 15 &  0.7077 & 149.5  & 0.9907 & 1.652 & 229.0977 & 90.5376 \\
 \cline{3-9}
 & & 25 &   0.7158  & 148.7  & 1.0021 & 1.645 & 230.8376 &  90.3666 \\
  \cline{3-9}
  & & 35 &  0.7247  & 147.8  & 1.0146 & 1.639 & 232.7550 &  90.1799 \\
  \cline{2-9}
   &  \multirow{3}{*}{0.1 (0.1)}  
& 15 &  0.7077 & 0.9662  & 0.9907 & 1.652 & 229.0977 & 0.5850 \\
 \cline{3-9}
 & & 25 &   0.7158  & 0.9608  & 1.0021 & 1.645 & 230.8376 &  0.5839 \\
  \cline{3-9}
  & & 35 &  0.7247  & 0.9548  & 1.0146 & 1.639 & 232.7550 &  0.5827 \\
    \cline{2-9}
  &  \multirow{3}{*}{0.8 (0.8)}  
& 15 &  0.7077 & 0.3416  & 0.9907 & 1.652 & 229.0977 & 0.2068 \\
 \cline{3-9}
 & & 25 &   0.7158  & 0.3397  & 1.0021 & 1.645 & 230.8376 &  0.2064 \\
  \cline{3-9}
  & & 35 &  0.7247  & 0.3376 & 1.0146 & 1.639 & 232.7550 &  0.2060 \\
\hline    
\hline                                
\end{tabular}
}
\end{table*}


\section{\label{sec:AppB} Mapping experiments to blob variables for NaPSS}

Consider a polyelectrolyte chain of molecular weight $M$, dissolved in a solvent at temperature $T$, with a temperature dependent relative permittivity $\varepsilon_\text{r} (T)$. If the monomer concentration in molar units is $c_\text{p}^\text{m}$, and the degree of ionization per chain is $\alpha$, then the number of counterions in solution per unit volume is, $\alpha \, c_\text{p}^\text{m} (N_\text{A} 1000)$, where $N_\text{A}$ is Avagadro's number. If the added salt \ce{A_{m}B_{n}}, with molar concentration $c_\text{s}^\text{m}$,   dissociates in solution according to,
\[ \ce{ A_{m}B_{n} -> m A^z+ + n B^z- } \]
where, $z_{+}$ and $z_{-}$ are the cation and anion valences, respectively, then the number of cations  \ce{A} per unit volume in solution is $c_{+} = m \, c_\text{s}^\text{m} (N_\text{A} 1000)$, and the number of anions  \ce{B} per unit volume is $c_{-} = n \, c_\text{s}^\text{m} (N_\text{A} 1000)$. It follows that the ionic strength $I$ of the solution is given by,
\begin{equation}
\label{eq:ionics}
I = \frac{1}{2} \left( m z_{+}^{2} + n z_{-}^{2} \right) c_\text{s}^\text{m} (N_\text{A} 1000) + \frac{1}{2} \, z_{p}^{2} \, \alpha \, c_\text{p}^\text{m} (N_\text{A} 1000) 
\end{equation}
where, $z_\text{p}$ is the counterion valence. 

The Bjerrum length  $l_\text{B}$ can be calculated from Eq.~(\ref{eq:lb}), while the Debye length is given by,
\begin{equation}
\label{eq:ld}
l_\text{D} = \left( 8 \pi \, l_\text{B} \, I \right)^{-\frac{1}{2}}
\end{equation}

Mapping the experimental system onto blob variables is straightforward once it is mapped onto the bare-model parameters $\{ N_\text{K}, f, {\hat l}_\text{B}, {\hat l}_\text{D} \}$. In order to do so, it is necessary to know the intrinsic persistence length of the polyelectrolyte chain, $\ell_{0}$, the molecular weight of a monomer, $M_\text{m}$, and the monomer length $b_\text{m}$. With this information, we can find, (i) the Kuhn step length, $b_\text{K} = 2 \ell_{0}$, (ii) the number of monomers in a chain, $N_\text{m} = M/ M_\text{m}$, (iii) the number of Kuhn steps, $N_\text{K} = (N_\text{m} b_\text{m}) / b_\text{K}$, and (iv) the number of charges per Kuhn step, $f = \alpha N_\text{m,K}$, where, $N_\text{m,K} =  b_\text{K}/ b_\text{m}$,  is the number of monomers per Kuhn step. It is also straight forward to find ${\hat l}_\text{B} = l_\text{B}/b_\text{K}$, and ${\hat l}_\text{D} = l_\text{D}/b_\text{K}$, at the given temperature and ionic strength. Once the bare-model parameters are known, the blob scaling variables, $\{X, Y, {\hat \xi}_\text{el} \}$ can be determined from Eqs.~(\ref{eq:x})--(\ref{eq:xi}).

Table~\ref{tableB} displays the blob scaling variables calculated with this procedure for sodium poly(styrene sulfonate) (NaPSS) at two different molecular weights, $M=20600$ Dalton and $M=206000$ Dalton. For NaPSS, the molecular weight of a monomer, $M_\text{m} =206$ Dalton, and the monomer length $b_\text{m}=2.5$\AA~\cite{Degiorgio1995}. The intrinsic persistence length of NaPSS is variously estimated~\cite{Degiorgio1995,Nishida2001a,Nishida2001b} as lying in the range, 9\AA $\, \leq \ell_{0} \leq \, $14\AA. For the purposes of this illustration, we choose a value, $\ell_{0} = 12.5$\AA, in order to get an integer number of monomers per Kuhn step. Similarly, a fixed value of $c_\text{p}^\text{m} = 10^{-6}$ mol/ltr is chosen as the monomer concentration, since at this concentration, according to the phase diagram in Ref.~\citenum{Nishida2001a}, the solution remains in the dilute regime for all the values of $N_\text{m}$ considered here. The degree of ionization (sulfonation) per chain for NaPSS is commonly assumed to be $\alpha =0.35$~\cite{Degiorgio1995,Nishida2001a,Nishida2001b}. Here, we choose a range of values between 0.01 and 0.35 in order to generate a range of values of the blob scaling variables $X$ and $Y$. For water~\cite{Malmberg1956}, $\varepsilon_\text{r} (t) = 87.740 - 0.40008 \, t +9.398 \times 10^{-4} \, t^{2} -1.410 \times 10^{-6} \, t^{3}$, where, $0^{\circ} \text{C} \leq t   \leq 100^{\circ} \text{C}$. We assume monovalent cations and anions for the added salt, with composition stochiometry $m=1$, and $n=1$. The salt concentration $c_\text{s}^\text{m}$ is varied in order to change the ionic strength, and consequently, the Debye length. Finally, three values of temperature are chosen that bracket typical room temperature values.

As mentioned earlier, in the current simulation, blob variable values in the range, $20 < X < 80$ and $0.01 \leq Y \leq 100$, have been considered. It is clear from Table~\ref{tableB}, that by a suitable choice of experimental parameters, it is possible to span this range of values of $X$ and $Y$, while maintaining the Manning parameter, $\gamma_{0} \lessapprox 1$. 

\bibliography{jcp}

\end{document}